\documentclass[11pt,english]{article}
\usepackage{amsmath,amssymb,amsthm}
\usepackage{multicol}
\usepackage[margin=1in]{geometry}
\usepackage{graphicx,color}
\usepackage{enumitem}
\usepackage{fullpage}
\usepackage[noblocks]{authblk}
\usepackage{tcolorbox}
\usepackage{babel}
\usepackage{wrapfig}
\usepackage{MnSymbol}
\usepackage{mdframed}

\usepackage[ruled,linesnumbered,vlined]{algorithm2e}
\usepackage{tablefootnote}
\usepackage{thm-restate}
\usepackage{bbding}
\usepackage{pifont}
\usepackage{nicefrac}

\definecolor{Darkblue}{rgb}{0,0,0.4}
\definecolor{Brown}{cmyk}{0,0.61,1.,0.60}
\definecolor{Purple}{cmyk}{0.45,0.86,0,0}
\definecolor{Darkgreen}{rgb}{0.133,0.543,0.133}

\usepackage[colorlinks,linkcolor=Darkblue,filecolor=blue,citecolor=blue,urlcolor=Darkblue,pagebackref]{hyperref}
\usepackage[nameinlink]{cleveref}

\usepackage[colorinlistoftodos,prependcaption,textsize=tiny]{todonotes}

\newif\ifdraft 
\draftfalse
\newcommand{\commentforlater}[1]{\ifdraft\todo[linecolor=blue,backgroundcolor=red!25,bordercolor=blue]{\textbf{VCA's comment for later:~} #1 }\fi}

\newtheorem{theorem}{Theorem}
\newtheorem{lemma}{Lemma}
\newtheorem{definition}{Definition}
\newtheorem{claim}{Claim}
\newtheorem{observation}{Observation}
\newtheorem{corollary}{Corollary}

\newtheorem{question}{Question}
\newtheorem{invariant}{Invariant}

\newcommand{\dm}{\mathrm{\textsc{Diam}}}

\newcommand{\poly}{\mathrm{poly}}
\newcommand{\polylog}{\mathrm{polylog}}

\newcommand{\tw}{\mathrm{tw}}
\newcommand{\pw}{\mathrm{pw}}

\newcommand{\R}{\mathbb{R}}
\newcommand{\N}{\mathbb{N}}
\newcommand{\T}{\mathcal{T}}

\newcommand{\la}{\leftarrow}
\newcommand{\ra}{\rightarrow}
\newcommand{\ssp}{\textsc{SSP}}
\newcommand{\im}{\textsc{Im}}

\newcommand{\BS}{\mathtt{BS}}

\newcommand{\cut}{\text{ \ding{34} }}
\newcommand{\Pt}{\mathtt{Pt}}
\newcommand{\supp}{\mathrm{supp}}

\newcommand{\calE}{\mathcal{E}}

\def\eps{\epsilon}

\DeclareMathAlphabet{\mathpzc}{OT1}{pzc}{m}{it}
\newcommand{\etal}{{et al. \xspace}}

\newlength{\dhatheight}

\newcommand {\ignore} [1] {}

\newcommand{\initOneLiners}{%
	\setlength{\itemsep}{0pt}
	\setlength{\parsep }{0pt}
	\setlength{\topsep }{0pt}
}
\newenvironment{OneLiners}[1][\ensuremath{\bullet}]
{\begin{list}
		{#1}
		{\initOneLiners}}
	{\end{list}}

\title{On Light Spanners, Low-treewidth Embeddings and Efficient Traversing in Minor-free Graphs}
\author{Vincent Cohen-Addad}
\affil{Google Research}
\author{Arnold Filtser}
\affil{Columbia University}
\author{Philip N. Klein}
\affil{Brown University}
\author{Hung Le}
\affil{University of Massachusetts at Amherst}

\date{}
\begin{document}
\maketitle
\begin{abstract}
	Understanding the structure of minor-free metrics, namely shortest path metrics obtained
	over a weighted graph excluding a fixed minor, has been an important research direction
	since the fundamental work of Robertson and Seymour.
	A fundamental idea that helps both to understand the structural properties of these metrics
	and lead to strong algorithmic results is to construct a ``small-complexity''
	graph that approximately preserves distances between pairs of points of the metric. We show the two following structural results for minor-free metrics:
	
	\begin{enumerate}
		\item Construction of a \emph{light} subset spanner. Given a  subset of vertices called terminals, and $\eps$,  in polynomial time we construct a subgraph that
		preserves all pairwise distances between terminals up to a multiplicative $1+\eps$ factor, of total weight at most $O_{\eps}(1)$ times the weight of the minimal Steiner tree spanning the terminals.
		\item Construction of a stochastic metric embedding into low treewidth graphs with expected additive distortion $\eps D$. 
		Namely, given a minor free graph $G=(V,E,w)$ of diameter $D$, and parameter $\eps$, we construct a distribution $\mathcal{D}$ over dominating metric embeddings into  treewidth-$O_{\eps}(\log n)$ graphs such that $\forall u,v\in V$, $\mathbb{E}_{f\sim\mathcal{D}}[d_H(f(u),f(v))]\le d_G(u,v)+\eps D$. 
	\end{enumerate}
	One of our important technical contributions is a novel framework that allows us to reduce \emph{both problems} to problems on simpler graphs of \emph{bounded diameter} that we solve using a new decomposition. Our results have the following algorithmic consequences: (1) the first efficient approximation scheme for subset TSP in minor-free metrics; (2) the first approximation scheme for vehicle routing with bounded capacity in minor-free metrics; (3) the first efficient approximation scheme for vehicle routing with bounded capacity on bounded genus metrics.	En route to the latter result, we design the first FPT approximation scheme for vehicle routing with
	bounded capacity on bounded treewidth graphs (parameterized by the treewidth). 
\end{abstract}

\newpage

\vfill
    {\small \setcounter{tocdepth}{2} \tableofcontents}
    \newpage
    \pagenumbering{arabic}

\section{Introduction}

Fundamental routing problems such as the Traveling Salesman Problem (TSP) and the Vehicle Routing Problem have been widely studied since the 50s.
Given a metric space, the goal is to find a minimum-weight collection of tours (only one for TSP) so as to meet a prescribed demand at some points of
the metric space. The research on these problems, from both practical and theoretical perspectives, has been part of the agenda of the operations research
and algorithm-design communities for many decades (see e.g.:  \cite{HR85,AKTT97,CFN85,Salazar03,BMT13,ZTXL15,LN15,ZTXL16}).
Both problems have been the source of inspiration for many algorithmic breakthroughs and, quite frustratingly, remain good examples of the limits of the
power of our algorithmic methods.

Since both problems are APX-hard in general graphs \cite{PY93,AKTT97}
and since the best known approximation for TSP remains the 40-year old
$\frac{3}{2}$-approximation of Christofides \cite{Chr76}, it has been
a natural and successful research direction to focus on
\emph{structured} metric spaces. Initially, researchers focused on achieving polynomial-time approximation schemes (PTASs) for  TSP
in planar-graphs~\cite{GKP95,AGKKW98} and Euclidean metrics~\cite{Arora97,Mitchell99}. Two themes emerged in the ensuing research: speed-ups and
generalization. 

 In the area of speed-ups, a long line of research on Euclidean TSP improved the running time  $n^{O(1/\epsilon)}$ of the innitial algorithm by Arora to linear time~\cite{BartalG13}.  In a parallel research thread,  Klein \cite{Klein05,Klein08} gave the
 first \emph{efficient PTAS}\footnote{A PTAS is an
 	\emph{efficient} PTAS (an EPTAS) if its running time is bounded by a
 	polynomial $n^c$ whose degree $c$ does not depend on $\eps$} for TSP in weighted planar graphs, a linear-time
 algorithm. 
  
 In the area of generalization, a key question was whether these results
 applied to more general (and more abstract) families of metrics.  One
 such generalization of Euclidean metrics is metrics of bounded
 doubling dimension. Talwar~\cite{Tal04} gave a \emph{quasi}-polynomial-time approximation scheme (QPTAS)  for this
 problem which was then improved to an EPTAS~\cite{Gottlieb15}. In minor-free metrics, an important generalization of planar metrics, Grigni~\cite{Grigni00} gave a QPTAS for TSP which was recently improved to EPTAS by Borradaile et al.~\cite{BLW17}.

 When the metric is that of a planar/minor-free graph, the problem of visiting
 every vertex is not as natural as that of visiting a given subset of
 vertices (the \emph{Steiner TSP} or \emph{subset TSP}) since the
 latter cannot be reduced to the former without destroying the graph
 structure. The latter problem turns out to be much harder than TSP in minor-free graphs, and
 in fact no approximation scheme was known until the recent PTAS for subset TSP by Le~\cite{Le20}.
 This immediately raises the question:
 
 \begin{question}\label{question:subsetTSP}
 	Is there an EPTAS for subset TSP in minor-free graphs?
 \end{question}

The purpose of this line of work is to understand what are the most general metrics
for which we can obtain approximation schemes for routing problems, and when it is the case
how fast can the approximation schemes be made.
Toward this goal, minor-free metrics have been a testbed of choice for generalizing
the algorithmic techniques designed for planar or bounded-genus graphs. Indeed, while minor-free metrics
offer very structured decompositions, as shown by the celebrated work of Robertson and
Seymour~\cite{RS03}, Klein et al.~\cite{KPR93}, and Abraham et al.~\cite{AGGNT19}
(see also~\cite{FT03,Fil19padded}), they do not exhibit a strong topological structure.
Hence, various strong results for planar metrics, such as the efficient approximation schemes
for Steiner Tree~\cite{BKM09} or Subset TSP~\cite{Klein06}, are not known to exist in minor-free metrics.

\begin{wrapfigure}{r}{0.5\textwidth}
		\vspace{-10pt}
		\resizebox{0.5\columnwidth}{!}{%
		\begin{tabular}{|l|l|l|l|}
			\hline
			\textbf{Space}                   & \textbf{Lightness}            & \textbf{TSP runtime}                             & \textbf{Reference}                            \\ \hline
			$(\mathbb{R}^{O(1)},\|\cdot\|_2)$ & $\eps^{-O(1)}$                & $2^{\eps^{-O(1)}}\cdot\tilde{O}(n)$             & \cite{RS98,LS19}       \\ \hline
			Doubling $O(1)$                 & $\eps^{-O(1)}$                & $2^{\eps^{-O(1)}}\cdot\tilde{O}(n)$             & \cite{Gottlieb15,BLW19}     \\ \hline
			Planar                           & $O(1/\eps)$               & $2^{O(1/\eps^2)}\cdot O(n)$                   & \cite{ADDJS93,Klein05} \\ \hline
			$K_{O(1)}$ free                  & $\tilde{O}({1}/{\eps^3})$ & $2^{\tilde{O}({1}/{\eps^4})}\cdot n^{O(1)}$ & \cite{DHK11,BLW17}     \\ \hline
		\end{tabular}
		}
		\vspace{-10pt}
\end{wrapfigure}

A common ingredient  to designing efficient PTAS for TSP  is the notion of \emph{light  spanner}: a weighted subgraph $H$
over the points of the original graph/metric space $G$ that preserves all pairwise distances   up to some $1+\eps$ multiplicative factor (i.e. $\forall u,v\in V(G),~d_H(u,v)\le (1+\eps)\cdot d_G(u,v)$). The \emph{lightness} of the spanner $H$ is the ratio between the total weight of $H$, to that of the Minimum Spanning Tree (MST) of $G$.  While significant progress has been made on understanding the structure of spanners (see the table), it is not the case for \emph{subset spanners}.  A subset spanner $H$ w.r.t.  a prescribed subset $K$ of vertices, called terminals, is a subgraph that preserves distances between terminals  up to a $1+\eps$ multiplicative factor (i.e. $\forall u,v\in K,~d_H(u,v)\le (1+\eps)\cdot d_G(u,v)$).   The \emph{lightness} of $H$ is the ratio between the weight of $H$ to the weight of a minimum Steiner tree \footnote{A Steiner tree is a connected subgraph containing all the terminals $K$. A minimum Steiner tree is a minimum-weight such subgraph; because cycles do no help in achieving connectivity, we can require that the subgraph be a tree.} w.r.t. $K$. While for light spanners the simple greedy algorithm is ``existentially optimal'' \cite{FS16}, in almost all settings, no such ``universal'' algorithm is known for constructing light subset spanners. In planar graphs, Klein \cite{Klein06} constructed  the first light subset spanner. Borradaile \etal \cite{BDT14} generalized Klein's construction to bounded-genus graphs. Unfortunately, generalizing these two results to minor-free metrics remained a major challenge since both approaches were heavily relying on topological arguments.
Recently, Le \cite{Le20} gave the first polynomial-time algorithm for computing a subset spanner with lightness $\poly(\frac{1}{\epsilon})\cdot\log|K|$ in $K_r$-minor-free graphs.  However, the following question remains a fundamental open problem, often mentioned in the literature \cite{DHK11,BDT14,BLW17,Le20}.
\begin{question}\label{question:light-subset-spanners}
	Does a subset spanner of lightness  $\poly(\frac{1}{\epsilon})$ exist in minor-free graphs?
\end{question}

A very related routing problem which is poorly understood even in structured metrics is the vehicle routing problem. Given a special vertex called the \emph{depot} and a
\emph{capacity} $Q$, the goal is to find a collection of subsets of the vertices of the graph each of size at most $Q+1$ such that (1) each vertex appears
in at least one subset, (2) each subset contains the depot, and (3) the sum of the lengths of the shortest tours visiting all the vertices of each subset is minimized.
This is a very classic routing problem, introduced in the late 50s by Dantzig and Ramser~\cite{dantzig1959truck}. While major progress
has been made on TSP during the 90s and 00s for planar and Eucliean metrics, the current understanding of vehicle routing is much less satisfactory.
In Euclidean space, the best known result is a QPTAS by Das and Mathieu \cite{DM15}, while the problem has been shown to be APX-hard for planar graphs
(in fact APX-hard for trees~\cite{Becker18})\footnote{More precisely, the problem where the demand at each vertex is arbitrary is known to be APX-hard
  on trees.} unless
the capacity $Q$ is a fixed constant -- note that the problem remains NP-hard in that case too (see~\cite{AKTT97}).
Given the current success of delivery platforms, the problem with constant capacity is still of high importance from an operations research perspective.
Hence, Becker et al.~\cite{BKS17} have recently given a quasi-polynomial approximation scheme for planar graphs, which was subsequently improved
to a running time of $n^{(Q\eps)^{-O(Q/\eps)}}$~\cite{BKS19}. The next question is:

\begin{question}\label{question:vhr-eptas-planar}
Does the vehicle routing with bounded capacity problem admit an EPTAS in planar and bounded genus graphs?
\end{question}

Since the techniques in previous work~\cite{BKS17,BKS19} for the vehicle routing problem rely on topological arguments, they are not extensible to minor-free graphs. In fact, no nontrivial approximation scheme was known for this problem in minor-free graphs. We ask:

\begin{question}\label{question:vhr-qptas-minor}
	Is it possible to design a QPTAS for the vehicle routing with bounded capacity problem minor-free graphs?
\end{question}

The approach of Becker et al. (drawing on~\cite{EKM14}) is through \emph{metric embeddings}, 
similar to the celebrated work of Bartal~\cite{Bar96} and
Fakcharoenphol et al.~\cite{FRT04} who showed how to embed
any metric space into a simple tree-like structure.
Specifically, Becker et al. aim at embedding the input metric space into a
``simpler''
target space, namely a graph of bounded treewidth, while (approximately)
preserving all pairwise distances. A major constraint arising in this
setting is that for obtaining approximation schemes, the distortion
of the distance should be carefully controlled.
An ideal scenario would be to embed $n$-vertex minor free graphs into
graphs of treewidth at most $O_\eps(\log n)$, while preserving the
pairwise distance up to a $1+\eps$ factor.
Unfortunately, as implied by the work of Chakrabarti \etal \cite{CJLV08}, there are $n$ vertex planar graphs such that every (stochastic) embedding into $o(\sqrt{n})$-treewidth graphs must incur  expected multiplicative distortion $\Omega(\log n)$ (see also~\cite{Rao99,KLMN04,AFGN18} for
embeddings into Euclidean metrics).

Bypassing the above roadblock,
Eisenstat \etal \cite{EKM14} and Fox-Epstein \etal \cite{FKS19} showed
how to embed planar metrics into bounded-treewidth graphs while preserving distances up
to a controlled \emph{additive} distortion.
Specifically,  given a planar graph $G$ and a parameter $\epsilon$, they showed how to construct a metric embedding into a graph $H$ of bounded treewidth such that all pairwise distances between pairs of vertices are preserved up to an additive $\eps D$ factor, where $D$ is the diameter of $G$. 
While $\eps D$ may look like a crude additive bound, it
is good enough for obtaining approximation schemes for some
classic problems such as $k$-center, and vehicle routing.
While Eisenstat \etal constructed an embedding into a graph of treewidth  $\poly(\frac1\eps)\cdot  \log n$, Fox-Epstein \etal constructed an embedding into a graph of treewidth $\poly(\frac1\eps)$, leading to the first PTAS for vehicle routing (with running time $n^{(Q/\eps)^{O(Q/\eps)}}$). Yet for minor-free graphs, or even bounded-genus graphs, obtaining such a result with any non-trivial bound on the treewidth is a major challenge; the embedding of   Fox-Epstein et al.~\cite{FKS19} heavily relies on planarity (for example by using the face-vertex incident graph). Therefore, prior to our work, the following question is widely open.

\begin{question}\label{question:embedding}
	Is it possible to (perhaps stochastically) embed a minor-free graph with diameter $D$ to a graph with treewidth $\polylog(n)$ and additive distortion at most $\epsilon D$?
\end{question}


\subsection{Main contribution}
We answer all the above questions by the affirmative. Our first main contribution is a ``truly'' \emph{light subset spanner} for minor-free metrics that bridges the gap for spanners between planar and minor-free metrics; this completely settles \Cref{question:light-subset-spanners}.
In the following, the $O_r$ notation hides factors in $r$, e.g. $x = O_r(m) \iff x \le m\cdot f(r)$ for some sufficiently large $m$ and
computable function $f$; and $\poly(x)$ is (some) polynomial function of $x$.

\begin{theorem}\label{thm:tsp-spanner}
	Given a $K_r$-minor-free graph $G$, a set of terminals $K\subseteq V(G)$, and a parameter $\eps\in(0,1)$, there is a polynomial time algorithm that computes a subset spanner with distortion $1+\eps$ and lightness $O_r(\poly(\frac{1}{\eps}))$.
\end{theorem}

Our second main contribution is a 
\emph{stochastic embedding} (see \Cref{def:stocastic}) of minor-free graphs into bounded-treewidth graphs with small expected additive distortion, obtaining
the first result of this kind for minor-free graphs and resolving \Cref{question:embedding} positively.

\begin{theorem}\label{thm:embedding-minor}
  Given an $n$-vertex $K_r$-minor-free graph $G$ of diameter $D$, and a parameter $\epsilon\in(0,1)$, in polynomial time one can construct a stochastic embedding from $G$ into graphs with treewidth $O_r(\frac{\log n}{\epsilon^2})$, and expected additive distortion $\eps D$.
\end{theorem}

While the embedding of planar graphs to low treewidth graphs by Fox-Epstein et al.~\cite{FKS19} is deterministic, our embedding in~\Cref{thm:embedding-minor} is stochastic. Thus, it is natural to ask whether randomness is necessary. We show in~\Cref{thm:LB} below that the embedding must be stochastic to guarantee (expected) additive distortion $\epsilon D$, for small enough $\epsilon $ (see \Cref{sec:emblowerbound} for details).

\begin{restatable}{theorem}{EmLowerBound}
	\label{thm:LB}
	There is an infinite graph family $\mathcal{H}$ of $K_6$-free graphs, such that for every $H\in \mathcal{H}$ with $n$ vertices and diameter $D$, every dominating embedding of $H$ into a treewidth-$o(\sqrt{n})$ graph has additive distortion at least $\frac{1}{20}\cdot D$.
\end{restatable}

For the more restricted case of a graph with genus $g$, we can construct a deterministic embedding without any dependence on the number of vertices.

\begin{theorem}\label{thm:embedding-genus}
	Given a genus-$g$ graph $G$ of diameter $D$, and a parameter $\epsilon\in(0,1)$, there exists an embedding $f$ from $G$ to
	a graph $H$ of treewidth $O_{g}(\poly(\frac{1}{\epsilon}))$ with additive distortion $\eps D$.
\end{theorem}

Next we describe the algorithmic consequences of our results.  First, we obtain an efficient PTAS for a Subset TSP problem in $K_r$-minor-free graphs for any fixed $r$, thereby completely answering \Cref{question:subsetTSP}. (See \Cref{appendix:subsetTSP} for details.)
\begin{theorem}\label{thm:tsp-eptas}
  Given a set of terminals $K$ in an $n$-vertex $K_r$-minor-free graph $G$ of,  there exists an algorithm with running time
  $2^{O_{r}(\poly(1/\epsilon))} n^{O(1)}$ that can find a tour visiting every vertex in $K$ of length at most $1+\epsilon$ times the length of
  the shortest tour. 
\end{theorem}

Second, we obtain the first polynomial-time approximation scheme for bounded-capacity vehicle routing in $K_r$-minor-free graphs.  

\begin{theorem}\label{thm:vhr-qptas}
  there is a randomized algorithm that,
  given an $n$-vertex  $K_r$-minor-free graph $G$ and an instance of bounded-capacity vehicle routing on $G$,  in time $n^{O_{\epsilon,Q,r}(\log\log n)}$
  returns a solution with expected cost at most $1+\epsilon$
  times the cost of the optimal solution. 
\end{theorem}
\Cref{thm:vhr-qptas} provides a definite answer to \Cref{question:vhr-qptas-minor}.  En route to this result, we design a new dynamic program for bounded-capacity vehicle routing on bounded-treewidth graphs that constitutes
the first approximation scheme that is fixed-parameter tractability in the treewidth (and also in $\eps$) for this class of graphs.  For planar graphs and bounded-genus graphs, this yields a $2^{\poly(\frac{1}{\epsilon})} n^{O(1)}$ approximation scheme and completely answers \Cref{question:vhr-eptas-planar}.

\begin{theorem}\label{thm:vhr-eptas}
  There is a randomized algorithm that, given a  graph $G$ with genus at most $g$ and an instance of bounded-capacity vehicle routing  on $G$, in time $2^{O_{g,Q}(\poly(1/\epsilon))} n^{O(1)}$ returns a solution whose expected cost at most $1+\epsilon$ times the cost of the optimal solution.
\end{theorem}

A major tool in our algorithm in \Cref{thm:vhr-eptas} is a new efficient dynamic programming for approximating bounded-capacity vehicle routing in bounded treewidth graphs. The best exact algorithm known for bounded-treewidth graphs has running time
$n^{O(Q\tw)}$\cite{BKS18}.  

\begin{theorem}	\label{thm:vehiclerouting-tw-dp}
  Let $\tw,\eps>0$. There is an algorithm that, for any instance of the vehicle routing problem
  $(G,Q,s)$
  such that $G$ has treewidth $\tw$ and $n$ vertices, 
  outputs a $(1+\eps)$-approximate solution in time
  $(Q\eps^{-1}\log n)^{O(Q\tw /\eps)} n^{O(1)}$.
\end{theorem}

We refer readers to \Cref{appendix:vhr} for details on \Cref{thm:vhr-qptas},  \Cref{thm:vhr-eptas}, and \Cref{thm:vehiclerouting-tw-dp}.

\subsection{Techniques} \label{sec:techniques}

In their seminal series of papers regarding minor free graph, Robertson and Seymour showed how to decompose a minor-free graph into four "basic components": \emph{surface-embedded graphs}, \emph{apices}, \emph{vortices} and \emph{clique-sums} \cite{RS03} (see \Cref{subsec:RobertsonSeymour} for details and definitions). 
Their decomposition suggested an algorithmic methodology, called the \emph{RS framework}, for solving a combinatorial optimization problem on minor-free graphs: solve the problem on planar graphs, and then generalize to bounded-genus graphs, to graphs embedded on a surface with few vortices, then deal with the apices, and finally extend to minor-free graphs. The RS framework has been successfully applied to many problems such as vertex cover, independent set and dominating set \cite{Grohe03,DHK05}. A common feature for these problems was that the graphs were unweighted, and the problems rather ``local''. This success can be traced back to the pioneering work of Grohe \cite{Grohe03} who showed how to handle graphs embedded on a surface with few vortices by showing that these graphs have linear local-treewidth.

However, there is no analogous tool that can be applied to fundamental connectivity problems such as Subset TSP, Steiner tree, and survivable network design. Therefore, even though  efficient PTASes for these problems were known for planar graphs \cite{Klein06,BKK07,BK08} for a long time, achieving similar results for any of them in minor-free graphs remains a major open problem. Inspired by the RS framework, we propose a multi-step framework for light subset spanner and embedding problems in minor-free graphs.

\paragraph{A multi-step framework}   The fundamental building block in our framework is  planar graphs each with a \emph{single vortex} with bounded diameter $D$, on which we solve the problems (Step 1 in our framework). We consider this as a major conceptual contribution as we overcome the barrier posed by vortices. We do so by introducing a hierarchical decomposition where each cluster in every level of the decomposition is separated from the rest of the graph by a constant number of shortest paths \emph{of the input graph}.
\footnote{One might hope that a similar decomposition can be constructed using the shortest-path separator of Abraham and Gavoille \cite{AG06} directly. Unfortunately, this is impossible as the length of the shortest paths in \cite{AG06} is unbounded w.r.t. $D$. Rather, they are shortest paths in different subgraphs of the original graph.}
Similar decomposition for planar graphs \cite{AGKKW98,Tho04} and bounded-genus graphs \cite{KKS11} has found many algorithmic applications \cite{AGKKW98,Thorup04,EKM14,KKS11}. Surprisingly, already for the rather restricted case of apex graphs,\footnote{A graph $G$ is an apex graph if there is a vertex $v$ such that $G\setminus v$ is a planar graph.} it is impossible to have such a decomposition. We believe that our decomposition is of independent interest. 

While it is clear that the diameter parameter $D$ is relevant for the embedding problem, a priori it is unclear why it is useful for the light-subset-spanner problem. As we will see later, the diameter comes from a reduction to subset local spanners (Le \cite{Le20}), while the assumption is enabled by using sparse covers \cite{AGMW10}.

In Step 2, we generalize the results to $K_r$-minor-free graphs. Step 2 is broken into several mini-steps. In Mini-Step 2.1,\footnote{In the subset spanner problem, there is an additional step where we remove the constraint on the diameter of the graph, and this becomes Step 2.0.} we handle the case of planar graphs with more than one vortex; we introduce a \emph{vortex-merging operation} to reduce to the special case in Step 1. In Mini-Step 2.2, we handle graphs embedded on a surface with multiple vortices. The idea is to cut along vortex paths to reduce the genus one at a time until the surface embedded part is planar (genus $0$), and in this case, Step 2.1 is applicable. In Mini-Step 2.3, we handle graphs embedded on a surface with multiple vortices and \emph{a constant number of apices}, a.k.a \emph{nearly embeddable graphs}. In Mini-Step 2.4, we show how to handle general $K_r$-minor-free graphs by dealing with clique-sums. 

In this multi-step framework, there are some steps that are simple to implement for one problem but challenging for the other. For example, implementing Mini-Step 2.3 is simple in the light subset spanner problem, while it is highly non-trivial for the embedding problem;  removing apices can result in a graph with unbounded diameter. Novel ideas are typically needed to resolve these challenges; we refer the reader to \Cref{sec:tech-ideas} for more technical details. 

We believe that our multi-step framework will find applications in designing PTASes for other problems in $K_r$-minor-free graphs, such as minimum Steiner tree or survivable network design. 

\paragraph{ An FPT approximation scheme for vehicle routing on low treewidth graphs } Our $(1+\eps)$-approximation for vehicle routing with bounded
capacity in bounded treewidth graphs relies on a dynamic program
that proceeds along the clusters of a branch decomposition\footnote{For
	simplicity, we work with branch decompositions}, namely
the subgraphs induced by the leaves of the subtrees of the
branch decomposition.
One key idea is to show that there exists a near-optimal solution such
that the number of tours entering (and leaving)
a given cluster with some fixed capacity $q \in [Q]$ can be rounded
to a power of $1+\tilde{\eps}$, for some
$\tilde{\eps}$ to be chosen later. To achieve this, we start from
the optimum solution and introduce
\emph{artificial paths}, namely paths that start at a vertex and go
to the depot (or from the depot to a vertex), without making any delivery
and whose only purpose is to help \emph{rounding} the number of
paths entering or
leaving a given cluster of the decomposition (i.e.: making it a power
of $1+\tilde{\eps}$). This immediately reduces the number
of entries in the dynamic programming table we are using, reducing
the running time of the dynamic program to the desired complexity.

The main challenge becomes to bound the number of artificial paths
hence created so as to show that the obtained solution has cost
at most $1+\eps$ times the cost of the optimum solution.
To do so, we design a charging scheme and prove that
every time a new path is created, its cost can be charged to the cost
of some  $\tilde{\eps}^{-1}$ paths of the original optimum solution.
Then, we ensure that each path of the original optimum solution does not
get charged more than $\eps$ times. This is done by showing by defining
that a path \emph{enters} (resp. \emph{leaves}) a cluster only if it
is making its next delivery  (resp. it has made its last delivery)
to a vertex inside. This definition helps limit the number of times
a path gets charged to $\tilde{\eps} = \eps/(Q \log n)$
but it also separates the underlying shortest path
metric from the structure of the graph: A path from vertices
$s_1,\ldots,s_k$ should not be considered entering any cluster of the
branch decomposition containing $s_i$ if it does not pick up its
next delivery (or has picked up its last delivery) within the
cluster of $s_i$. This twist demands a very careful design of the
dynamic program by working with distances rather than explicit paths.

Then, our dynamic program works as follows: The algorithm computes
the best solution at a given cluster $C$ of the
decomposition, for any prescribed number of tours
(rounded to a power of $1+\tilde{\eps}$)
entering and leaving $C$.
This is done by iterating over all pairs
of (pre-computed) solutions for the child clusters of $C$
that are consistent with (namely, that potentially \emph{can} lead to)
the prescribed number of tours entering and leaving at $C$.
Given consistent solutions for the child cluster,
the optimal cost of combining them
(given the constraints on the number of
tours entering at $C$) is then computed through a min-cost max-flow
assignment.

\section{Proof Overviews} \label{sec:tech-ideas}

\subsection{Light subset spanners for minor-free metrics}
In this section, we give a proof overview and review the main technical ideas for the proof of \Cref{thm:tsp-spanner}. A subgraph $H$ of a graph $G$ is called a \emph{subset} $L$-\emph{local} $(1+\eps)$-spanner of $G$ with respect to a set $K$ of terminals if:        
\begin{equation*}
\forall t_1,t_2\in K \mbox{ ~s.t.~ } d_G(t_1,t_2) \le L \qquad \mbox{it holds that ~~} d_H(t_1,t_2) \leq (1+\eps)\cdot d_G(t_1,t_2) 
\end{equation*}

\noindent Our starting point is the following reduction of Le~\cite{Le20}.

\begin{theorem}[Theorem 1.4~\cite{Le20}]\label{thm:reduction-Le}
	Fix an $\eps\in(0,1)$. Suppose that for any $K_r$-minor-free weighted graph $G=(V,E,w)$, subset $K\subseteq V$ of terminals, and parameter $L > 0$, there is a subset $L$-local $(1+\eps)$-spanner w.r.t. $K$ of weight at most $O_{r}(|K|\cdot L\cdot\poly(\frac{1}{\eps}))$.	 For any terminal set, $G$ admits a subset $(1+\eps)$-spanner with lightness $O_r(\poly(\frac{1}{\eps}))$.
\end{theorem}

\noindent Our main focus is to construct a light subset $L$-local spanner. 
\begin{restatable}{proposition}{MinorFreeSubset}
	\label{prop:ell-close-spanner}
	For any edge-weighted $K_r$-minor-free graph $G=(V,E,w)$, any subset $K\subseteq V$ of terminals, and any parameter $L > 0$, there is a subset $L$-local $(1+\eps)$-spanner for $G$ with respect to $K$ of weight $O_{r}(|K|\cdot L\cdot\poly(\frac{1}{\eps}))$.
\end{restatable}

\Cref{thm:tsp-spanner} follows directly by combining \Cref{thm:reduction-Le} with \Cref{prop:ell-close-spanner}. Our focus now is on proving \Cref{prop:ell-close-spanner}.
The proof is divided into two steps: 
in step 1 we solve the problem on the restricted case of planar graphs with bounded diameter and a single vortex.
Then, in step 2, we reduce the problem from $K_r$-minor-free graphs to the special case solved in step 1.
\paragraph{Step 1: Single vortex with bounded diameter}
The main lemma in step 1 is stated below; the proof appears in \Cref{sec:oneVortex}.
We define a \emph{single-vortex} graph $G = G_{\Sigma} \cup W$ as a graph whose edge set can
be partitioned into two parts $G_{\Sigma}, W$ such that $G_{\Sigma}$ induces a plane graph and $W$ is a vortex of width  \footnote{The \emph{width} of the vortex is the width
	of its path decomposition.\label{foot:VortexWidth}} at most $h$ glued to some face of $G_{\Sigma}$.
\begin{restatable}[Single Vortex with Bounded Diameter]{lemma}{SingleVortexBoundedDiam}
  \label{lm:one-vortex-Bounded-diam}
  Consider a single-vortex 
  graph $G = G_{\Sigma}\cup W$ with diameter $D = O_h(L)$,	
  where $G_{\Sigma}$ is planar, and $W$ is a vortex of width at most $h$ glued to a face of $G_{\Sigma}$.
   For any terminal set $K$, there exists a subset $L$-local $(1+\eps)$-spanner for $G$ with respect to $K$ of weight $O_{h}(|K|L\cdot\poly(\frac{1}{\eps}))$.
\end{restatable}
	The basic idea in constructing the spanner for \Cref{lm:one-vortex-Bounded-diam} is to use shortest-path
        separators to recursively break down the graph into clusters while maintaining the distance from every terminal
        to the boundaries of its cluster.  Let $k$ be the number of terminals.
	The idea is to construct a hierarchical tree of clusters of depth $O(\log k)$ where each terminal-to-boundary-vertex path
        is well approximated. An elementary but inefficient approach to
        obtain such a result is add a single-source spanner (\Cref{lm:ss-spanner}) from each terminal $t$ to every shortest path
        (at distance at most $L$) in each one of the separators in all the recursive levels. As a result, the spanner will consist
        of $O_h(k\log k)$ single-source spanners of total weight $O_{h}(L\cdot k\log k\cdot\poly(\frac{1}{\eps}))$, as obtained in \cite{Le20}.
        Thus, a very natural and basic question is whether minor-free graphs have enough structure so that we can do avoid
        the $\log k$ factors coming from the depth of the hierarchy. We show that this is indeed the case.
	
	The problem with the previous approach is that for each hierarchical cluster $\varUpsilon$ in the decomposition, the total weight of the edges added to the spanner is proportional to the number of terminals in $\varUpsilon$ (and is thus $k\log k\cdot L$ in total, considering the entire process).
	Our approach is the following: instead of adding single-source spanners from a terminal to paths in $O(\log k)$ separators, we add bipartite spanners (\Cref{lm:bipartite-spanner}) from the paths in a newly added separator to all the separator paths in the boundary of the current cluster.
	A \emph{bipartite spanner} is a set of edges that preserve all pairwise distances between two paths such that its weight is proportional to the distance between the paths and their lengths. The hope is to pay only $L$ for each hierarchical cluster $\varUpsilon$, regardless of the number of terminals it contains.
    This approach has two main obstacles:
	(1) the number of paths in the boundary of a cluster at depth $t$ of the recursion can be as large as $\Omega_h(t)$, implying that the total number of bipartite spanners added is $\Omega_h(k\log k)$ -- and we would not have gained anything compared to the elementary approach, and
	(2) the weight of the shortest paths is unbounded. While initially the diameter and thus the length of shortest paths is bounded by $O_h(L)$, in the clusters created recursively, after deleting some paths there is no such bound. Note that in the approach that used the single-source spanners this was a non-issue, as for single-source spanner (\Cref{lm:ss-spanner}), the length of the shortest path $P$ does not matter. However, the weight of a bipartite spanner (\Cref{lm:bipartite-spanner}) depends on the weight of the paths it is constructed for.
	
	We resolve both these issues by recursively constructing separators with a special structure. According to Abraham and Gavoille \cite{AG06}, a separator can be constructed using a fundamental vortex cycle constructed between two vortex paths induced by an arbitrary tree (see \Cref{def:vortex-path}).
	We construct a spanning tree $T$ as a shortest-path tree rooted in the perimeter vertices. Every separator then consists of two shortest paths from perimeter vertices to vertices in the embedded part of the graph, and at most two bags. The important property is that for every cluster $G_\Upsilon$ we encounter during the recursion, $T\cap G_\Upsilon$ is a spanning tree of $G_\Upsilon$. As a result, all the shortest paths we use for the separators throughout the process are actual shortest paths in $G$. In particular, their length is bounded, and thus issue (2) is resolved.
	In order to resolve issue (1), we control the number of shortest paths in the boundary of a cluster in our decomposition using a more traditional approach.
        Specifically, in some recursive steps, we aim for a reduction in the number of paths in the boundary instead of a reduction
        in the number of terminals.

\paragraph{Step 2: From minor-free to single vortex with bounded diameter.}
We generalize the spanner construction of Step 1 to minor-free graphs using the Robertson-Seymour decomposition.
We have five sub-steps, each generalizing further (at the expense of increasing the weight of the spanner by an additive term  $O_{h}(k L\cdot\poly(\frac{1}{\eps}))$).

Thus, consider the construction proposed in Step 1.
In the first sub-step, we remove the assumption on the bounded diameter and make our spanner construction work for arbitrary
planar graphs with a single vortex.
The approach is as follows: Break a graph with unbounded diameter to overlapping clusters of diameter $O_h(L)$ such that every pair of vertices at
distance at most $L$ belongs to some cluster, and each vertex belong to at most $O_h(1)$ clusters. This is done using Abraham \etal
sparse covers \cite{AGMW10}. 
Then construct a spanner for each cluster separately by applying the approach of Step 1, namely \Cref{lm:one-vortex-Bounded-diam}, and
return the union of these spanners. More concretely, we prove the following lemma, whose proof appears in
\Cref{subsec:diameterReduction}.
\begin{restatable}[Single Vortex]{lemma}{LemmaOneVortexUbnoundedDiam}
  \label{lm:one=Vortex-Unbounded-diam}
  Consider a graph $G = G_{\Sigma}\cup W$	
  where $G_{\Sigma}$ is planar, and $W$ is a vortex of width at most $h$ glued to a face of $G_{\Sigma}$. For any terminal set $K$, there exists a subset $L$-local $(1+\eps)$-spanner for $G$ with respect to $K$ of weight $O_{h}(|K|L\cdot\poly(\frac{1}{\eps}))$.
\end{restatable}

In the second sub-step, we generalize to planar graphs with at most $h$ vortices of width $^{\ref{foot:VortexWidth}}$ $h$.  
The basic idea is to ``merge'' all vortices into a single vortex of width $O(h^2)$. This is done by repeatedly deleting a shortest path
between pairs of vortices, and ``opening up'' the cut to form a new face. The two vortices are then ``merged'' into a single vortex -- in other
words, they can be treated as a single vortex by the algorithm obtained at the first sub-step.
This is repeated until all the vortices have been ``merged'' into a single vortex, at which point \Cref{lm:one=Vortex-Unbounded-diam} applies.
Here we face a quite important technical difficulty: when opening up a shortest path between two vortices, we may alter shortest paths between
pairs of terminals (e.g.: the shortest path between two terminals intersects the shortest path between our two vortices, in which case
deleting the shortest path between the vortices destroys the shortest path between the terminals).
To resolve this issue, we compute a single-source spanner from each terminal to every nearby deleted path, thus controlling 
the distance between such terminal pairs in the resulting spanner.

The above idea is captured in the following lemma, whose proof appears in \Cref{subsec:reducingVortices}.
\begin{restatable}[Multiple Vortices]{lemma}{NearlyEmbdablSpannerNoApicesNoGenus}
  \label{lm:planar-with-many-vortices}
  Consider a graph $G = G_{\Sigma}\cup  W_1\cup\dots \cup W_{h'}$,  where $G_{\Sigma}$ is planar, $h'\le h$, and each $W_i$ is a vortex of width at most $h$ glued to a face of $G_{\Sigma}$.
  For any terminal set $K$, there exists a subset $L$-local $(1+\eps)$-spanner for $G$ with respect to $K$  of weight $O_{h}(|K|L\cdot\poly(\frac{1}{\eps}))$.
\end{restatable}

In our third sub-step, we generalize to graphs of bounded genus with multiple vortices.
The main tool here is ``vortex paths'' from \cite{AG06}. Specifically, we can remove two vortex paths and reduce the genus by one (while increasing the number of vortices).
Here each vortex path consists of essentially $O_h(1)$ shortest paths.
We apply this genus reduction repeatedly until the graph has genus zero. The graph then has $O(g)$ new vortices. Next, we apply \Cref{lm:planar-with-many-vortices} to create a spanner.
The technical difficulty of the previous step arises here as well: There may be shortest paths between pairs of terminals that intersect
the vortex paths. We handle this issue in a similar manner.
The proof appears in \Cref{subsec:removingGenus}.

\begin{restatable}[Multiple Vortices and Genus]{lemma}{NearlyEmbdablSpannerNoApices}
  \label{lm:nearly-embed-spanner-no-apices}
  Consider a graph $G = G_{\Sigma}\cup W_1\cup\dots \cup W_{h'}$ 	
  where $G_{\Sigma}$ is (cellularly) embedded on a surface $\Sigma$ of genus at most $g=O(h)$,  $h'\le h$, and each $W_i$ is a vortex of width at most $h$ glued to a face of $G_{\Sigma}$. For any terminal set $K$, there exists a subset $L$-local $(1+\eps)$-spanner for $G$ with respect to $K$ of weight $O_{h}(|K|L\cdot\poly(\frac{1}{\eps}))$.
\end{restatable}

In our fourth sub-step, we generalize to nearly $h$-embeddable graphs.
That is, in addition to genus and vortices, we also allow $G$ to have at most $h$ apices. 
The spanner is constructed by first deleting all the apices and applying \Cref{lm:nearly-embed-spanner-no-apices}. Then, in order to compensate for
the deleted apices, we add a shortest path from each apex to every terminal at distance at most $L$.
The proof appears in \Cref{subsec:removingApices}.	
\begin{restatable}[Nearly $h$-Embeddable]{lemma}{NearlyEmbdablSpanner}
  \label{lm:nearly-embed-spanner}
  Consider a nearly $h$-embeddable graph $G$ with a set $K$ of $k$ terminals. There exists an $L$-local $(1+\eps)$-spanner for $K$ of weight $O_{h}(kL\cdot\poly(\frac{1}{\eps}))$.
\end{restatable}

Finally, in our last sub-step, we generalize to minor-free graphs, thus proving \Cref{prop:ell-close-spanner}. 
Recall that according to \cite{RS03} a minor graph can be decomposed into a clique-sum decomposition, where each node in the decomposition
is nearly $h$-embeddable.
Our major step here is transforming the graph $G$ into a graph $G'$ that preserves all terminal distances in $G$,
while having at most $O(k)$ bags in its clique-sum decomposition.
This is done by first removing leaf nodes which are not ``essential'' for any terminal distance, and then shrinking long paths in the decomposition where all
internal nodes have degree two and (roughly) do not contain terminals.
Next, given $G'$, we make each vertex that belongs to one of the cliques in the clique-sum decomposition into a terminal. The new number of terminals is
bounded by $O_h(k)$. The last step is simply to construct an internal spanner for each
bag separately using \Cref{lm:nearly-embed-spanner}, and return the union of the constructed spanners.
The proof (of \Cref{prop:ell-close-spanner}) appears in \Cref{subsec:EliminatingCliqueSum}.

\subsection{Embedding into low-treewidth graphs}\label{subsec:tech-tw-emb}
At a high level, we follow the same approach as for the subset spanner. Due to the different nature of the constructed structures, and the different distortion guarantees, there are some differences that raise significant challenges.

Our first step is to generalize the result of Fox-Epstein et al.~\cite{FKS19} to graph of bounded genus. 
Basically, our approach is the same as for the subset spanner: we decompose the graph into simpler and simpler pieces by removing shortest paths. Here, instead of deleting a path, we will use a \emph{cutting lemma}. However, in this setting it is not clear how to use single-source or bipartite spanners to compensate for the changes to the shortest-path metric due to path deletions, since these spanners may have large treewidth.
Instead, we will \emph{portalize} the cut path.   That is, we add an $\eps D$-net\footnote{An $r$-net of a set $A$, is a set $N\subset A$ of vertices all at distance at least $r$ from each other, and such that every $v\in A$ has a net point $t\in N$ at distance at most $r$. If $A$ is a path of length $L$, then for every $r$-net $N$, $|N|=O(\frac{L}{r})$.\label{foot:net}}
of the path to every bag of the tree decomposition of the host graph. Clearly, this strategy has to be used cautiously since
it immediately increases the treewidth significantly.

Apices pose an interesting challenge. Standard techniques to deal with apices consist in removing them from the graph, solve the problem on
the remaining graph
which is planar, and add back the apices later~\cite{Grohe03,DHK05}. However, in our setting, removing apices can make the diameter of the resulting graph, say $G'$, become arbitrarily larger than $D$ and thus, it seems hopeless to embed $G'$ into a low treewidth graph with an additive
distortion bounded by $D$. This is where randomness comes into play: we use padded decomposition~\cite{Fil19padded} to randomly partition $G'$ into pieces of (strong) diameter  $D' = O(\frac{D}{\epsilon})$. We then embed each part of the partition (which is planar) separately into graphs of bounded treewidth with additive distortion $\eps^2 D' = O(\eps D)$,
add back the apices by connecting them to all the vertices of all the bounded treewidth graphs
(and so adding all of them to each bag of each decomposition) and obtain a graph with
bounded treewidth and an \emph{expected} additive distortion $\epsilon D$.

Our next stop on the road to minor-free metrics is to find bounded treewidth embeddings of clique-sums of bounded genus graphs with apices.

Suppose that $G$ is decomposed into clique-sums of graphs $G_1,G_2, \ldots, G_k$. We call each $G_i$ a \emph{piece}. A natural idea is to embed each $G_i$ into a low treewidth graph $H_i$, called the \emph{host graph} with a tree decomposition $\mathcal{T}_i$, and then combine all the tree decompositions together. Suppose that $G_1$ and $G_2$ participate in the clique-sum decomposition of $G$ using the clique $Q$. To merge $G_1$ and $G_2$, we wish to have an embedding from $G_i$ to $H_i$, $i = 1,2$, that \emph{preserves} the clique $Q$ in the clique-sum of $G_1$ and $G_2$. That is, the set of vertices $\{ f_i(v) | v \in Q\}$ induces a clique in $H_i$ (so that there will be bag in the tree decomposition of $H_i$ containing $f(Q)$). However, it is impossible to have such an embedding even if all $G_i$'s are planar
\footnote{To see this, suppose that $G$ is clique-sums of a graph $G_0$ with many other graphs $G_1,G_2\ldots, G_s$ in a star-like way, where $G_0$ has treewidth polynomial in $n$, and every edge of $G_0$ is used for some clique sum. If $H_0$ preserves all cliques, it contains $G_0$ and thus has treewidth polynomial in $n$.}.
To overcome this obstacle, we will allow each vertex in $G_i$ to have multiple images in $H_i$. Specifically, we introduce \emph{one-to-many} embeddings. Note that given a one-to-many embedding, one can construct a classic embedding by identifying each vertex with an arbitrary copy.
\begin{definition}[One-to-many embedding]
  An embedding $f:G\rightarrow2^H$ of a graph $G$ into a graph $H$ is a \emph{one-to-many embedding} if for every $v\in G$, $f(v)$ is a non empty set of vertices in $H$, where the sets $\{f(v)\}_{v\in G}$ are disjoint.
  
  We say that $f$ is \emph{dominating} if for every pair of vertices $u,v\in G$, it holds that $d_G(u,v)\le \min_{u'\in f(u),v'\in f(v)}d_H(u',v')$. 
  We say that $f$	has additive distortion $\eps D$ if it is dominating and $\forall u,v\in G$ it holds that $\max_{u'\in f(u),v'\in f(v)}d_H(u',v')\le d_G(u,v)+\eps D$.
  Note that, as for every vertex $v\in G$, $d_G(v,v)=0$, having additive distortion $\eps D$ implies that all the copies in $f(v)$ are at distance at most $\eps D$ from each other.
  
  A stochastic one-to-many embedding is a distribution $\mathcal{D}$ over dominating one-to-many embeddings. We say that a stochastic one-to-many embedding has expected additive distortion $\eps D$ if $\forall u,v\in G$ it holds that $\mathbb{E}[\max_{u'\in f(u),v'\in f(v)}d_H(u',v')]\le d_G(u,v)+\eps D$.
\end{definition}
We can show that in order to combine the different one-to-many embeddings of the pieces $G_1,\dots,G_s$, it is enough that for every clique $Q$
we will have a bag $B$ containing at least one copy of each vertex in $Q$. Formally, 
\begin{definition}[Clique-preserving embedding]
  A one-to-many embedding $f:G\rightarrow2^H$ is called clique-preserving embedding if for every clique $Q \in G$,
  there is a clique $Q'$ in $H$ such that for every vertex $v \in Q$, $f(v)\cap Q'\ne\emptyset$.
\end{definition}
While it is impossible to preserve all cliques in a one-to-one embedding, it is possible to preserve all cliques in a one-to-many embedding; this is one of our major conceptual contributions. One might worry about the number of maximal cliques in $G$. However, since $G$ has constant degeneracy, the number of maximal cliques is linear~\cite{ELS10}.
Suppose that $f$ is clique-preserving, and let $\mathcal{T}$ be some tree-decomposition of $H$. Then for every clique $Q$ in $G$, there is a bag of $\mathcal{T}$ containing a copy of (the image of) $Q$ in $H$.

We now have the required definitions, and begin the description of the different steps in creating the embedding. 
The most basic case we are dealing with directly is that of  a planar graph with a single vortex and diameter $ D$ into  a graph of treewidth $O(\frac{\log n}{\epsilon})$ and additive distortion $\epsilon  D$.  The high level idea is to use vortex-path separator to create a hierarchical partition tree  $\tau$ as in \Cref{sec:oneVortex}. The depth of the tree will be $O(\log n)$. To accommodate for the damage caused by the separation, we \emph{portalize} each vortex-path in the separator.
That is for each such path $Q$, we pick an $\eps D$-net $^{\ref{foot:net}}$
$N_Q$ of size $O(\frac{1}{\epsilon})$.  The vertices of $N_Q$ called \emph{portals}. Since each node of $\tau$ is associated with a constant number of vortex-paths, there are at most $O(\frac{1}{\epsilon})$ portals corresponding to each node of $\tau$.  Thus, if we collect all portals along the path from a leaf to the root of $\tau$, there are $O(\frac{\log n}{\epsilon})$ portals. 
We create a bag for each leaf $\Upsilon$ of the tree $\tau$. In addition for each bag we add the portals corresponding to nodes along the path from the root to $\Upsilon$. The tree decomposition is then created w.r.t. $\tau$. 
Finally, we need to make the embedding clique-preserving. Consider a clique $Q$, there will be a leaf $\Upsilon_Q$ of $\tau$ containing a sub-clique $Q'\subseteq Q$, while all the vertices in $Q\setminus Q'$ belong to paths in the boundary of $\Upsilon_Q$. We will create a new bag containing (copies) of all the vertices in $Q$ and all the corresponding portals. The vertices of $Q'$ will have a single copy in the embedding, while the distortion of the vertices $Q'\subseteq Q$ will be guaranteed using a nearby portal.

\begin{restatable}[Single Vortex with Bounded Diameter]{lemma}{embPlanarVortex}
	\label{lm:emb-planar-vortex}
	Given a single-vortex graph $G = G_{\Sigma}\cup W$ where the vortex $W$ has width $h$.
        There is a one-to-many, clique-preserving embedding $f$ from $G$ to a graph $H$ with
        treewidth $O(\frac{h\log n}{\epsilon})$ and additive distortion $\epsilon D$ where $D$ is diameter of $G$. 
\end{restatable}

We then can extend the embedding to planar graphs with multiple vortices using the \emph{vortex merging} technique (\Cref{subsec:reducingVortices}), and then to graphs embedded on a genus-$g$ surface with multiple vortices by cutting along vortex-paths. 
The main tool here is a \emph{cutting lemma} described in \Cref{sec:cut} which bound the diameter blowup after each cutting step. At this point, the embedding is still deterministic. The proofs appear in \Cref{sec:EmbedManyVortices} and \Cref{sec:genus-minor} respectively.
\begin{restatable}[Multiple Vortices]{lemma}{emVortexReduce}
	\label{lm:em-vortexReduce}
	Consider a graph $G = G_{\Sigma}\cup  W_1\cup\dots \cup W_{v(G)}$ of diameter $D$,  where $G_{\Sigma}$ can be drawn on the plane, and each $W_i$ is a vortex of width at most $h$ glued to a face of $G_{\Sigma}$, and $v(G)$ is the number of vortices in $G$. There is a one-to-many, clique-preserving embedding $f$ from $G$ to a graph $H$ of treewidth at most $\frac{h2^{O(v(G))}\log n}{\epsilon}$ with additive distortion $\epsilon D$.
\end{restatable}
\begin{restatable}[Multiple Vortices and Genus]{lemma}{embedGenusVortex}
	\label{lm:embed-genus-vortex}
	Consider a graph $G = G_{\Sigma}\cup  W_1\cup\dots \cup W_{v(G)}$ of diameter $D$,  where $G_{\Sigma}$ is (cellularly) embedded on a surface $\Sigma$ of genus $g(G)$, and each $W_i$ is a vortex of width at most $h$ glued to a face of $G_{\Sigma}$. There is a one-to-many clique-preserving embedding $f$ from $G$ to a graph $H$ of treewidth at most $\frac{h2^{O(v(G)g(G))}\log n}{\epsilon}$ with additive distortion $\epsilon D$.
\end{restatable}

We then extend the embedding to  graphs embedded on a genus-$g$ surface with multiple vortices and apices (a.k.a. nearly embeddable graphs). The problem with apices, as pointed out at the beginning of this section, is that the diameter of the graph after removing apices could be unbounded in terms of the diameter of the original graph. Indeed, while the embedding in \Cref{lm:embed-genus-vortex} is deterministic, it is not clear how to \emph{deterministically} embed a nearly embeddable graph  into a bounded treewidth graph with additive distortion $\epsilon  D$. We use padded decompositions~\cite{Fil19padded}  to decompose the graph into clusters of strong diameter $O(D/\eps)$, embed each part separately, and then combine all the embeddings into a single graph. Note that separated nodes will have additive distortion as large as $2D$, however, this will happen with probability at most $O(\eps)$. 
To make this embedding clique-preserving, we add to each cluster its neighborhood. Thus some small fraction of the vertices will belong to multiple clusters.
As a result, we obtain a one-to-many stochastic embedding with expected additive distortion $\epsilon D$. The proof appears in \Cref{sec:EmbedApices}.
\begin{restatable}[Nearly $h$-Embeddable]{lemma}{embedNearlyEmbeddable}
	\label{lm:embed-nearly-embeddable}
	Given a nearly $h$-embeddable graph $G$ of diameter $D$, there is a one-to-many stochastic clique-preserving embedding into graphs with treewidth $O_h(\frac{\log n}{\epsilon^2})$ and expected additive distortion $\epsilon D$. Furthermore, every bag of the tree decomposition of every graph in the support contains (the image of) the apex set of $G$.
\end{restatable}

Finally we are in the case of general minor free graph  $G = G_1 \oplus_h G_2 \oplus_h \ldots \oplus_h G_s$. We sample an embedding for each $G_i$ using \Cref{lm:embed-nearly-embeddable} to some bounded treewidth graph $H_i$.  As all these embeddings are clique-preserving, there is a natural way to combine the tree decompositions of all the graphs $H_i$ together.
Here we run into another challenge: we need to guarantee that the additive distortion caused by merging tree decompositions is not too large. To explore this challenge, let us  consider the clique-sum  decomposition tree $\mathcal{T}$ of $G$: each node of $\mathcal{T}$ corresponds uniquely to $G_i$ for some $i$, and that $G$ is obtained by clique-summing all adjacent graphs  $G_i$ and $G_j$ in $\mathcal{T}$. Suppose that $\mathcal{T}$ has a (polynomially) long path $\mathcal{P}$ with hop-length $p$. Then, for a vertex $u$ in the graph corresponding to one end of $\mathcal{P}$ and a vertex $v$ in the graph corresponding to another end of $\mathcal{P}$, the additive distortion between $u$ and $v$ could potentially $p \epsilon  D$ since every time the shortest path between $u$ and $v$ goes through a graph $G_i$, we must pay additive distortion $\epsilon  D$ in the embedding of $G_i$. When $p$ is polynomially large, the additive distortion is polynomial in $n$.  We resolve this issue by the following idea:(1) pick a \emph{separator piece} $G_i$ of $\mathcal{T}$ ($G_i$ is a separator of $\mathcal{T}$ if each component $\mathcal{T}\setminus G_i$ has at most $2/3$ the number of pieces of $\mathcal{T}$), (2) recursively embed pieces in subtrees of $\mathcal{T}\setminus G_i$ and  (3) add the join set between $G_i$ and each subtree, say $\mathcal{T}'$ of $\mathcal{T}\setminus G_i$ to all bags of the tree decomposition corresponding to $\mathcal{T}'$. We then can show that this construction incurs another additive $\log n$ factor in the treewidth while insuring a total additive distortion of $\eps D$. Hence the final tree decomposition has width $O(\frac{\log n}{\epsilon^2})$.
The proof of  \Cref{thm:embedding-minor} appears in \Cref{sec:EmbedGeneralMinor}.

An interesting consequence of our one-to-many embedding approach is that the host graphs $H$ will contain Steiner points. That is, its vertex set will be greater than $V$. We do not know whether it is possible to obtain the properties of \Cref{thm:embedding-minor} while embedding into $n$-vertex graphs.
In this context, the Steiner point removal problem studies whether it is possible to remove all Steiner points while preserving both pairwise distance and topological structure \cite{Fil19SPR,Fil20scattering}. Unfortunately, in general, even if $G$ is a tree, a multiplicative distortion of $8$ is necessary \cite{CXKR06}.
Nevertheless, as Krauthgamer \etal \cite{KNZ14} proved, given a set $K$ of $k$ terminals in a graph $H$ of treewidth $\tw$, we can embed the terminal set $K$ isometrically (that is with multiplicative distortion $1$) into a graph with $O(k\cdot\tw^3)$ vertices and treewidth $\tw$. It follows that we can ensure that all embeddings in the support of the stochastic embedding in \Cref{thm:embedding-minor} are into graphs with $O_r(n\cdot \frac{\log^3 n}{\eps^6})$ vertices.

\section{Related work}\label{sec:related}

\paragraph{TSP in Euclidean and doubling metrics} Arora~\cite{Arora97} and Mitchell~\cite{Mitchell99} gave polynomial-time approximation schemes (PTASs) for TSP (Arora's algorithm is a PTAS for any fixed dimension). Rao and Smith~\cite{RS98} gave
an $O(n \log n)$ approximation scheme for bounded-dimension Euclidean
TSP, later improved to linear-time by Bartal and
Gottlieb~\cite{BartalG13}.  For TSP in doubling metrics,  Talwar~\cite{Tal04} gave a QPTAS;  Bartal \etal \cite{BGK16} gave a PTAS; and Gotlieb~\cite{Gottlieb15} gave efficient PTAS.

\paragraph{TSP and subset TSP in minor-closed families} For TSP problem in planar graphs,  Grigni \etal \cite{GKP95} gave the first (inefficient) PTAS for \emph{unweighted} graphs;  Arora \etal \cite{AGKKW98} extended Grigni \etal \cite{AGKKW98} to weighted graphs;  Klein \cite{Klein05} designed the first EPTAS by introducing the contraction decomposition framework.  Borradaile \etal \cite{BDT14} generalized Klein's EPTAS to bounded-genus graphs. 
The first PTAS for $K_r$-minor-free graph was desgined by Demaine \etal \cite{DHK11} that improved upon the QPTAS by Grigni \cite{Grigni00}. Recently, Borradaile \etal \cite{BLW17} obtained an EPTAS for TSP in $K_r$-minor-free graphs by connstructing light spanners; this work completed a long line of research on approximating classical TSP in $K_r$-minor-free graphs. 

For subset TSP, Arora \etal \cite{AGKKW98} designed the first QPTAS for weighted planar graphs. Klein \cite{Klein06} obtained the first EPTAS for subset TSP in planar graphs by constructing a light planar subset spanner. Borradaile \etal \cite{BDT14} generalized Klein's subset spanner construction to bounded-genus graphs, thereby obtained an EPTAS. Le \cite{Le20} designed the first (inefficient) PTAS for subset TSP in minor-free graphs.  Our \Cref{thm:tsp-eptas} completed this line of research.

\paragraph{Light (subset) spanners} Light and sparse spanners were introduced for distributed computing \cite{Awerbuch85,PS89,ABP91}. Since then, spanners attract ever-growing interest; see ~\cite{AGSHJKS19} for a survey. Over the years, light spanners with constant lightness have been shown to exist in Euclidean metrics~\cite{RS98,LS19}, doubling metrics~\cite{Gottlieb15,BLW19}, planar graphs~\cite{ADDJS93}, bounded genus graphs~\cite{Grigni00} and minor-free graphs~\cite{BLW17}.  For subset spanners, relevant results include subset spanners with constant lightness for planar graphs by Klein~\cite{Klein06}, for bounded genus graphs by Borradaile \etal~\cite{BDT14}. Le~\cite{Le20} constructed subset spanners with lightness $O(\log |K|)$ for minor-free graphs.

	\paragraph{Capacitaed vehicle routing} There is a rich literature on the capacitated vehicle routing problem. When $Q$ is arbitrary, the problem becomes extremely difficult as there is no known PTAS for any non-trivial metric. For $\mathbb{R}^2$, there is a QPTAS by Mathieu and Das for $\mathbb{R}^2$~\cite{DM15} and for tree metrics, there is a (tight) $\frac{4}{3}$-approximation algorithm by Becker~\cite{Becker18}. In general graphs,  Haimovich and Rinnooy Kan~\cite{HR85} designed a $2.5$-approximation algorithm. 
	
	In Euclidean spaces, better results were known for restricted values of $Q$: PTASes in $\mathbb{R}^2$ for $Q  = O( 2^{\log^{O_{\epsilon}(1)}n})$ by a sequence of work~\cite{HR85,AKTT97,ACL09} and for $Q = \Omega(n)$ by Asano et al.~\cite{ AKTT97}; a PTAS in $\mathbb{R}^d$  for $Q = O(\log n^{1/d})$ by Khachay and Dubinin~\cite{KD16}. 
	
	For constant $Q$, progress has been made on designing approximation schemes for various minor-closed families of graphs. In recent work, Becker et al.~\cite{BKS19} designed a PTAS for planar graphs. Becker et al.~\cite{BKS17} gave a QPTAS for planar and bounded-genus graphs. 
	
	Other relevant works include a PTAS for graphs of bounded highway dimension and constant $Q$~\cite{BKS18}, a bicriteria PTAS for tree metrics and arbitrary $Q$~\cite{BP19}, and an exact algorithm for treewidth-$\tw$ graphs with running time $O(n^{\tw Q})$~\cite{BKS18}.

	\section{Preliminaries} \label{sec:prelim}
	$O_r$ notation hides factors in $r$, e.g. $O_r(m)=O(m)\cdot f(r)$ for some function $f$ of $r$.
	
	We consider connected undirected graphs $G=(V,E)$ with edge weights
	$w_G: E \to \R_{\ge 0}$. Additionally, we denote $G$'s vertex set and edge set by $V(G)$ and $E(G)$, respectively. Let $d_{G}$ denote the shortest path metric in
	$G$, i.e., $d_G(u,v)$ equals to the minimal weight of a path from $u$ to $v$. Given a vertex $v$ and a subset of vertices $S$, $d_G(v,S)=\min_{u \in S}d_G(v,u)$ is the distance between $v$ and $S$. If $v \in S$, then $d_G(v,S) = 0$. When the graph is clear from the context, we simply use $w$ to refer to $w_G$, and $d$ to refer to $d_G$. 
	$G[S]$ denotes the induced subgraph by $S$. We define the \emph{strong}\footnote{The \emph{weak} diameter of $S$ is $ \max_{u,v \in S}d_{G}(u,v)$.} diameter of $S$, denoted by $\dm(S)$,  to be $\max_{u,v \in S}d_{G[S]}(u,v)$.
	For a subgraph $H$ of $G$, $w_G(H)=\sum_{e\in E(H)}w_G(e)$ denotes the total weight of all the edges in $H$.

	For two paths $P_1,P_2$ where the last vertex of $P_1$ is the first vertex of $P_2$. We denote by $P_1\circ P_2$ the 
	concatenation of $P_1$ and $P_2$.  We denote by $P[u,v]$ a subpath between $u$ and $v$ of $P$.
	
	We say a subset of vertices $N$ is a \emph{$r$-net} of $G$ if the distance between any two vertices of $N$ is at least $r$ and for every $x \in V(G)$, there exists $y \in N$ such that $d_G(x,y) \leq r$.
	
	Given a subset $U\subseteq V$ of vertices, a \emph{Steiner tree} of $U$ is an acyclic subgraph of $G$ such that all the vertices in $U$ belong to the same connected component. A \emph{Minimum Steiner tree} is a subgraph of minimum weight among all such subgraphs (it is not necessarily unique).
	Given a subset $U$ of terminals, a \emph{subset $t$-spanner w.r.t. $U$} is a subgraph $H$ that preserves the distances between any pair of terminals, up to a multiplicative factor of $t$, i.e.,  $\forall u,v\in U$, $d_H(u,v)\le t\cdot d_G(u,v)$. Note that as $H$ is a subgraph, it necessarily holds that $d_G(u,v)\le d_H(u,v)$.    The lightness of $H$ is the ratio of its weight $w(H)$ to the weight of a minimum Steiner tree of $U$.
 
	A metric embedding is a function $f:G\rightarrow H$ between two graphs $G$ and $H$.
	We say that metric embedding $f$ is \emph{dominating} if for every pair of vertices $u,v\in G$, it holds that $d_G(u,v)\le d_H(f(u),f(v))$. 
	\begin{definition}[Stochastic embedding]\label{def:stocastic}
		\sloppy A stochastic embedding, is a distribution $\mathcal{D}$ over dominating embeddings $f$. We say that a stochastic embedding has expected additive distortion $\eps D$, if  $\forall u,v\in G$ it holds that $\mathbb{E}_{f\sim\mathcal{D}}[d_H(f(u),f(v))]\le d_G(u,v)+\eps D$.
	\end{definition}
	\subsection{Robertson-Seymour decomposition of minor-free graphs}\label{subsec:RobertsonSeymour}
	
   In this section, we review notation used in graph minor theory by Robertson and Seymour. Readers who are familiar with Robertson-Seymour decomposition can skip this section. Basic definitions such as tree/path decomposition and treewidth/pathwidth are provided in \Cref{appendix:additionalNotation}. 
   	 
	Informally speaking, the celebrated theorem of Robertson and Seymour (\Cref{thm:RS}, \cite{RS03}) said that any minor-free graph can be decomposed into a collection of graphs \emph{nearly embeddable} in the surface of constant genus, glued together into a tree structure by taking \emph{clique-sum}.  To formally state the Robertson-Seymour decomposition, we need additional notations.

	A \emph{vortex} is a  graph $G$ equipped with a pah decomposition $\{X_1,X_2,\ldots, X_t\}$ and a sequence of $t$ designated vertices $x_1,\ldots, x_t$, called the \emph{perimeter} of $G$, such that each $x_i \leq X_i$ for all $1\leq i \leq t$. The \emph{width} of the vortex is the width of its path decomposition.
	We say that a vortex $W$ is \emph{glued} to a face $F$ of a surface embedded graph $G$ if $W\cap F$ is the perimeter of $W$ whose vertices appear consecutively along the boundary of $F$. 
	
	\paragraph{Nearly $h$-embeddability}  A graph $G$ is nearly $h$-embeddable if  there is a set of at most $h$ vertices $A$, called \emph{apices}, such that $G\setminus A$ can be decomposed as $G_{\Sigma}\cup \{W_1, W_2,\ldots, W_{h}\}$ where $G_{\Sigma}$ is (cellularly) embedded on a surface $\Sigma$ of genus at most $h$ and each $W_i$ is a vortex of width at most $h$ glued to a face of $G_{\Sigma}$.

	\paragraph{$h$-Clique-sum} A graph $G$ is a $h$-clique-sum of two graphs $G_1,G_2$, denoted by $G  = G_1\oplus_h G_2$, if there are two cliques of size exactly $h$ each such that $G$ can be obtained by  identifying vertices of the two cliques and remove some clique edges of the resulting identification. 
	
	Note that clique-sum is not a well-defined operation since the clique-sum of two graphs is not unique due to the clique edge deletion step. We now can state the decomposition theorem.

	\begin{theorem}[Theorem~1.3~\cite{RS03}] \label{thm:RS} There is a constant $h = O_r(1)$ such that any $K_r$-minor-free graph $G$ can be decomposed into a tree $\mathcal{T}$ where each node of $\mathcal{T}$ corresponds to a nearly $h$-embeddable graph such that $G = \cup_{X_iX_j \in E(\mathcal{T})} X_i \oplus_h X_j$.
	\end{theorem} 
	
	By slightly abusing notation, we use the term \emph{nodes} of $\mathcal{T}$ to refer to both the nodes and the         graphs corresponding to the nodes of $\mathcal{T}$. Note that nodes of $\mathcal{T}$ may not be
        subgraphs of $G$, as in the clique-sum, some edges of a node, namely some edges of a nearly $h$-embeddable subgraph associated to a node,
        may not be present in $G$. However, for any edge $(u,v)$ between two vertices of a node, say $X$, of $\mathcal{T}$, that are not present in $G$,
        we add edge $(u,v)$ to $G$ and set its weight to be $d_G(u,v)$. It is immediate that this does not
        change the Robertson-Seymour decomposition of the graph, nor its shortest path metric.
        Thus, in the decomposition of the resulting graph, the clique-sum operation does not remove any edge.
        This is an important point to keep in mind as in what follows, we will remove some nodes out of $\mathcal{T}$ while guaranteeing that
        the shortest path metric between terminals is not affected.

\subsection{Vortex paths}
	Throughout the paper, we will use the notion of \emph{vortex-path}, which was first introduced by Abraham and Gavoille~\cite{AG06}.

\begin{definition}[Vortex-path~\cite{AG06}]\label{def:vortex-path} Given a vortex embedded graph $G = G_{\Sigma}\cup W_1\cup W_2\ldots \cup W_{v(G)}$, a \emph{vortex-path} between two vertices $u,v$, denoted by $\mathcal{V}[u,v]$, is a subgraph of $G$ that can be written as $\mathcal{V}[u,v] = P_0\cup X_1\cup Y_1\cup P_1 \cup \ldots \cup X_\ell \cup Y_\ell\cup P_{\ell}$ such that:
	\begin{itemize}[nolistsep,noitemsep]
		\item[(a)]  $P_i$ is a path of $G_{\Sigma}$ for all $0\leq i\leq \ell$.
		\item[(b)] For all $1\leq i\leq \ell$, $X_i$ and $Y_i$ are two bags of the same vortex, denoted $\mathcal{W}(X_i)$.
		\item[(c)] For any $1\leq i\not= j\leq \ell$, $\mathcal{W}(X_i) \not = \mathcal{W}(X_j)$.
		\item[(d)] $P_0$ ($P_{\ell}$)  is a path from $u$ ($v$) to a perimeter in $X_1$ ($Y_\ell$). $P_i$ is a path from a perimeter vertex in $Y_{i}$ to a perimeter vertex in $X_{i+1}$, $1\leq i\leq \ell-1$. No path $P_i$ contains a perimeter vertex as an internal vertex for any $i \in [0,\ell]$.
	\end{itemize}
	Each path $P_i$ is called a \emph{segment} of $\mathcal{V}$.
\end{definition}

\begin{figure}[]
	\centering{\includegraphics[width=1\textwidth]{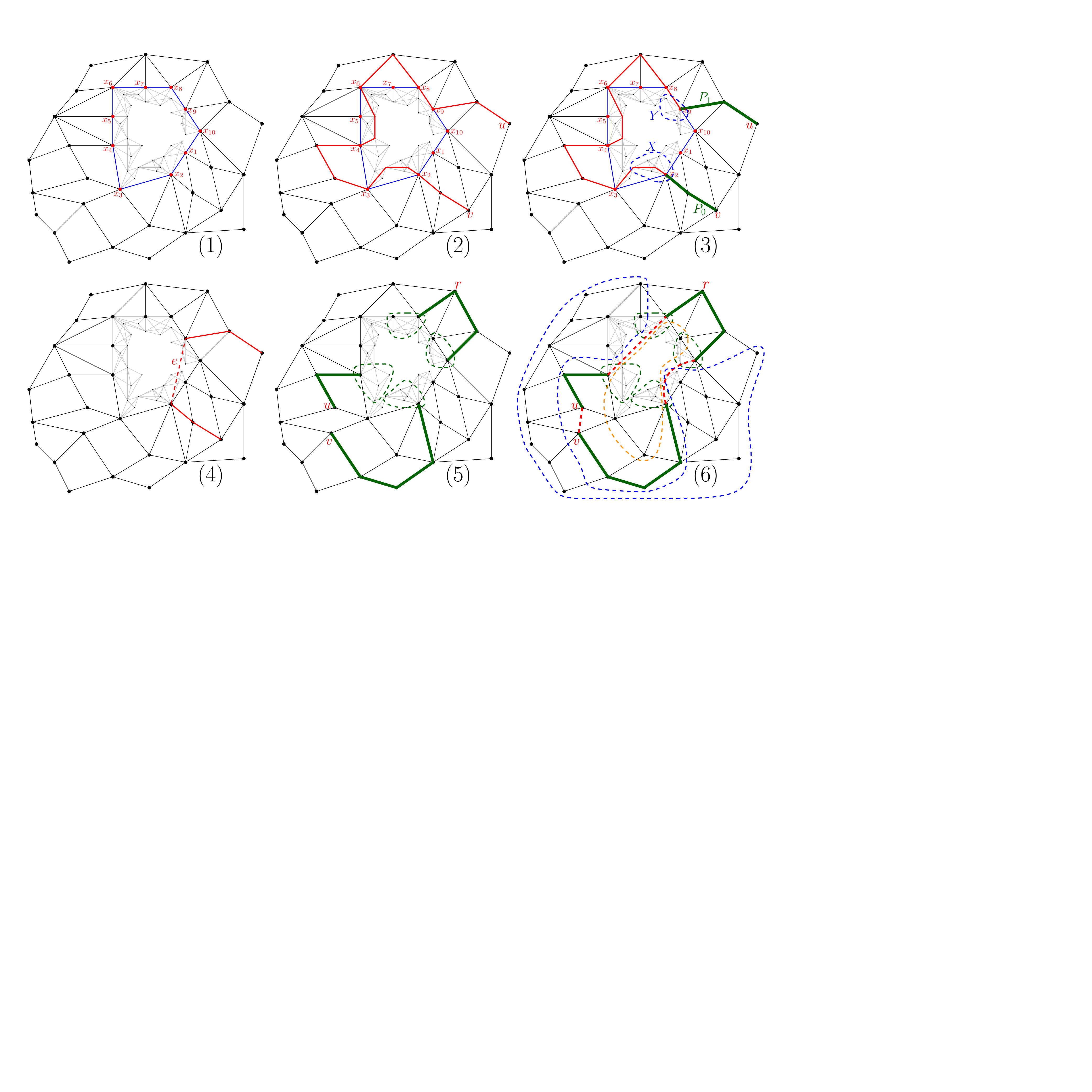}} 
	\caption{\label{fig:vortexDef}\small \it 
		In part (1) displayed a planar graph with a single vortex. The vortex embedded on a face colored in blue. The perimeter vertices are colored in red. The vertices belonging to the vortex but not to the planar graph displayed by smaller dots (while their edges in gray).\newline
		In part (2) a path $P$ from $v$ to $u$ is displayed, in red.\newline
		In part (3) displayed a vortex path $\mathcal{V}[v,u]=P_0\cup X\cup Y\cup P_1$ which induced by $P$. Here $P_0,P_1$ displayed in green, where $P_0$ is the prefix of $P$ from $v$ to $x_2$, and $P_2$ is the suffix of $P$ from $x_9$ to $u$. $X$ (resp. $Y$) which is encircled by a dashed blue line, is the bag $X_2$ ($X_9$) associated with the perimeter vertex $x_2$ ($x_9$).\newline
		In part (4) displayed in red the projection $\bar{\mathcal{V}}$ of the vortex path $\mathcal{V}$. $\bar{\mathcal{V}}$ consist of $P_0,P_1$, and an imaginary edge $e$ (dashed) between $x_2$ to $x_9$. \newline
		In part (5) displayed a fundamental vortex cycle $C$, where it's embedded is in green. All the vertices in $\mathcal{W}(C)$ are encircled by a blue line.\newline
		In part (6) we display the close curve $\bar{C}$ induced by $C$. $e_C$, as well as the other imaginary edges are displayed by a red dashed lines. The interior of $\bar{C}$ is encircled by an orange dashed line, while the exterior is encircled by an purple dashed line.
	}
\end{figure}

When the endpoints of a vortex-path $\mathcal{V}[u,v]$ are not relevant in our discussion, we would omit the endpoints and simply denote it by $\mathcal{V}$. 
The \emph{projection} of the  vortex-path  $\mathcal{V}[u,v]=P_0\cup X_1\cup Y_1\cup P_1 \cup \ldots \cup X_\ell \cup Y_\ell\cup P_{\ell}$ denoted by $\bar{\mathcal{V}}$ is a path formed by $P_0\circ e_1\circ P_1\circ e_2\circ\dots\circ e_{\ell}\circ P_\ell$ where $e_i$ is an (imaginary) extra edge added to $G_\Sigma$ between the perimeter vertex of $X_i$ and the perimeter vertex of $Y_i$, and embedded inside the cellular face upon which the vortex $W_i$ is glued. We observe that even though $\mathcal{V}$ may not be a path of $G$, its projection 
$\bar{\mathcal{V}}$ is a curve of $\Sigma$. See \Cref{fig:vortexDef} for a simple example and see \Cref{fig:vortex-path-complex} in \Cref{appendix:additionalNotation} for a more complex one.

Consider a path $Q=(u=v_0,v_1,\dots,v_r=v)$ in $G$ with two endpoints in the embedded part $G_{\Sigma}$. $Q$ \emph{induces} a vortex path $\mathcal{V}[u,v]$ defined as follows:
Start a walk on $Q$ until you first encounter a vertex $v_{i_1}$ that belongs to a vortex $W_{i_1}$. Let $u_{i_1}$ be the last vertex in $Q$ belonging to $W_{i_1}$. Note that necessarily  $v_{i_1},u_{i_1}$ are perimeter vertices in $W_{i_1}$, denote them $x^{i_1}_{j_1}
,x^{i_1}_{l_1}$ respectively. We continue and define $v_{i_2}$ to be the first vertex (after $u_{i_1}$) belonging to some vortex $W_{i_2}$, and $u_{i_2}$ being the last vertex in $Q\cap W_{i_2}$. $x^{i_2}_{j_2}
,x^{i_2}_{l_2}$ are defined in the natural manner. We iteratively define $\left(v_{i_3}=x^{i_3}_{j_3},W_{i_3},u_{i_3}=x^{i_3}_{l_3}\right),\left(v_{i_4}=x^{i_4}_{j_4},W_{i_4},u_{i_4}=x^{i_4}_{l_4}\right),\dots$ until the first index $i_s$ such that there is no vertex $v_{i_{s+1}}$ after $u_{i_s}$ belonging to a vortex.
The respective induced vortex path is defined as $P_0\cup X_1\cup Y_1\cup P_1 \cup \ldots \cup X_\ell \cup Y_s\cup P_{s}$ where $P_0=(v_0,\dots,v_{i_1})$, $P_q=(u_{i_q},\dots,v_{i_{q+1}})$, $X_q$ (resp. $Y_q$) is a bag in $W_{i_q}$ associated with the perimeter vertex $x^{i_q}_{j_q}$ (resp. $x^{i_q}_{l_q}$), and $P_{s}= Q[u_{i_s}, v]$. See \Cref{fig:vortexDef}.

Suppose next that $G$ has genus $0$. Specifically, that $G = G_{\Sigma}\cup  W_1\cup\dots \cup W_{h'}$,  where $G_{\Sigma}$ can be drawn on the plane, $h'\le h$, and each $W_i$ is a vortex of width at most $h$ glued to a face of $G_{\Sigma}$.
Fix some drawing of $G_{\Sigma}$ on the plane, let $T_r$ be an arbitrary spanning tree of $G$ rooted at $r$.

A \emph{fundamental vortex cycle} $C$ of $T$ is a union of vortex paths $\mathcal{V}[r,v]\cup \mathcal{V}[r,u]$, induced by two paths $Q_1,Q_2$, both starting at the root $r$, end at $u,v\in G_\Sigma$, such that either $u,v$ are neighbors in $G_\Sigma$, or a curve could be added between $u$ to $v$ without intersecting any other curve in the drawing $G_{\Sigma}$. Denote this edge/imaginary curve by $e_C$. 
We call the union of the projections, $\bar{\mathcal{V}}[r,v]\cup \bar{\mathcal{V}}[r,u]$ the \emph{embedded} part of $C$.
Adding $e_C$ to the embedded part, $\bar{\mathcal{V}}[r,v]\cup \bar{\mathcal{V}}[r,u]\cup e_C$ induces a close curve $\bar{C}$ which is associated with $C$. 
Removing the fundamental vortex cycle $C$ from $G$ partitions $G\setminus C$ into two parts, interior $\mathcal{I}$ and exterior $\mathcal{E}$. 
The embedded part $G_{\Sigma}$ is partitioned to interior $\mathcal{I}\cap G_\Sigma$ and exterior $\mathcal{E}\cap G_\Sigma$, w.r.t. the closed curve $\bar{C}$ associated with $C$. For every vertex $z$ belonging to the vortex only ($z\in W\setminus G_{\Sigma}$), which was not deleted, let $X_i$ be an arbitrary bag containing $z$. Note that $X_i$ is not one of the bags belonging to the fundamental vortex cycle. In particular, even though it might be deleted, the perimeter vertex $x_i$ belongs either to the interior or the exterior of $\bar{C}$.
If $x_i$ is in the interior, respectively exterior, part of $G \setminus C$, then
vertex $z$ joins the interior $\mathcal{I}$, respectively exterior $\mathcal{E}$, part of $G\setminus C$.
Note that cycle vertices $C$ belong to neither to the interior or the exterior.

\begin{claim}\label{clm:CisSeperator}
	$\mathcal{I},C,\mathcal{E}$ form a partition of $G$. Further, there are no edges between $\mathcal{I}$ and $\mathcal{E}$.
\end{claim}
\begin{proof}
	Let $u\in \mathcal{E}$, and $v\in\mathcal{I}$. Assume for contradiction that they are neighbors in $G$, denote $e=(u,v)$. We continue by case analysis,
	\begin{OneLiners}
		\item  Suppose $e\in G_\Sigma$. In this case $e$ must cross the closed curve $\bar{C}$, a contradiction.
		\item Suppose $e\notin G_\Sigma$. Then $u,v$ must belong to the same vortex $W$ (they might be perimeter vertices, however $e$ must belong to $W$). 
		Denote by $I_{u}=\{i\mid u\in X_i\}$ and  $I_{v}=\{i\mid v\in X_i\}$ the set of indices belonging to bags containing $u$ and $v$, respectively.
		By the definition of path decomposition, there are integers $a_u,b_u,a_v,b_v$ such that $I_{u}=[a_u,b_u]$ and $I_{v}=[a_v,b_v]$. 
		As $u\in \mathcal{E}$ and  $v\in \mathcal{I}$, there are indices $i_u\in [a_u,b_u]$, $i_v\in [a_v,b_v]$ such that $x_{i_u}$ is outside $\bar{C}$, while $x_{i_v}$ is inside $\bar{C}$. W.l.o.g. $i_u\le i_v$. The curve $\bar{C}$ must intersect the path $(x_{i_u},x_{i_u+1},\dots, x_{i_v})$ at a perimeter vertex $x_{i_c}$, where the entire bag $X_{i_c}$ belongs to the fundamental vortex cycle $C$. As $u,v$ do not belong to $C$, it must hold that $i_c\notin [a_u,b_u]\cup[a_v,b_v]$, implying that $a_u\le i_u\le b_u<i_c<a_v\le i_v\le b_v$. Thus $I_{u}\cap I_{v}=\emptyset$, a contradiction to the assumption that $u$ and $v$ are neighbors in $W$.	
	\end{OneLiners}
	
\end{proof}

We will use the following lemma, which is a generalization of the celebrated Lipton-Tarjan planar separator theorem \cite{LT79,Tho04}, to planar graphs with vortices.
This is a slightly different \footnote{Originally \cite{AG06} used three vortex-paths to separate the graph into components of weight at most $\mathcal{W}/2$ each. Here we use two vortex-paths, but each component has at weight at most $\frac{2\mathcal{W}}{3}$ instead. Additionally, \cite{AG06} is more general and holds for an arbitrary number of vortices.} version of Lemma 6 in~\cite{AG06} (Lemma 10 in the full version).	
\begin{lemma}[\cite{AG06}]\label{lm:AG06one-vortex-separator} Consider a graph $G = G_{\Sigma}\cup  W_1\cup\dots \cup W_{h'}$,  where $G_{\Sigma}$ can be drawn on the plane and $W_1\cup\dots \cup W_{h'}$ are vortices  glued to a faces of $G_{\Sigma}$. Let $T_r$ be a spanning tree of $G$ rooted at $r$, and a weight function $\omega:V\ra\R_+$ over the vertices. Set $\mathcal{W}=\sum_{v\in V}\omega(v)$ to be the total vertex weight of $G$. Then there is a fundamental vortex cycle $C$, such that both the interior and exterior in $G\diagdown C$ has vertex weight at most $\frac{2\mathcal{W}}{3}$, i.e., $\sum_{v\in\mathcal{I}}\omega(v),\sum_{v\in\mathcal{E}}\omega(v)\le \frac{2\mathcal{W}}{3}$.
\end{lemma}

\section{Light Subset Spanners for Minor-Free Metrics}

In our construction, we will use \emph{single-source spanners} and \emph{bipartite spanners} as black boxes. These concepts were introduced initially by Klein~\cite{Klein06} for planar graphs, and then generalized to general graphs by Le~\cite{Le20}.

\begin{lemma}[Single-source spanners~\cite{Le20}]\label{lm:ss-spanner} Let $p$ be a vertex and $P$ be a shortest path in an edge-weighted graph $G$. Let $d(p,P) = R$. There is a subgraph $H$ of $G$ of weight at most $8R\epsilon^{-2}$ that can be computed in polynomnial time such that:
	\begin{equation} \label{eq:ss-spanner}
	d_G(p,x) \leq d_{P\cup H}(p,x) \leq (1+\epsilon) d_G(p,x) \qquad\forall x \in P
	\end{equation}
\end{lemma}

\begin{lemma}[Bipartite spanners~\cite{Le20}]\label{lm:bipartite-spanner}
	Let $W$ be a path and $P$ be a shortest path in an edge-weighted graph $G$. Let $R = \min_{v \in W}d_G(v, P)$ be the distance between $W$ and $P$. Then, there is a subgraph $H$ constructible in polynomnial time such that $$d_{H\cup P}(p,q) \leq (1+\eps)d_{G}(p,q) \qquad \forall p \in W, q \in P$$ and  $w(H) =  O(\eps^{-3})w(W) + O(\eps^{-2})R$.
\end{lemma}

Lemma~\ref{lm:ss-spanner} is extracted from Lemma 4.2 in~\cite{Le20}, and Lemma~\ref{lm:bipartite-spanner} is extracted from Corollary 4.3 in~\cite{Le20}.  Given a shortest path $P$, we denote by $\ssp(t,P, G)$ a single-source spanner from $t$ to $P$ with stretch $(1+\eps)$ constructed using \Cref{lm:ss-spanner}.
Given a parameter $L>0$, denote
\[
\ssp(t,P,G,L)=\begin{cases}
\ssp(t,P,G) & d_{G}(t,P)\le L\\
\emptyset & d_{G}(t,P)>L
\end{cases}
\]
That is, in case $d_{G}(t,P)\le L$, $\ssp(t,P,G,L)=\ssp(t,P,G)$, while otherwise it is an empty-set. 
Similarly, given a path $W$ and a shortest path $P$, let $\BS(P,W,G)$ be a bipartite spanner $P$ to $W$ with stretch $(1+\eps)$
constructed using \Cref{lm:bipartite-spanner}.

\subsection{Step (1): Planar graphs with a single vortex and bounded diameter, proof of \Cref{lm:one-vortex-Bounded-diam}}\label{sec:oneVortex}
We begin by restating the main lemma of the section:
\SingleVortexBoundedDiam*
	
This section contains a considerable amount of notations. \Cref{appendix:key} contains a summary of all the definitions and notations
used in the section. The reader is encouraged to refer to this index while reading.
	Recall that a \emph{vortex} is a  graph equipped with a path decomposition $\{X_1,X_2,\ldots, X_t\}$ and a sequence of $t$ designated vertices $x_1,\ldots, x_t$, called the \emph{perimeter} of the vortex, such that $\forall i\in[t]$, $x_i \in X_i$. The \emph{width} of the vortex is the width of its path decomposition.
	We say that a vortex $W$ is \emph{glued} to a face $F$ of a surface embedded  graph $G_{\Sigma}$ if $W\cap F$ is the perimeter of $W$ whose vertices appear consecutively along the boundary of $F$. 
	
	\paragraph{Graph preprocessing.} In order to simplify the spanner construction and its proof, we modify the graph as follows.
    We add an auxiliary vertex $\tilde{x}$, with weight-$D$ edges to all the vertices in the vortex $W$, where $D$ is the diameter of the graph. 
	In the drawing $G_\Sigma$, we add an arc between the perimeter vertices $x_1,x_t$ and draw $\tilde{x}$ somewhere along this arc.
        
    $\tilde{x}$ is added to the vortex, which is now considered to have perimeter $\tilde{x},x_1,\ldots, x_t$. Note that only the edges $\{\tilde{x},x_1\},\{\tilde{x},x_2\}$ are added to the embedded part.
    The bag associated with $\tilde{x}$ is the singleton $\{\tilde{x}\}$. Every other perimeter vertex $x_i$, has an associated bag $X_i\cup \{\tilde{x}\}$.
    See \Cref{fig:vortexDef} for illustration of the modification. 
	As a result, we obtain a planar graph with a single vortex of width at most $h+1$. Note that the diameter is still bounded by $D$.
	We abuse notation and call this graph $G$, its drawing (i.e. planar part) $G_\Sigma$, and its vortex $W$.
    In the following, we show how to construct a subset spanner of this graph. A subset spanner of this graph immediately yields
    a subset spanner of the original graph, we can simply discard $\tilde{x}$ and the resulting subset spanner would indeed be a subset spanner
    of the original graph. To see this, observe that for any pair of terminals $t,t'$ their shortest path in the subset spanner of $G$
    does not go through $\tilde{x}$ since otherwise their distance would be greater than $2D\ge 2\cdot d_G(t,t')$.
	\begin{figure}[]
		\centering{\includegraphics[width=.8\textwidth]{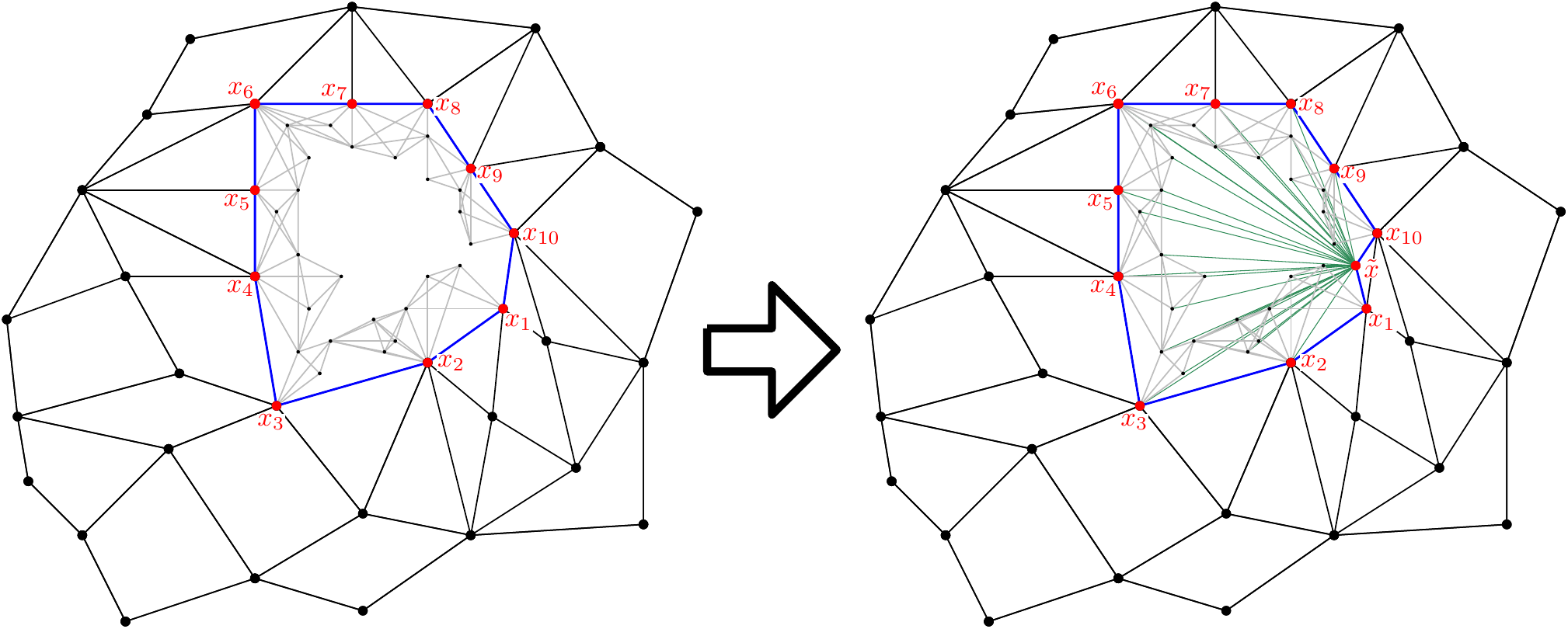}}
		\caption{\label{fig:vortexModification}\small \it 
			The modification step illustrated. On the left there is a planar graph with a single vortex, glued to a face colored in blue. The perimeter vertices are colored in red, while all the edges not in the embedded part are colored in gray. On the right is illustrated a modified graph, where we add a new vertex $\tilde{x}$ with edges towards all other vortex vertices. The perimeter vertices, and the face upon which the vortex is glued, are updated accordingly.
		}
	\end{figure}

      Next, we construct a tree that will be used later to create separators. Let 
        $T_\Sigma$ be a shortest path tree of $G_\Sigma$ rooted in $W$. Note that $T_\Sigma$ has $t+1$ connected components. Furthermore, every path in $T_\Sigma$ from a perimeter vertex $x_j$ to a vertex $v\in G_\Sigma$ will be fully included in $G_\Sigma$, and will have length at most $D=O_h(L)$. 
	We extend $T_W$ to $T_{\tilde{x}}$, a spanning tree of $G$ by adding an edge from $\tilde{x}$ to every vortex vertex, formally $T_{\tilde{x}}=T_\Sigma\cup\{(\tilde{x},v)\mid v\in W\setminus\{\tilde{x}\}\}$. We think of $T_{\tilde{x}}$ as a  spanning tree rooted at $\tilde{x}$.
	This choice of root and tree, induces a restricted structure on vortex paths and fundamental vortex cycles. Specifically, consider a path $Q=(\tilde{x}=v_0,\dots,v_q=v)$ from the root $\tilde{x}$ to a vertex $v\in G_\Sigma$ and let  $\mathcal{V}[\tilde{x},v]=P_0\cap X\cup Y\cup P_1$ be the induced vortex path. 
	As there are no edges from $\tilde{x}$ towards $G_\Sigma\setminus W$, necessarily $v_1=x_i$ is a perimeter vertex (there are no other neighbors of $\tilde{x}$ on a path towards a vertex in $G_\Sigma$). It holds than $P_0=(\tilde{x})$ is a singleton path, $X=\tilde{X}$, $Y=X_i$, and $P_1=(x_i=v_1,v_2,\dots,v_q=v)$.
	Furthermore, consider a fundamental vortex cycle $C$ that consists of the two vortex paths $\mathcal{V}[\tilde{x},v],\mathcal{V}[\tilde{x},u]$.
	Then $C$ actually contains $\tilde{x}$, two vortex bags $X_i,X_j$, and two paths from in $T_\Sigma$ from $x_i,x_j$
        to vertices $v,u$ in $G_\Sigma$, of length $O_h(L)$.
	
	\paragraph{Hierarchical tree construction}
	We recursively apply \Cref{lm:AG06one-vortex-separator} to hierarchically divide the vertex set $V$ into disjoint subsets. 
	Specifically, we have a hierarchical tree $\tau$ of sets with origin $V$. Each node in $\tau$ is associated with a subset $\Upsilon$, and a graph $G_\Upsilon$, which contains $\Upsilon$ and in addition vertices out of $\Upsilon$.
	We abuse notation and denote the tree node by $\Upsilon$. In the same level of $\tau$, all the vertex sets will be disjoint, while the same vertex might belong to many different subgraphs. 
	We maintain the following invariant:
	\begin{invariant}\label{inv:SubgraphSingleVortex}
	  The vertex set of the graph $G_\Upsilon$ is a subset of $G$.
	  It contains a single vortex $W_\Upsilon$, and a drawing in the plane which coincides with that of $G_\Sigma$.
	  Each perimeter vertex $x^\Upsilon_i$ of $W_\Upsilon$ is also a perimeter vertex $x_{i'}$ in $G_\Sigma$.
      Furthermore, the bag $X^\Upsilon_i$ associated with $x^\Upsilon_i$ equals to $X_{i'}\cap G_\Upsilon$, the bag associated with $x_{i'}$. 
	  Finally, there exists a set $\calE$ of perimeter edges that have been added by the algorithm such that
      the graph $G_\Upsilon - \calE$ is an induced subgraph of $G$.
	\end{invariant}

	The root vertex $\tilde{x}$ belongs to all the subgraphs $G_\Upsilon$ of all the tree nodes $\Upsilon\in \tau$.
	Consider the subgraph  $T_\Upsilon=T_{\tilde{x}}\cap G_\Upsilon$ rooted at $\tilde{x}$.
	We also maintain the following invariant:
	\begin{invariant}\label{inv:tree}
		$T_\Upsilon$ is a spanning tree of $G_\Upsilon$.
	\end{invariant}
	It follows from \Cref{inv:SubgraphSingleVortex} and \Cref{inv:tree}, that in similar manner to $T_{\tilde{x}}$, every fundamental vortex
        cycle consists of $\tilde{x}$, two bags and two paths of length $O_h(L)$ originated in perimeter vertices.
	Given a fundamental cycle $C$, we denote by $\mathcal{P}(C)$ the set of at most $2(h+1)+1$ paths from which $C$ is composed.
        We abuse notation here and treat the vertices in the deleted bags as singleton paths.

	In each hierarchical tree node $\Upsilon$, if it contains between $1$ to $2(h+1)$ terminals (that is $1\le |\Upsilon\cap K|\le 2(h+1)$), it is defined
        as a leaf node in $\tau$. Otherwise, we use \Cref{lm:AG06one-vortex-separator} to produce a fundamental vortex
        cycle $C_\Upsilon$, w.r.t. $T_\Upsilon$ and a weight function to be specified later.
	Using the closed curve $\bar{C}_\Upsilon$ induced by $C_\Upsilon$, the 
	set $\Upsilon$  is partitioned to interior $\Upsilon^{\mathcal{I}}$ and exterior $\Upsilon^{\mathcal{E}}$. $\Upsilon^{\mathcal{I}}$ and
        $\Upsilon^{\mathcal{E}}$ are the children of $\Upsilon$ in $\tau$ (unless they contain no terminals, in which case they are discarded).

	Note that the graph $G_\Upsilon$ contains vertices out of $\Upsilon$. Thus the exterior and interior of  $C_\Upsilon$ in $G_\Upsilon$ may
        contain vertices out of $\Upsilon$. Nonetheless, $\Upsilon^{\mathcal{E}},\Upsilon^{\mathcal{I}}$ consists of subsets of $\Upsilon$. Formally, they
        are defined as the intersection of $\Upsilon$ with the exterior and interior of  $C_\Upsilon$ in $G_\Upsilon$, respectively.
	By the definition  of $T_\Upsilon$ and assuming \Cref{inv:tree} indeed holds,  we deduce:
	\begin{observation}\label{obs:FundamentalPathLenght}
	  Every path $Q\in \mathcal{P}(C_\Upsilon)$ is one of the following:
	  \begin{enumerate}
	    \initOneLiners
	  \item A path $Q$ in $T_{\tilde{x}}$ from a perimeter vertex $x_i$ to a vertex $v\in G_\Upsilon\setminus W_\Upsilon$. In particular $Q$ is a shortest path in $G$ of length $O_h(L)$.\label{type:simple}
	  \item A singleton vortex vertex $u\in W_\Upsilon$.\label{type:singleton}
	  \end{enumerate}
	\end{observation}
        
	Denote by $\mathcal{C}_\Upsilon$ the set of all the fundamental vortex cycles removed from the ancestors of $\Upsilon$ in $\tau$. 
	Denote by $\bar{\mathcal{C}}_\Upsilon$ the set of paths constituting the fundamental vortex cycles in $\mathcal{C}_\Upsilon$.
	Note that \Cref{obs:FundamentalPathLenght} also holds for all the ancestors of $\Upsilon$ in $\tau$.
	Each path $Q\in \bar{\mathcal{C}}_\Upsilon$ will have a \emph{representative vertex} $v_Q$. Specifically, for a path $Q$ of type (\ref{type:simple}), set $v_Q=v$, while for a singleton path $Q=(u)$ (path of type (\ref{type:singleton})) set $v_Q=u$.

	Finally, we define $\mathcal{P}_\Upsilon\subseteq \bar{\mathcal{C}}_\Upsilon$, the subset of shortest paths that are added to $G_\Upsilon$.
    Intuitively, $Q\in \bar{\mathcal{C}}_\Upsilon$ joins $\mathcal{P}_\Upsilon$ if it has a neighbor in $\Upsilon$. However, we would like to avoid double counting that might appear due to intersecting paths.
    Formally this is a recursive definition. For $\Upsilon=V$, $\mathcal{P}_V=\emptyset$ and $G_V=G$. Consider $\mathcal{P}_{\Upsilon}$ and $G_{\Upsilon}$.
   	We define next $\mathcal{P}_{\Upsilon^{\mathcal{E}}}$ and $G_{\Upsilon^{\mathcal{E}}}$ (which will also imply the definition of $T_{\Upsilon^{\mathcal{E}}}$). $\mathcal{P}_{\Upsilon^{\mathcal{I}}}$ and $G_{\Upsilon^{\mathcal{I}}}$ are defined symmetrically.
    $\mathcal{P}_{\Upsilon^{\mathcal{E}}}$ will contain all the paths in $\mathcal{P}(C_\Upsilon)$ (the at most $2(h+1)+1$ paths composing $C_\Upsilon$). In addition, we add to $\mathcal{P}_{\Upsilon^{\mathcal{E}}}$  every path $Q\in\mathcal{P}_{\Upsilon}$ such that the 
    representative vertex $v_Q$ belongs to the exterior of $C_\Upsilon$ in $G_\Upsilon$.
    
    The graph $G_{\Upsilon^{\mathcal{E}}}$ is defined as the graph induced by the vertex set $\Upsilon^{\mathcal{E}}$ and all the vertices belonging to paths in $\mathcal{P}_{\Upsilon^{\mathcal{E}}}$. 
    In addition, in order to maintain the vortex intact, we add additional edges between the perimeter vertices. Specifically, suppose that the perimeter vertices in $G_\Upsilon$ are $\tilde{x},x_1,\dots,x_q$, while only $\tilde{x},x_{i_1},\dots,x_{i_l}$ belong to $G_{\Upsilon^{\mathcal{E}}}$. Then we add the edges $\{x_{i_1},x_{i_2}\},\{x_{i_{s-1}},x_{i_{s}}\},\dots, \{x_{i_{s-1}},x_{i_s}\}$ (unless this edges already exist),
    where the weight of $\{x_{i_{j}},x_{i_{j+1}}\}$ is $d_G(x_{i_{j}},x_{i_{j+1}})$. 
    
	We maintain the following invariant:
	\begin{invariant}\label{inv:holes}
	  $|\mathcal{P}_\Upsilon|\le 12\cdot (h+1)$.
	\end{invariant}
	
	It is straightforward that \Cref{inv:SubgraphSingleVortex} is maintained.
	\begin{claim}\label{clm:InvTreeMaintained}
		\Cref{inv:tree} is maintained.
	\end{claim}
	\hspace{-18pt}\emph{Proof.~}
		We will show that $T_{\Upsilon^{\mathcal{E}}}$ is a tree, the argument for $T_{\Upsilon^{\mathcal{I}}}$ is symmetric. 
		It is clear that $T_{\Upsilon^{\mathcal{E}}}$ is acyclic, thus it will be enough to show that it is connected. As $\tilde{x}$ is part of the fundamental vortex cycle $C_\Upsilon$, $\tilde{x}\in G_{\Upsilon^{\mathcal{E}}}$, it thus belongs to $T_{\Upsilon^{\mathcal{E}}}$.
		We show that every vertex contains a path towards $\tilde{x}$.
		Consider a vertex $u\in G_{\Upsilon^{\mathcal{E}}}$.
		First note that if $u\in W_{\Upsilon^{\mathcal{E}}}$, then $(\tilde{x},u)\in T_{\Upsilon}\cap G_{\Upsilon^{\mathcal{E}}}=T_{\Upsilon^{\mathcal{E}}}$.
		Next, if $u\in G_{\Upsilon^{\mathcal{E}}}\setminus \Upsilon^{\mathcal{E}}$, then $u$ is a part of some path $Q_u\in \mathcal{P}_{\Upsilon^{\mathcal{E}}}$ from $u$ to a perimeter vertex $x_{i_u}$. Here $Q_u\cup (r,x_{i_u})\subseteq T_{\Upsilon}\cap G_{\Upsilon^{\mathcal{E}}}=T_{\Upsilon^{\mathcal{E}}}$, thus we are done.
		
		For the last case ($u\in \Upsilon^{\mathcal{E}}\setminus W_{\Upsilon^{\mathcal{E}}}$), let $x_i,x_j$ be the two perimeter vertices such that the fundamental vortex cycle $C_\Upsilon$ contains the paths $Q_i,Q_j$ starting at $x_i,x_j$. \footnote{It is possible that $x_i=x_j$. In this case, we will abuse notation and treat $Q_i,Q_j$ as different paths.\label{foot:diffPaths}}
		Let $Q_u\subseteq T_\Upsilon$ be a path from a perimeter vertex towards $u$ (exist by the induction hypothesis).
		Note that the paths $\tilde{x}\circ Q_i,\tilde{x}\circ Q_j,\tilde{x}\circ Q_v$ are all paths in a tree $T_\Upsilon$. In particular, while they might have a mutual prefix, once they diverge, the paths will not intersect again. 
		Denote $\tilde{x}\circ Q_u=(\tilde{x}=v_0,v_1,\dots,v_q=u)$.
		Let $v_s\in \tilde{x}\circ Q_u$ be the vertex with maximal index intersecting $Q_i\cup Q_j$.
		All the vertices $(\tilde{x}=v_0,v_1,\dots,v_s)$ belongs to the fundamental vortex cycle, implying that they belong to $G_{\Upsilon^{\mathcal{E}}}\cap T_{\Upsilon}$.
		
		\begin{wrapfigure}{r}{0.19\textwidth}
			\begin{center}
				\vspace{-20pt}
				\includegraphics[width=0.19\textwidth]{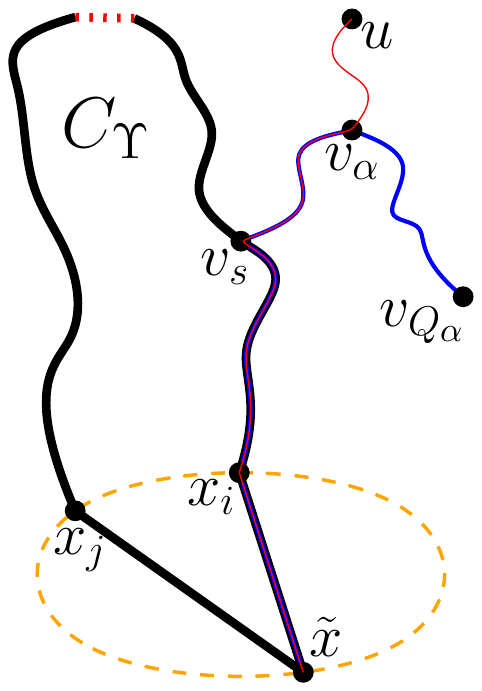}
				\vspace{-15pt}
			\end{center}
			\vspace{-15pt}
		\end{wrapfigure}
		As $u$ is in the exterior of $\bar{C}_\Upsilon$ (the closed curve associated with $C_\Upsilon$), the entire path $(v_{s+1},\dots,v_q=u)$ is in the exterior of $\bar{C}_\Upsilon$.
		First suppose that all the vertices $v_{s+1},\dots,v_q=u$ belong to $\Upsilon$. It follows that $v_{s+1},\dots,v_q=u$
		belong to $\Upsilon^{\mathcal{E}}$, and therefore also to $G_{\Upsilon^{\mathcal{E}}}\cap T_{\Upsilon}=T_{\Upsilon^{\mathcal{E}}}$.
		
		Finally, assume that not all of $\{v_{s+1},\dots,v_q=u\}$ belong to $\Upsilon$. Let $\alpha$ be the maximal index such that $v_\alpha\in G_\Upsilon\setminus\Upsilon$. It follows that there is a path $Q_\alpha\in\mathcal{P}_\Upsilon$ such that $v_\alpha\in Q_\alpha$. Note that the intersection of $Q_\alpha$ with the fundamental cycle $C_\Upsilon$ equals to $(\tilde{x}=v_0,v_1,\dots,v_\alpha)$. It follows that both $v_\alpha$ and $v_{Q_\alpha}$ (the representative vertex of $Q_\alpha$) belong to the exterior of $\bar{C}_\Upsilon$, implying $Q_\alpha\in \mathcal{P}_{\Upsilon^{\mathcal{E}}}$.
		
		We conclude that all the vertices along $Q_u$ belong to $G_{\Upsilon^{\mathcal{E}}}$. The claim follows.\qed

	\vspace{10pt}Next we define the weight function $\omega$ which we used to invoke \Cref{lm:AG06one-vortex-separator}. There are two cases. First, if $|\mathcal{P}_\Upsilon|\le 10\cdot (h+1)$, then $\omega(t)=1$ iff $t$ is a terminal in $\Upsilon$, and otherwise $\omega(t)=0$.
	In the second case (when $|\mathcal{P}_\Upsilon|> 10\cdot h$), initially the weight of all the vertices is $0$. For every $Q\in\mathcal{P}_\Upsilon$,  the weight of $v_Q$, its representative vertex, will increase by $1$.
	
	\begin{claim}\label{clm:InvHolesMaintained}
		\Cref{inv:holes} is maintained. Furthermore, if $|\mathcal{P}_\Upsilon|> 10\cdot (h+1)$ then $|\mathcal{P}_{\Upsilon^{\mathcal{E}}}|,|\mathcal{P}_{\Upsilon^{\mathcal{I}}}|\le 10\cdot (h+1)$.
	\end{claim}
	\begin{proof}
		Given that $|\mathcal{P}_{\Upsilon}|\le 12\cdot (h+1)$ we will prove that $|\mathcal{P}_{\Upsilon^{\mathcal{E}}}|\le 12\cdot (h+1)$.
		The argument for $\mathcal{P}_{\Upsilon^{\mathcal{I}}}$ is symmetric.
		For $\Upsilon=V$, $|\mathcal{P}_V|=0$ and $|\mathcal{P}_{V^{\mathcal{E}}}|\le2(h+1)+1$. For every other $\Upsilon$, as $\{\tilde{X}\}\in\mathcal{P}_\Upsilon$, it is clear that $$|\mathcal{P}_{\Upsilon^{\mathcal{E}}}|\le |\mathcal{P}_{\Upsilon}|+|\mathcal{P}(C_\Upsilon)\setminus\{\tilde{x}\}|\le |\mathcal{P}_{\Upsilon}|+2(h+1)~.$$ 
		Thus in the first case, where $|\mathcal{P}_{\Upsilon}|\le 10\cdot (h+1)$, clearly $|\mathcal{P}_{\Upsilon^{\mathcal{E}}}|\le12\cdot (h+1)$.
		Otherwise, the total weight of all the vertices is $|\mathcal{P}_{\Upsilon}|\le 12\cdot (h+1)$. By \Cref{lm:AG06one-vortex-separator}, the exterior $C_\Upsilon$ in $G_\Upsilon$ contains at most $\frac{2}{3}\cdot |\mathcal{P}_{\Upsilon}|\le \frac{2}{3}\cdot 12\cdot (h+1)=8\cdot (h+1)$ representative vertices. As $|\mathcal{P}(C_\Upsilon)|$ contains at most $2(h+1)$ new paths, we conclude $|\mathcal{P}_{\Upsilon^{\mathcal{E}}}|\le 10\cdot (h+1)+2(h+1)= 12\cdot (h+1)$, as required.
	\end{proof}

	Before we turn to the construction of the spanner, we observe the following crucial fact regarding the graph $G_\Upsilon$.
	\begin{claim}\label{clm:GUpsilon}
		Consider a set $\Upsilon\in \tau$. Let $u$ be some vertex such that $u$ has a neighbor in $\Upsilon$. Then $v\in G_\Upsilon$.
	\end{claim}
	\begin{proof}
		The proof is by induction on the construction of $\tau$.
		Suppose the claim holds for $\Upsilon$, we will prove it for $\Upsilon^{\mathcal{E}}$ (the proof for  $\Upsilon^{\mathcal{I}}$ is symmetric).
		Consider a pair of neighboring vertices $v,u$ where $v\in \Upsilon^{\mathcal{E}}$. As $\Upsilon^{\mathcal{E}}\subseteq \Upsilon$, $v\in \Upsilon$. Thus by the induction hypothesis $u\in G_\Upsilon$.
		If $u\in \Upsilon^{\mathcal{E}}$ or $u\in C_\Upsilon$, then trivially  $u\in G_{\Upsilon^{\mathcal{E}}}$, and we are done.
		
		Otherwise, according \Cref{clm:CisSeperator} there are no edges between the exterior and interior of $C_\Upsilon$. Thus $u$ belongs to the exterior of $C_\Upsilon$ in $G_\Upsilon$.
		Further, as $u\notin \Upsilon^{\mathcal{E}}\cup C_\Upsilon$ it must be that $u\notin\Upsilon$. In particular, $u$ belongs to a path $Q_u\in \mathcal{P}_\Upsilon$.
		We proceed by case analysis.
                \begin{itemize}
			\item First assume that $u$ belong to the embedded part of $G_\Upsilon$.
			Here $Q_u$ is a path from a perimeter vertex $x_l$ towards a representative vertex $v_{Q_u}$.
			Let $x_i,x_j$ be the two perimeter vertices such that the fundamental vortex cycle $C_\Upsilon$
                        contains the paths $Q_i,Q_j$ starting at $x_i,x_j$. $^{\ref{foot:diffPaths}}$
			Note that the paths $\tilde{x}\circ Q_i,\tilde{x}\circ Q_j,\tilde{x}\circ Q_u$ are all paths in a tree $T_\Upsilon$. In particular, while $Q_i, Q_j, Q_u$ might have mutual prefix, once they diverge, the paths will not intersect again. 
			It follows that the suffix of the path $Q_u$ from $u$ to the representative $v_{Q_u}$ will not intersect $Q_i,Q_j$ (as otherwise $u\in C_\Upsilon$). As $u$ is in the exterior of $C_\Upsilon$,  $v_{Q_u}$ will also belong to the exterior. Thus $u\in Q_u\subseteq G_{\Upsilon^{\mathcal{E}}}$.
			\item Second, assume that $u$ does not  belong to the embedded part. It follows that $Q_u=\{u\}$. As $v_{Q_u}=u$ belongs to the exterior of $C_\Upsilon$, 
			it follows that $u\in Q_u\subseteq G_{\Upsilon^{\mathcal{E}}}$.
	        \end{itemize}

	\end{proof}

	\paragraph{Construction of the spanner $H$, and bounding its weight}
	For each node $\Upsilon\in \tau$ of the hierarchical tree, we will construct a spanner $H_\Upsilon$. The final spanner $H=\cup_{\Upsilon\in\tau}H_\Upsilon$ is the union of all these spanners.
	We  argue that $\tau$ contains $O(k)$ nodes, and that for every $\Upsilon\in \tau$, $w(H_\Upsilon)=O_{h}(L\cdot\poly(\frac{1}{\eps}))$. It then follows
        that $w(H)=O_{h}(kL\cdot\poly(\frac{1}{\eps}))$.
	
	First consider a leaf node $\Upsilon\in \tau$, let $K_\Upsilon=\Upsilon\cap K$ be the set of terminals in $\Upsilon$. As $\Upsilon$ is a leaf, $|K_\Upsilon|= O(h)$.
	For every shortest path $Q\in \mathcal{P}_{\Upsilon}$ and terminal $t\in K_\Upsilon$, we add to $H_\Upsilon$ a $(1+\eps)$ single source spanner from $t$ to $Q$ (w.r.t. $G$)  using \Cref{lm:ss-spanner}. Additionally, for every pair of terminals $t,t'\in K_\Upsilon$, we add the shortest path from $t$ to $t'$  in $G$ to $H_\Upsilon$. Formally,
	\[
	H_{\Upsilon}=\cup_{t\in K_{\Upsilon}}\cup_{Q\in\ensuremath{\mathcal{P}_{\Upsilon}}\cup K_{\Upsilon}}\mathtt{SSP}(t,Q,G)~,
	\]
	where we abuse notation and treat vertices as singleton paths.
	As each $Q \in\mathcal{P}_{\Upsilon}$ is a shortest path in $G$, and the distance from every $t\in K$ is bounded by the diameter $D=O_h(L)$ of $G$, by \Cref{lm:ss-spanner} $w(\mathtt{SSP}(t,Q,G))=O_{h}(L\cdot\poly(\frac{1}{\eps}))$.
	We conclude that $w(H_{\Upsilon})=\sum_{t\in K_{\Upsilon}}\sum_{Q\in\ensuremath{\mathcal{P}_{\Upsilon}}\cup K_{\Upsilon}}w\left(\mathtt{SSP}(t,Q,G)\right)=O_{h}(L\cdot\poly(\frac{1}{\eps}))$, where we used \Cref{inv:holes} to bound the number of addends by $O(h^2)$.
	
	For the general case ($\Upsilon$ is an internal node), recall that we have a fundamental vortex cycle $C_\Upsilon$, which consist of at most $O(h)$ paths $\mathcal{P}(C_\Upsilon)$.
	First, we add all the paths in $\mathcal{P}(C_\Upsilon)$ to $H_\Upsilon$.
	Next, for every pair of shortest paths $Q\in \mathcal{P}_{\Upsilon}\cup \mathcal{P}(C_\Upsilon)$ and $Q'\in \mathcal{P}(C_\Upsilon)$, we add to $H_\Upsilon$ a $(1+\eps)$ bipartite spanner between $Q$ and $Q'$ (w.r.t. $G$)  using \Cref{lm:bipartite-spanner}. Formally,
	\[
	H_{\Upsilon}=\mathcal{P}(C_{\Upsilon})\cup\left(\cup_{Q\in\mathcal{P}(C_{\Upsilon})}\cup_{Q'\in\ensuremath{\mathcal{P}_{\Upsilon}}\cup\mathcal{P}(C_{\Upsilon})}\BS(Q,Q',G)\right)~.
	\]
	As each path in the union is a shortest path in $G$, so a graph with diameter $O_h(L)$, we have
        by \Cref{lm:bipartite-spanner} that $\forall Q,Q',\,w\left(\BS(Q,Q',G)\right)=O_{h}(L\cdot\poly(\frac{1}{\eps}))$.
	\sloppy We conclude that 
	$w(H_{\Upsilon})\le\sum_{Q\in\mathcal{P}(C_{\Upsilon})}w\left(Q\right)+\sum_{Q\in\mathcal{P}(C_{\Upsilon})}\sum_{Q'\in\ensuremath{\mathcal{P}_{\Upsilon}}\cup\mathcal{P}(C_{\Upsilon})}w\left(\BS(Q,Q',G)\right)=O_{h\cdot\poly(\frac{1}{\eps})}(L)$
	, 
	where we used \Cref{inv:holes} to bound the number of addends by $O(h^2)$.
	
	Next, we bound the number of nodes in $\tau$.
	We say that a node $\Upsilon'\in\tau$ is a grandchild of $\Upsilon\in \tau$ if there is a node $\Upsilon''$ which is the child of $\Upsilon$, and the parent of  $\Upsilon'$. 
	Consider some node $\Upsilon$. It follows from \Cref{clm:InvHolesMaintained} that either $|\mathcal{P}_\Upsilon|\le 10\cdot (h+1)$ or for both its children $|\mathcal{P}_{\Upsilon^{\mathcal{E}}}|,|\mathcal{P}_{\Upsilon^{\mathcal{I}}}|\le 10\cdot (h+1)$. In particular, either in $\Upsilon$, or in both its children the number of terminals drops by a $\frac{2}{3}$ factor. We conclude that if $\Upsilon'$ is a grandchild of $\Upsilon$ then $|K_{\Upsilon'}|\le\frac23 |K_{\Upsilon}|$.

	For the sake of analysis, we will divide $\tau$ into two trees, $\tau_{O}$ and $\tau_{E}$. $\tau_{E}$ (resp. $\tau_{O}$) contains all the nodes of even (resp. odd ) depth. There is an edge between $\Upsilon$ to $\Upsilon'$ if $\Upsilon'$ is a grandchild $\Upsilon$.
	Consider $\tau_{E}$. Note that the number of leafs is bounded by $k$ (as they are all disjoint and contain at least one terminal). Further, if an internal node $\Upsilon$ (other than the root $V$) has degree $2$ in $\tau_O$, it follows that $\Upsilon$ has a single grandchild in $\tau$. There is some terminal in $\Upsilon\setminus\Upsilon'$ (as  $|K_{\Upsilon'}|\le\frac23 |K_{\Upsilon}|$). As there are $k$ terminals, we conclude that the number of degree $2$ nodes is bounded by $k$. It follows that $\tau_{E}$ has $O(k)$ nodes. A similar argument will imply that $\tau_{O}$ has $O(k)$ nodes. It follows that $\tau$ has $O(k)$ nodes. We conclude
	\[
	w(H)\le\sum_{\Upsilon\in\tau}w(H_{\Upsilon})=O_{h}(kL\cdot\poly(\frac{1}{\eps}))~.
	\]

	\paragraph{Bounding the stretch}
	The following claim  will be useful for bounding the stretch between terminals:
	\begin{claim}\label{clm:TerminalToInactive}
		Consider an internal node $\Upsilon\in\tau$, and a terminal vertex $t\in \Upsilon$. For every fundamental vortex cycle vertex $u\in C_{\Upsilon}$, it holds that $d_H(t,u)\le (1+\eps)d_G(t,u)$.
	\end{claim}
	\begin{proof}
		Let $Q^u\in \mathcal{P}(C_\Upsilon)$ be the shortest path belonging to the fundamental vortex cycle that contains $u$.
		Let $Q_{t,u}=\{t=v_0,v_1,\dots,v_s=u\}$ be the shortest path from $t$ to $u$ in $G$.
		The proof is by induction on $s$ (the number of hops in the path).  We proceed by case analysis.
		\begin{itemize}
		\item Suppose that not all the vertices in $Q_{t,u}\setminus\{u\}$ belong to $\Upsilon$.
                  Let $v_i$ be the vertex with minimal index not in $\Upsilon$.
		  As $v_{i-1}\in \Upsilon$, by \Cref{clm:GUpsilon} $v_i\in G_\Upsilon$. In particular, there is some path $Q^{v_i}\in\mathcal{P}_\Upsilon$ such that $v_i\in Q^{v_i}$, where $Q^{v_i}$ is a path belonging to a fundamental vortex cycle removed in an ancestor of $\Upsilon$ in $\tau$. By the induction hypothesis, $d_H(t,v_i)\le (1+\eps)d_G(t,v_i)$.
		  During the construction of $H_\Upsilon$, we added $\BS(Q^u,Q^{v_i},G)$, a bipartite spanner between the paths $Q^u,Q^{v_i}$, to $H_\Upsilon$. Thus 
		  $d_H(t,u)\le d_H(t,v_i)+d_H(v_i,u)\le(1+\eps)(d_G(t,v_i)+d_G(v_i,u))=(1+\eps)d_G(t,u)$.
		\item Otherwise (all vertices in $Q_{t,u}\setminus\{u\}$ belong to $\Upsilon$), suppose that there is some vertex $v_i \neq u$ that belongs to $C_{\Upsilon}$.
		  By the induction hypothesis, $d_H(t,v_i)\le (1+\eps)d_G(t,v_i)$.
		  By the construction of $H_\Upsilon$, $d_{H_\Upsilon}(v_i,u)\le(1+\eps)d_G(v_i,u)$. It follows that
		  $d_H(t,u)\le(1+\eps)d_G(t,u)$.
		\item Otherwise (all vertices in $Q_{t,u}\setminus\{u\}$ belong to $\Upsilon$, and $u$ is the only vertex in $Q_{t,u}$
                  belonging to $C_{\Upsilon}$), if there exists a future hierarchical step such that in a node $\Upsilon'$,
                  some vertices in $Q_{t,u}\setminus\{u\}$ belongs to the fundamental vortex cycle $C_{\Upsilon'}$.
                  Then, let $\Upsilon'$ be the first such set (that is the closest to $\Upsilon$ w.r.t. $\tau$).
                  Let $v_i\in C_{\Upsilon'}$. By the minimality of $\Upsilon$, all the vertices $\{t=v_0,\dots,v_i,\dots,v_{s-1}\}$ belong to $\Upsilon'$.
		  By the induction hypothesis,  $d_H(t,v_i)\le (1+\eps)d_G(t,v_i)$. Further, as $v_{s-1}\in \Upsilon'$, by \Cref{clm:GUpsilon} $v_s\in G_{\Upsilon'}$. In particular the spanner $H_{\Upsilon'}$ has stretch $(1+\eps)$ between $v_i$ and $u$ (as we added a bipartite spanner between two paths containing $v_i,u$).  It follows that $d_H(t,u)\le(1+\eps)d_G(t,u)$.
		\item Otherwise, (all vertices in $Q_{t,u}\setminus\{u\}$ belong to $\Upsilon$, and $u$ is the only vertex in $Q_{t,u}$ belonging to a fundamental cycle in $\Upsilon$, and in all future hierarchical steps).
                  Then, all vertices in $Q_{t,u}\setminus\{u\}$ belong to some leaf node $\Upsilon'\in \tau$.
                  By \Cref{clm:GUpsilon} $v_s\in G_{\Upsilon'}$. In particular $u$ belongs to some path in $\mathcal{P}_{\Upsilon'}$.
                  During the construction, we added
                  to $H_{\Upsilon'}$ a single source spanner from $t$ to this path.  It follows that $d_H(t,u)\le(1+\eps)d_G(t,u)$.
		\end{itemize}
	\end{proof}
	Consider a pair of terminals $t,t'$ with a shortest path $Q_{t,t'}$ is $G$. 
	If $t$ and $t'$ end up together in a leaf node $\Upsilon\in \tau$, than we added a shortest path between them to $H_\Upsilon$. Thus $d_H(t,t')=d_G(t,t')$.
	Otherwise, let $v\in Q_{t,t'}$ be the first vertex which was added to a fundamental vortex cycle during the construction of $\tau$ (first w.r.t. the order defined by $\tau$). By \Cref{clm:TerminalToInactive} it holds that	
	$$d_H(t,t')\le d_H(t,v_i)+d_H(v_i,t')\le(1+\eps)(d_G(t,v_i)+d_G(v_i,t'))=(1+\eps)d_G(t,t')~,$$
	hence the bound on the stretch.

	\subsection{Step (2.0): Unbounded diameter, proof of \Cref{lm:one=Vortex-Unbounded-diam}}\label{subsec:diameterReduction}
	We start by restating the lemma we will prove in this subsection.
	\LemmaOneVortexUbnoundedDiam*
	The \emph{strong diameter} \footnote{On the other hand, the \emph{weak diameter} of a cluster $A\subseteq V$ equals to the maximal distance between a pair of vertices $u,v\in A$ in the original graph. Formally $\max_{\{u,v\in A\}}d_{G}(u,v)$. See \cite{Fil19padded,Fil20scattering} for further details on sparse covers and related notions.} of a cluster $A\subseteq V$ equals to the maximal distance between a pair of vertices $u,v\in A$ in the induced graph $G[A]$. Formally $\max_{\{u,v\in A\}}d_{G[A]}(u,v)$.
	The main tool we will use here is \emph{sparse covers}.
\begin{definition}[Sparse Cover]
Given a weighted graph $G=(V,E,w)$, a collection of clusters $\mathcal{C} = \{C_1,..., C_t\}$ is called a $(\rho,s,\Delta)$-strong sparse cover if the following conditions hold.
\begin{enumerate}
\item Bounded diameter: The strong diameter of every $C_i\in\mathcal{C}$ is bounded by $\rho\cdot\Delta$.\label{condition:RadiusBlowUp}
\item Padding: For each $v\in V$, there exists a cluster $C_i\in\mathcal{C}$ such that $B_G(v,\Delta)\subseteq C_i$.
\item Overlap: For each $v\in V$, there are at most $s$ clusters in $\mathcal{C}$ containing $v$.		
\end{enumerate}
We say that a graph $G$ admits a $(\rho,s)$-strong sparse cover scheme, if for every parameter $\Delta>0$ it admits a $(\rho,s,\Delta)$-strong sparse cover. A graph family $\mathcal{G}$ admits a $(\rho,s)$-strong sparse cover scheme, if every $G\in\mathcal{G}$ admits a $(\rho,s)$-strong sparse cover scheme.
\end{definition}
Abraham et al. \cite{AGMW10} constructed strong sparse covers.
\begin{theorem}[\cite{AGMW10}]\label{AGMW10Covers}
Every weighted graph excluding  $K_{r,r}$ as a minor admits an $(O(r^2), 2^{O(r)}\cdot r!)$-strong sparse cover scheme constructible in polynomial time.
\end{theorem}

Consider a graph $G$ as in the lemma with terminal set $K$ and parameters $\eps,L$. Note that $G$ is $K_{h+2,h+2}$ free.
Using \Cref{AGMW10Covers} let $\{C_1,..., C_t\}$ be an $(O(h^2), 2^{O(h)}\cdot h!,L)$ sparse cover for $G$. Note that each cluster $C_i$ has strong diameter $O(h^2)\cdot L=O_h(L)$. For every $i$, using \Cref{lm:one-vortex-Bounded-diam}, let $H_i$ be a $(1+\eps)$-spanner for $G[C_i]$, w.r.t. terminal set $K_i=C_i\cap K$ and parameters $L,\eps$.
Set $H=\bigcup H_i$. Then $H$ has weight 
\[
w(H)\le\sum_{i}w(H_{i})\le\sum_{i}O_{h}(|K_{i}|\cdot L)=O_{h}(k\cdot L\cdot2^{O(h)}\cdot h!)=O_{h}(k\cdot L)~,
\]
	where the first equality follows as every terminal is counted at most $2^{O(h)}\cdot h!$ times in the sum.
	
	We argue that $H$ preserves all terminal distances up to $L$.
	Consider a pair of terminals $t,t'$ such that $d_G(t,t')\le L$. There is a cluster $C_i$ such that
        the ball of radius $L$ around $t$ contained in $C_i$. In particular the entire shortest path from $t$ to $t'$ is contained in $C_i$. We conclude
	$$d_H(t,t')\le d_{H_i}(t,t')\le (1+\eps)\cdot d_{G[C_i]}(t,t')=(1+\eps)\cdot d_{G}(t,t')~.$$	
		
	\subsection{Step (2.1): Reducing vortices, proof of \Cref{lm:planar-with-many-vortices}}\label{subsec:reducingVortices}
	We start by restating \Cref{lm:planar-with-many-vortices}:
	\NearlyEmbdablSpannerNoApicesNoGenus*
	This subsection is essentially devoted to proving the following lemma:
	\begin{lemma}\label{lm:vorexReduce}
		Consider a graph $G = G_{\Sigma}\cup  W_1\cup\dots \cup W_{\ell}$,  where $G_{\Sigma}$ is drawn on plane, and each $W_i$ is a vortex of width at most $h$ glued to a face of $G_{\Sigma}$.
		Then given a terminal set $K$ of size $k$, and parameter $L>0$, there is an induced subgraph $G'$ of $G$ and a spanning subgraph $H_{vo}$ of $G$ such that:
		\begin{OneLiners}
		\item $G'$ can be drawn on the plane with a single vortex of width at most $h$.
		\item $w(H_{vo})\le O(k\ell L\cdot\poly(\frac{1}{\eps}))$.
		\item For every pair of terminals $t,t'\in K$ at distance at most $L$, either $d_{G'}(t,t')=d_G(t,t')$ or $d_{H_{vo}}(t,t')\le (1+\eps)d_G(t,t')$.
		\end{OneLiners}
	\end{lemma}
	Given \Cref{lm:vorexReduce},  \Cref{lm:planar-with-many-vortices} easily follows.
	\begin{proof}[Proof of \Cref{lm:planar-with-many-vortices}]
		We begin by applying \Cref{lm:vorexReduce} on the graph $G$. As a result we receive the graphs $G',H_{vo}$, where $G'$ has a single vortex of width at most $h$, $H_{vo}$ has weight $O_{h}(k\ell L)\cdot\poly(\frac{1}{\eps})$, and for every pair of terminals at distance up to $L$, either  $d_{G'}(t,t')=d_G(t,t')$ or $d_{H_{vo}}(t,t')\le (1+\eps)d_G(t,t')$.
		
		Next, we apply \Cref{lm:one=Vortex-Unbounded-diam} on $G'$ and receive a $(1+\eps)$-subset spanner $H'$ of weight $O_{\eps}(kL\cdot\poly(\frac{1}{\eps}))$ that preserves all terminal distances up to $L$ (w.r.t. $G'$). Set $H=H'\cup H_{vo}$. Note that $H$ has weight $O_{h}(kL\cdot\poly(\frac{1}{\eps}))$. Let $u,v\in K$ be a pair of terminals at distance at most $L$. Then either $d_{H'}(t,t')\le(1+\eps)d_{G'}(t,t')=(1+\eps)d_G(t,t')$, or $d_{H_{vo}}(t,t')\le (1+\eps)d_G(t,t')$, implying the lemma.
	\end{proof}

Before turning to the proof of \Cref{lm:vorexReduce}, we introduce the \emph{vortex merge} procedure.
Given two vortices $W,W'$, a \emph{proper vortex path} $P$ is a path in $G$ between a vertex $v\in W$ to a vertex $u\in W'$ such that all but the first and last vertices of $P$ belong to the planar part $G_\Sigma$, and $P$ does not contain any vortex vertices. Given a proper vertex path $P=\{v=v_0,\dots,v_s=u\}$ from $v\in W$ to $u\in W'$, the following operation is called \emph{vortex merge} w.r.t. $P$:
Let $X_1,\dots,X_t$ (respectively $Y_1,\dots,Y_{t'}$) be a path decomposition of $W$ (respectively $W'$) with perimeter $x_1,\dots,x_t$ ($y_1,\dots,y_{t'}$) such that $v=x_i$ ($u=y_{j}$).
Let $F_W$ and $F_{W'}$ be respectively the face containing the vertices of $W$ and $W'$ in the planar part.
The operation consists in \emph{cutting open} the surface along the path $P$ (see e.g.:~\cite{Klein08}
for a formal definition), resulting in two copies of the vertices
of the path $P_1,P_2$. Then for each copy $P_i$, delete all the edges that have exactly one endpoint that belongs to
$P_i \cup \{x_{i-1}, x_{i+1}, y_{i-1}, y_{i+1}\}$, and finally contract all
the edges in $P_i$. This yields edges between either
$\{x_{i-1},y_{j-1}\}$, and $\{x_{i+1},y_{j+1}\}$, or $\{x_{i-1},y_{j+1}\}$, $\{x_{i+1},y_{j-1}\}$ -- in the following we assume that
the former happened.
Since contraction and deletion preserve the genus, the genus of the embedded part has not increased.

\begin{figure}[]
\centering{\includegraphics[scale=0.9]{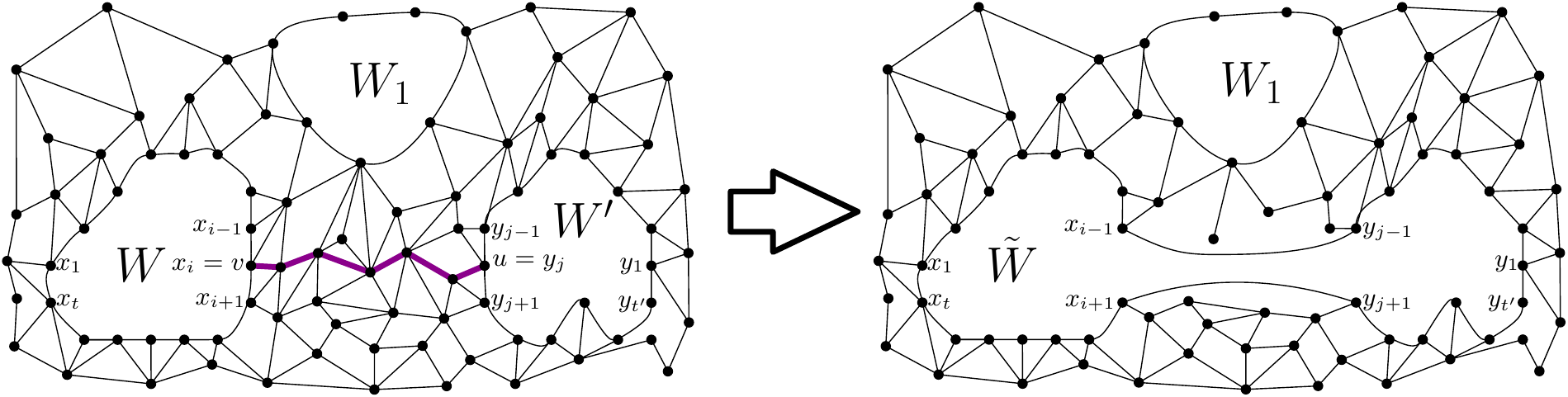}} 
\caption{\label{fig:VortexMerge}\small \it 
The graph $G$, who has three vortices $W,W',W_1$, is illustrated on the left. The (purple) path $P$ is a proper vortex path between $W,W'$. On the right, we illustrate the vortex merge that creates a new graph $G'$ by deleting the path $P$ and adding two new edges.
}
\end{figure}

See \Cref{fig:VortexMerge} for illustration.
The process produces a new vortex, glued to the face $x_1,\dots,x_{i-1},y_{j-1},\dots,y_1,y_{t'},\dots,y_{j+1},x_{i+1}\dots,x_t$, which will be the perimeter vertices. The bags in the path decomposition will be $X_1,\dots,X_{i-1},Y_{j-1},\dots,Y_1,Y_{t'},\dots,Y_{j+1},X_{i+1}\dots,X_t$, respectively (the newly added edges will be part of the embedded graph).

\begin{observation}\label{obs:vortexMerge}
  Consider a graph $G = G_{\Sigma}\cup  W_1\cup\dots \cup W_{\ell}$,  where $G_{\Sigma}$ is drawn on the plane
  and each $W_i$ is a vortex of width at most $h$ glued to a face of $G_{\Sigma}$.
  Let $P$ be a proper vortex path between $W_\ell$ to $W_{\ell-1}$, and consider the graph $G'$ created by a vortex merge w.r.t. $P$.
  Then $G'$ has planar drawing with vortices $W_1,\ldots, W_{\ell-2},\tilde{W}_{\ell-1}$ all of width at most $h$. 
\end{observation}

We are now ready to prove the main lemma of the subsection:
\begin{algorithm}[h]
\caption{\texttt{Contracting Vortices}}	\label{alg:vortex}
\DontPrintSemicolon
\SetKwInOut{Input}{input}\SetKwInOut{Output}{output}
\Input{Graph $G=(V,E,w)=G_\Sigma\cup W_1\cup\dots\cup W_\ell$, $k$ terminals $K\subseteq V$}
\Output{Induced graph $G'$ and subgraph $H_{vo}$ of $G$}
\BlankLine

Let $G_1 \la G$, $H_{vo}\la\emptyset$, $\im\la\emptyset$\;
\For{$j=1$ to $\ell-1$}{
Let $P_j$ be the shosrtest proper vortex path in $G_j$ betweem $x_i\in W$ and $y_j\in W'$, let $\mathcal{P}_j$ be the set of paths consisting of $P_j$ and all the singleton paths consisting of vertices in $X_i\cup Y_j$
\footnotemark \;
$H_{vo}\la H_{vo}\cup\left(\cup_{t\in K\cap G_j}\cup_{P\in\mathcal{P}_j} \ssp(t,P,G_j,L)\right)$\;
Create a new graph $G_{j+1}$ by preforming a vortex merge w.r.t. $P_j$ in $G_j$\;
Add to $\im$ the (at most $2$) new edges $G_{j+1}\setminus G_j$\;
}
Remove from $G_\ell$ and from $H_{vo}$ all the edges in $\im$\label{line:removeIm}\;
\Return $G_\ell$ and $H_{vo}$\;
\end{algorithm}
\footnotetext{$P_j$ can be picked as the minimal path having its endpoints in two different vortices. By minimality, $P_j$ necessarily will be a proper vortex path.}
\begin{proof}[Proof of \Cref{lm:vorexReduce}]
The algorithm for constructing $G'$ and $H_{vo}$ is described in \Cref{alg:vortex}. Consider the graphs $G_\ell, H_{vo}$ just before the removal of the set $\im$ in \Cref{line:removeIm}.
By induction and \cref{obs:vortexMerge}, we have that $G_\ell$ has a single vortex of  width at most $h$. 
Further,  $G'=G_\ell\setminus\im$ can also be drawn on the plane with a single vortex of the same width.

Next we bound the weight of $H_{vo}$. In each of the $\ell-1\le h$ rounds of \Cref{alg:vortex} we added at most $k\cdot (2h+1)$ single source spanners, each of weight $O(L\cdot\poly(\frac{1}{\eps}))$. It follows that $w(H_{vo})=O(k\ell L\cdot\poly(\frac{1}{\eps}))$. 

Finally we bound the stretch. Consider a pair of terminals $t,t'\in K$ at distance at most $d_G(t,t')\le L$. Let $P_{t,t'}$ be the shortest path between $t$ to $t'$ in $G$.
If no vertex of $P_{t,t'}$ was deleted during the execution of \Cref{alg:vortex}, then $d_{G'}(t,t')=d_G(t,t')$ and the we are done.
Otherwise, let $s$ be the minimal index where some vertex $v\in P_{t,t'}$ belongs to a path $P_v\in\mathcal{P}_s$, and thus was deleted. By minimality, it holds that $d_{G_s}(t,P_v)\le d_{G_v}(t,v)=d_{G}(t,v)\le L$. Similarly $d_{G_s}(t',P_v)\le L$. Using the properties of $\ssp(t,P_v,G_s,L)\cup \ssp(t',P_v,G_s,L)$, we conclude
\[
d_{H_{vo}}(t,t')\le d_{\ssp(t,P_{v},G_{s},L)}(t,v)+d_{\ssp(t',P_{v},G_{s},L)}(t',v)\le(1+\eps)\left(d_{G_{s}}(t,v)+d_{G_{s}}(t',v)\right)=(1+\eps)d_{G_{s}}(t,t')~.
\]	
Note that no edge of $\im$ incident on edges of $H_{vo}$, as all these edges have weight greater $\Omega_\eps(L)$. Thus the inequality holds also in $H_{vo}\setminus\im$. The lemma now follows.

	\end{proof}

	\subsection{Step (2.2): Cutting out genus, proof of \Cref{lm:nearly-embed-spanner-no-apices}}\label{subsec:removingGenus}
	We begin by restating \Cref{lm:nearly-embed-spanner-no-apices}.
	\NearlyEmbdablSpannerNoApices*
	Given a graph with an embedded part and some width $h$ glued vortices, we denote by $g(G)$ the genus of the surface embedded part, and by  $v(G)$ the number of vortices.
	Most of this section is essentially devoted to proving the following lemma:
\begin{lemma}\label{lm:genus}
	Consider a graph $G = G_{\Sigma}\cup W_1\cup\dots \cup W_{v(G)}$, 	
	where $G_{\Sigma}$ is (cellularly) embedded on a surface $\Sigma$ of genus $g(G)$, and each $W_i$ is a vortex of width at most $h$ glued to a face of $G_{\Sigma}$. Then given a terminal set $K$ of size $k$ and parameter $L>0$, there is an induced subgraph $G'$ of $G$ and a spanning subgraph $H_g$ of $G$ such that:
	\begin{OneLiners}
	\item $G'$ has genus $g(G')=0$ and at most $v(G')\le v(G)+g(G)$ vortices, all of width at most $h$.
	\item $w(H_g)\le O\left(kL\cdot g(G)\left(g(G)+v(G)\right)\right)\cdot\poly(\frac{1}{\eps})$.
	\item For every pair of terminals $t,t'\in K$ at distance at most $L$  either $d_{G'}(t,t')=d_G(t,t')$ or $d_{H_g}(t,t')\le (1+\eps)d_G(t,t')$.
	\end{OneLiners}
	\end{lemma}
	
	Given \Cref{lm:genus}, \Cref{lm:nearly-embed-spanner-no-apices} easily follows.
	\begin{proof}[Proof of \Cref{lm:nearly-embed-spanner-no-apices}]
		Apply  \Cref{lm:genus}  on $G$, and let $G',H_g$ be the output. Construct a $(1+\eps)$-subset spanner $H'$ for $G'$ using \Cref{lm:planar-with-many-vortices}. Set $H=H'\cup H_g$. Then $H$ has weight $O_{h}(kL\cdot\poly(\frac{1}{\eps}))$. 
		To bound the stretch, let $t,t'\in K$ be a pair of terminals at distance at most $L$. Then either $d_{H'}(t,t')\le(1+\eps)d_{G'}(t,t')=(1+\eps)d_G(t,t')$, or $d_{H_g}(t,t')\le (1+\eps)d_G(t,t')$, implying the lemma.
	\end{proof}

	Recall the definition of vortex-path (\Cref{def:vortex-path}). Essentially one can think of a vortex-path as a union of $V(G)+1+2\cdot v(G)\cdot h=O(v(G)\cdot h)$ paths (see \Cref{sec:oneVortex} for details).
	A vortex-path is called a \emph{shortest-vortex-path} if all the paths it contains are shortest paths in $G$.
	We will use the following \emph{Cutting Lemma} of Abraham and Gavoille \cite[Lemma 6, full version]{AG06}.

\begin{lemma}[Cutting Lemma~\cite{AG06}]\label{lm:cuttingAG06} Given a $h$-nearly embeddable graph $G$, there are (efficiently computable) two shortest-vortex-paths $\mathcal{V}_1,\mathcal{V}_2$ of $G$ such that the graph $G' = G\setminus (\mathcal{V}_1\cup \mathcal{V}_2)$ is $h$-nearly embeddable and has $v(G') \leq v(G)+1$ and $g(G')\le g(G)-1$.
\end{lemma}

Intuitively, \Cref{lm:cuttingAG06} says that we can reduce the genus of $G$ by removing two vortex-paths at the expense of increasing the number of vortices by $1$, without affecting their width.  
Given the characterization above, an immediate corollary is that given a graph $G$ that embeds on a surface of genus $g(G)$, with $v(G)$ vortices of genus $g$, we can remove a set $\mathcal{P}$ of $O(v(G)\cdot h)$ shortest paths to obtain a graph $G'=G\setminus \mathcal{P}$, where $v(G') \leq v(G)+1$ and $g(G') = g(G)-1$.
We proceed now to proving \Cref{lm:genus}.

\begin{proof}[Proof of \Cref{lm:genus}]
In \Cref{alg:genus} below we repeatedly apply \Cref{lm:cuttingAG06}, and remove paths until the remaining graph has genus $0$.

\begin{algorithm}[h]
\caption{\texttt{Genus Reduction}}	\label{alg:genus}
\DontPrintSemicolon
\SetKwInOut{Input}{input}\SetKwInOut{Output}{output}
\Input{Embedded graph $G=(V,E,w)$ with genus $g(G)$, $v(G)$ vortices, and terminal set  $K$}
\Output{Induced graph $G'$ and subgraph $H_g$ of $G$}
\BlankLine

Let $G_0 \la G$, $H_g\la\emptyset$, $j\la 0$\;
\While{$g(G_j)>0$}{
Let $\mathcal{P}_j$ be the set of $O(h\cdot v(G_j))$ shortest paths as guaranteed  by \Cref{lm:cuttingAG06} w.r.t. $G_j$\;
Add to $H_g$ the set $\cup_{t\in K}\cup_{P\in \mathcal{P}_j}\ssp(t,P, G_j, L)$\;
$G_{j+1}\la G_j\setminus \mathcal{P}_j$\;
$j\la j+1$\;
}
\Return $G'$ and $H_g$\;
\end{algorithm}
By induction it holds that $g(G_j)\le g(G)-j$ and $v(G_j)\le v(G)+j$. Let $J$ be final value of $j$. Then $J\le g(G)$. Thus $g(G_J)=0$ and $v(G_J)\le v(G)+g(G)$. Furthermore,
\begin{align*}
w(H_{g})\le\sum_{j=0}^{J-1}\sum_{t\in K}\sum_{P\in\mathcal{P}_{j}}w\left(\ssp(t,P,G_{j},L)\right) & \le\sum_{j=0}^{J-1}O(kv(G_{j})h\cdot L)\cdot\poly(\frac{1}{\eps})\\
& =O\left(kL\cdot g(G)\left(g(G)+v(G)\right)\right)\cdot\poly(\frac{1}{\eps})~.
\end{align*}
Finally, we prove the stretch guarantee. Consider a pair of terminals  $t,t'\in K$ such that $d_G(t,t')\le L$.
Let $P_{t,t'}$ be the shortest path from $t$ to $t'$ in $G$. If $P_{t,t'}\subseteq G'$, then $d_{G'}(t,t')=d_G(t,t')$ and the we are done.
Otherwise, let $j$ be the minimal index such that $P_{t,t'}\cap \mathcal{P}_j\neq \emptyset$ , and let $v\in P_{t,t'}$ be some vertex that belongs to a path $Q\in \mathcal{P}_j$.
By minimality, it holds that $d_{G_j}(t,Q)\le d_{G_j}(t,v)=d_{G}(t,v)\le L$. Similarly $d_{G_j}(t',Q)\le  d_{G_j}(t',v)=d_{G}(t',v)\le L$. We conclude
\[
d_{H_{g}}(t,t')\le d_{\ssp(t,P_{s},G_{j},L)}(t,v)+d_{\ssp(t',P_{s},G_{j},L)}(t',v)\le(1+\eps)\left(d_{G_{j}}(t,v)+d_{G_{j}}(t',v)\right)=(1+\eps)d_{G}(t,t')~.
\]
\end{proof}

	\subsection{Step (2.3): Removing apices, proof of \Cref{lm:nearly-embed-spanner}}\label{subsec:removingApices}
	\NearlyEmbdablSpanner*
	\begin{proof}
		Consider an $h$-nearly embeddable graph $G$. Let $A$ be the set of apices. By definition $|A| \leq h$. 
		Set $H_A\la \emptyset$. For any apex $a\in A$ and terminal $t\in K$ such that $d_G(a,t)\le L$, we add the shortest $a-t$ path in $G$ to $H_A$. Then $w(H_A)\le k\cdot h\cdot L$. 
		Let $G'=G[V\setminus A]$ be the graph $G$ after we removed all the apices. Create a subset spanner  $H'$ for $G'$ using  \Cref{lm:nearly-embed-spanner-no-apices}.
		Set $H=H'\cup H_A$ to be a subset spanner for $G$. Then $w(H)\le O_{h}(kL)\cdot\poly(\frac{1}{\eps})+ khl =O_{h,\eps}(kL)\cdot\poly(\frac{1}{\eps})$. 
		We argue that $H$ preserves all terminal distances up to $L$. Consider a pair of terminals $t,t'\in K$ such that $d_G(t,t')\le L$. If the shortest path between $t$ to $t'$ contains an apex $a$, then $d_H(t,t')\le d_{H_A}(t,a)+d_{H_A}(a,t')=d_{G}(t,a)+d_{G}(a,t')=d_{G}(t,t')$. Otherwise, 
		$d_H(t,t')\le d_{H'}(t,t')\le (1+\eps)d_{G'}(t,t')= (1+\eps)d_{G}(t,t')$.
	\end{proof}

	\subsection{Step (2.4): Eliminating clique-sums, proof of \Cref{prop:ell-close-spanner}}\label{subsec:EliminatingCliqueSum}
	In this subsection we generalize to minor free graphs.
	\MinorFreeSubset*
The main technical step here is in proving the following lemma, which will allow us to assume that the clique sum decomposition has only $O(k)$ nodes.
\begin{lemma}\label{lm:tree-reduction} 
	Consider a weighted graph $G=(V,E,w)$ with a set $K\subseteq V$ of $k$ terminals,  and a clique-sum decomposition $\T$ of $G$ s.t. $G = \cup_{X_iX_j \in E(\mathcal{T})} X_i \oplus_h X_j$.
	Then there is a graph $G'=(V'\subseteq V,E,w)$ containing all the terminals, with a clique-sum decomposition $\T'$ of $G'$ s.t. $G' = \cup_{X_iX_j \in E(\T')} X_i \oplus_h X_j$. The weight function $w$ of $G'$ gives weight $d_G(u,v)$ to each edge $\{u,v\}$. It holds that:
	\begin{OneLiners}
		\item The number of nodes in $\T'$ is $O(k)$.
		\item For each node $X\in \T'$, either $X\in \T$ or $X$ contains at most $2h$ vertices.
		\item For every pair of terminals $t,t'$, it holds that $d_{G}(t,t')=d_{G'}(t,t')$.
	\end{OneLiners}
\end{lemma}
We proceed directly to proving \Cref{prop:ell-close-spanner}. 
\begin{proof}[Proof of \Cref{prop:ell-close-spanner}]
	According to the \cite{RS03} \Cref{thm:RS}, $G$ can be decomposed into a tree $\mathcal{T}$ where each node of $\mathcal{T}$ corresponds a nearly $h$-embeddable subgraph, such that $G = \cup_{X_iX_j \in E(\mathcal{T})} X_i \oplus_h X_j$. 
	We apply \Cref{lm:tree-reduction} and receive graph $G'$ with clique-sum decomposition $\T'$ as above.
	Set $K'$ to be the set of all vertices in $G'$ which belong to at least one clique in the set of clique-sums used by $\T'$.
	Formally, $K'=\left\{v\in G'\mid \exists X_iX_j \in E(\mathcal{T}')\mbox{ such that }v\in X_i\cap X_j\right\}$. Note that we add at most $h$ vertices to $K'$ per each edge of $\mathcal{T}'$, it thus holds that $|K'|\le h\cdot(|\mathcal{T}'|-1)=O_h(k)$. Set $\hat{K}=K\setminus K'$ to be the set of terminals out of $K'$.
	For each node $X\in\T'$, set $\hat{K}_X=\hat{K}\cap X$ and $K'_X=K'\cap X$.
	If $X\in\T$ let $H_X$ be a subset spanner constructed using \Cref{lm:nearly-embed-spanner} w.r.t. the terminal set $K'_X\cup \hat{K}_X$. Note that $H_X$ is a subgraph of $G$, of weight $O_{h}(|K'_X\cup \hat{K}_X|\cdot L)\cdot\poly(\frac{1}{\eps})=O_{h}(|\hat{K}_X|\cdot L+L)\cdot\poly(\frac{1}{\eps})$.
	Else ($X\notin\T$), then $X$ contains at most $2h$ vertices. 
	For every $v,u\in X$, if $d_G(u,v)\le L$, set $P^L_{v,u}$ to be an arbitrary shortest path from $u$ to $v$ in $G$, else ($d_G(u,v)> L$), set $P^L_{v,u}=\emptyset$.
	Let $H_X=\cup_{u,v\in X}P^L_{v,u}$ to be a subgraph of $G$ that contains all the shortest paths between $X$ vertices at distance at most $L$ in $G$. Note that the weight of $H_X$ is bounded by $O(h^2\cdot L)=O_h(L)$. 
	Set $H=\cup_{X\in\T'}H_X$. 
	
	It is clear that $H_{\T'}$ is a subgraph of $G$.
	We first bound the weight of $H_{\T'}$,
	\begin{align*}
	w(H) & =\sum_{X\in\T'}w(H_{X})=\sum_{X\in\T'}O_{h}(|\hat{K}_{X}|\cdot L+L)\cdot\poly(\frac{1}{\eps})\\
	& =O_{h}(|\T'|\cdot L)\cdot\poly(\frac{1}{\eps})+O_{h}(L)\cdot\poly(\frac{1}{\eps})\cdot\sum_{X\in\T'}|\hat{K}_{X}|=O_{h}(kL)\cdot\poly(\frac{1}{\eps})~,
	\end{align*}
	where the last equality follows as $\cup_{X\in\T'}\hat{K}_{X}=\hat{K}$, and $\left\{ \hat{K}_{X}\right\} _{X\in\T'}$ are disjoint.

	Next, we argue that $H_{\T'}$ preserves terminals distances up to $L$. 
	Consider a pair of terminals $t,t'\in K\cup K'$ at distance at most $L$.
	Let $P_{t,t'}=\{t=v_0,v_1,\dots,v_s=t'\}$ be a shortest path between $t$ to $t'$ in $G'$. Let $\mathcal{I}=\{0=i_0<i_1<\dots<i_q=s\}\subseteq[0,s]$ be a minimal set of indexes such that for every $j\in [q]$ there is a node $X_j\in\T'$ such that $\{v_{i_{j-1}},v_{i_{j-1}+2},\dots ,v_{i_{j}}\}\subseteq X_j$. 
	By the minimality of $\mathcal{I}$, it necessarily holds that $v_{i_1},v_{i_2},\dots, v_{i_q-1}\in K'$. Furthermore, as $P_{t,t'}$ is a shortest path it holds that 
	$d_{G'[X_{j}]}(v_{i_{j-1}},v_{i_{j}})=d_{G'}(v_{i_{j-1}},v_{i_{j}})$. As $w(P_{t,t'})\le L$,  the spanner $H_{X_{j}}$ has distortion $1+\eps$ w.r.t. the pair $v_{i_{j-1}},v_{i_{j}}$
	We conclude 
	\begin{align*}
	d_{H}(t,t') & \le\sum_{j=1}^{q}d_{H_{\T'}}(v_{i_{j-1}},v_{i_{j}})\le\sum_{j=1}^{q}d_{H_{X_{j}}}(v_{i_{j-1}},v_{i_{j}})\\
	& \le(1+\eps)\cdot\sum_{j=1}^{q}d_{G'[X_{j}]}(v_{i_{j-1}},v_{i_{j}})=(1+\eps)\cdot d_{G'}(t,t')=(1+\eps)\cdot d_{G}(t,t')~.
	\end{align*}
\end{proof}

The rest of this subsection is devoted to proving \Cref{lm:tree-reduction} .	
The modification of $G$ into $G'$, and of $\T$ into $\T'$ is described in \Cref{alg:cliqueSum}. Initially $\T'\leftarrow\T$ and $G'\leftarrow G$. The algorithm has two steps.
In the first step, we ensure that the number of leaves is bounded by $k$. This is done by repeatedly deleting non-essential leaf nodes from $\T'$. 
In the second step, we bound the number of nodes of degree two in $\T'$. This is done by deleting redundant paths where all the internal nodes have degree two. Here, however, it will be necessary to add a new node to $\T'$ to compensate for the deleted ones.

\begin{algorithm}[h]
	\caption{\texttt{Clique-Sum Modification}}	\label{alg:cliqueSum}
	\DontPrintSemicolon
	\SetKwInOut{Input}{input}\SetKwInOut{Output}{output}
	\Input{Graph $G=(V,E,w)$, terminals $K\subseteq V$, clique-sum decomposition $\T$ of $G$ into clique-sums}
	\Output{Subgraph $G'$ of $G$ and a clique-sum decomposition $\T'$ of $G'$.}
	\BlankLine
	
	Let $\T' \leftarrow \T$, $G' \leftarrow G$, $\nu \leftarrow \emptyset$.\hspace{87pt} {\color{Darkgreen}\small /** $\nu:K\hookrightarrow\T$ is a function, initially undefined}\;	
	\While {there is a leaf $l\in\T'$ such that $\nu^{-1}(l)=\emptyset$}{
		\If{$l$ contains a terminal $t$ such that $\nu(t)$ is undefined}
		{Set $\nu(t)=l$\;
		}
		\Else{
			Delete $l$ from $\T'$, and delete from $G'$ all the vertices that belong only to $l$
			\;
		}	
	}
	\ForEach{terminal $t\in K$ for which $\nu(t)$ is undefined}
	{Pick arbitrary node $X\in\T'$ containing $t$. Set $\nu(t)=X$.}
	A node $X\in\T'$ is called \emph{redundant} if it has degree $2$ in $\T'$ and $\nu^{-1}(X)=\emptyset$.\; A path $P=\{X_0,X_1,\dots,X_s\}$ in $\T'$ is called \emph{redundant}, if all the internal nodes $\{X_1,\dots,X_{s-1}\}$ are redundant.\;
	\ForEach{maximal redundant path $P=\{X_0,X_1,\dots,X_s\}$}
	{Let $K_0\subseteq X_0$ (resp. $K_s\subseteq X_s$) be the set used for the clique sum $X_0\oplus_h X_1$ (resp. $X_{s-1}\oplus_h X_s$)\;
		Remove $\{X_1,\dots,X_{s-1}\}$ from $\T'$. Delete all vertices $v$ in $G'$ that belong only to nodes in $\{X_1,\dots,X_{s-1}\}$.\;
		Add new node $X$ with $K_1\cup K_2$ as vertices, and the complete graph between them as edges (the weight of $\{v,u\}$ will be $d_G(u,v)$). Add to  $\T'$ the edges $\{X_0,X\},\{X_s,X\}$. 			
		Add the respective edges to $G'$.\;
	}
	\Return Graph $G'$ and clique-sum decomposition $\T'$\;
\end{algorithm}

By \Cref{alg:cliqueSum}, it is straightforward that $\T'$ is a decomposition of $G'$. The second property (that for every $X\in \T'$ either $X\in\T$ or $|X|\le 2h$) is also obvious.
In \Cref{clm:cliqueSumReductionDistancePreserved} we prove that all the terminal distances are preserved exactly by $\T'$, while in \Cref{clm:cliqueSumReductionNumberOfNodesBound} we prove that $\T'$ contains $O(k)$ nodes.
\begin{claim}\label{clm:cliqueSumReductionDistancePreserved}
	For every pair of terminals $t,t'\in K$, $d_G(t,t')=d_{G'}(t,t')$.
\end{claim}
\begin{proof}
	The proof is by induction on the construction of $G'$ and $\T'$ following \Cref{alg:cliqueSum}. Initially $G=G'$ so the claim obviously holds.
	There are two types of modifications that occur: (1) deletion of a leaf node. (2) Replacement of a redundant path by a single new node. Let $G'$ be the graph at some stage with decomposition $\T'$. Suppose that $\tilde{G}$ with decomposition $\tilde{\T}$ is obtained from $G',\T'$ by a single modification step (of type (1) or (2)).	
	Following the algorithm, it is clear that no terminal vertex is ever deleted (as it necessarily belongs to some non-redundant node). 
	Furthermore, if there is a pair of neighboring vertices, $v,u\in G'$ who belong to $\tilde{G}$ then they also neighbors in $\tilde{G}$. This holds because if they both belong to a deleted node, then they are necessarily part of some clique in the clique sum together, and will remain there.
	Let $P_{t,t'}=\{t=v_0,\dots,v_s=t'\}$ be the shortest path from $t$ to $t'$ in $G'$, with the minimal number of hops (that is minimizing $s$ among all shortest paths). By the induction hypothesis, $d_G(t,t')=d_{G'}(t,t')$.
	If no vertex of $P_{t,t'}$ is deleted, then obviously $d_{\tilde{G}}(t,t')=d_{G'}(t,t')$, and we are done.
	Else let $v_i,v_j\in P_{t,t'}$ be the vertices with the minimal and maximal indices among the deleted vertices, respectively. 
	
	We first deal with modification of type (1), deletion of a leaf note $X\in \T'$. We argue that no vertex of $P_{t,t'}$ is deleted.
        Assume toward contradiction that this is not the case.
	As $v_i,v_j$ are deleted, they belong to $X$ and no other node in $\T'$. By the  minimality of $v_{i}$, $v_{i-1}$ was not deleted. Similarly, $v_{j+1}$ was also not deleted. However, as they are neighbors of deleted vertices, they necessarily belong to the clique sum part in $X$. In particular, $v_{i-1},v_{j+1}$ are neighbors in $G'$. Implying that the path $\{t=v_0,\dots,v_{i-1},v_{j+1},\dots,v_s=t'\}$ has weight $\le d_G(t,t')$ but less hops than $P_{t,t'}$, a contradiction. 
	
	Next, we deal with modification of type (2), deletion of a redundant path $P=\{X_0,X_1,\dots,X_s\}$.
	As $v_{i-1},v_{j+1}$ were not deleted, but have deleted neighbors, they necessarily belong to the clique parts in $X_0$ or $X_s$ (the one responsible for joining to $X_1,X_{s-1}$). In particular, $v_{i-1},v_{j+1}$ belong to the newly created node $X$, and are neighbors in $\tilde{G}$. We conclude that 
	$$d_{\tilde{G}}(t,t')\le d_{G'}(t,v_i)+d_{G'}(v_i,v_j)+d_{G'}(v_j,t')=d_{G'}(t,t')=d_{G}(t,t')~.$$	
\end{proof}
\begin{claim}\label{clm:cliqueSumReductionNumberOfNodesBound}
	The number of nodes in $\T'$ is $O(k)$.
\end{claim}
\begin{proof}
	We make the following notation for $\T'$:
	$N$ is the total number of nodes,
	$l$ denotes the number of leafs,
	$a$ denotes the number of degree $2$ nodes for which $\nu^{-1}(X)\ne\emptyset$,
	$b$ denotes the number of degree $2$ nodes for which $\nu^{-1}(X)=\emptyset$,
	Finally $c$  denotes the number of nodes of degree at least $3$.
	
	Recall that for every leaf $X\in \T'$ it holds that $\nu^{-1}(X)\ne\emptyset$ (as otherwise it would've been removed). Thus $l+a\le k$.
	Furthermore, we called nodes $X\in\T'$ of degree $2$ where $\nu^{-1}(X)=\emptyset$ \emph{redundant} and removed all paths consisting of such nodes. It follows that there are no pairs of adjacent  redundant nodes. In particular, the number of redundant nodes is bounded by half the number of edges, thus $b\le \frac{N-1}{2}$.
	Using the sum-of-degree formula we conclude,
	\begin{align*}
	& 2N-2=\sum_{X\in \T'}\deg(X)\ge l+2(a+b)+3c=l+2(a+b)+3(N-l-a-b)=3N-2l-a-b\\
	& N\le l+(l+a)+b-2\le2k+\frac{N-1}{2}-2\\
	& N\le4k-5~.
	\end{align*}
\end{proof}

\paragraph{Remark.}
	\begin{wrapfigure}{r}{0.2\textwidth}
	\begin{center}
		\vspace{-20pt}
		\includegraphics[width=0.2\textwidth]{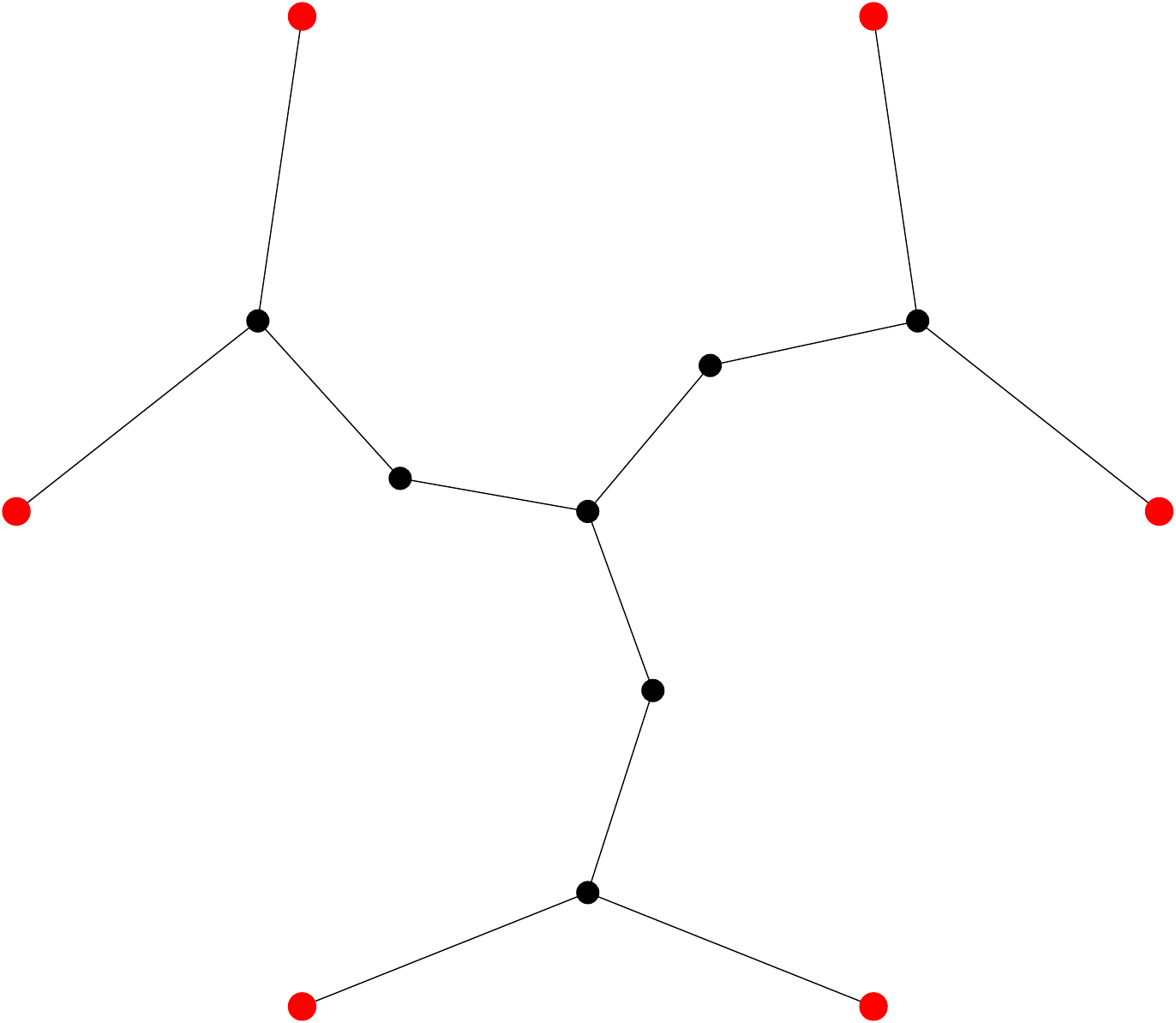}
		\vspace{-15pt}
	\end{center}
	\vspace{-15pt}
\end{wrapfigure}
The analysis of the number of nodes in the decomposition $\T'$ created by \Cref{alg:cliqueSum} is tight, as illustrated by the figure on the right.
Indeed, suppose that only the leaf nodes (colored in red) contain terminals, one each. Then the number of terminals is $k=6$. However, the number of nodes in the decomposition is $19=4\cdot k-5$, and none of them will be removed in \Cref{alg:cliqueSum}.

\section{Embedding Minor-Free Graphs into Small Treewidth Graphs} \label{sec:embedding}

We refer readers to ~\Cref{subsec:tech-tw-emb} for an overview of the argument.

\subsection{Step (1): Planar graphs with a single vortex, proof of \Cref{lm:emb-planar-vortex}}\label{sec:embed-vortex}
We begin by restating the main lemma of the section:
\embPlanarVortex*

Our construction here follows the same steps as the proof of \Cref{lm:one-vortex-Bounded-diam} in \Cref{sec:oneVortex}.
The main difference is that we aim for a clique-preserving embedding, and thus the embedding will be one-to-many.
Our first step is to construct the same hierarchical partition tree $\tau$ as in  \Cref{sec:oneVortex},
where the set of terminal is the entire set of vertices (i.e. $K=V$).
Recall that each node $\Upsilon\in \tau$ is associated with some cluster $\Upsilon\subseteq V$, and some subgraph $G_\Upsilon$ of $G$. There is a spanning tree $T_{\tilde{x}}$ of $G$, such that for every $\Upsilon\in\tau$, $T_\Upsilon=T_{\tilde{x}}\cap G_\Upsilon$ is a spanning tree of $G_\Upsilon$ (see \Cref{inv:SubgraphSingleVortex} and \Cref{inv:tree}).
We used a fundamental vortex cycle $C_\Upsilon$ in $G_\Upsilon$ w.r.t. $T_\Upsilon$ to partition $\Upsilon$ into two parts $\Upsilon^{\mathcal{I}},\Upsilon^{\mathcal{E}}$ and apply this recursively.
The fundamental cycle $C_\Upsilon$ consists of at most $2(h+1)$ shortest paths denoted $\mathcal{P}(C_\Upsilon)$, all of length at most $D$ (\Cref{obs:FundamentalPathLenght}).
By \Cref{clm:InvHolesMaintained}, and the choice of fundamental vortex cycles, the number of terminals, i.e. vertices, in a cluster $\Upsilon$ drops in every two steps of the hierarchy $\tau$. It follows that the depth of $\tau$ is bounded by $O(\log n)$. \footnote{Actually there is no need to control for the size of $\mathcal{P}_\Upsilon$ here. Thus a simpler rule for choosing $C_\Upsilon$ could be applied (compared to \Cref{sec:oneVortex}). For continuity considerations, we will not take advantage of this.}

For a node $\Upsilon\in \tau$, denote by $\tilde{\mathcal{P}}_\Upsilon$ the set of all paths, in all the fundamental vortex cycles $\mathcal{P}(C_{\Upsilon'})$ in all the ancestors $\Upsilon'$ of $\Upsilon$ in $\tau$. Note that the set $\mathcal{P}_\Upsilon$ defined during the proof of \Cref{lm:emb-planar-vortex} is only a subset of $\tilde{\mathcal{P}}_\Upsilon$.
As each fundamental vortex cycle consist of at most $2(h+1)$ paths, and $\tau$ has depth $O(\log n)$, if follows that $|\tilde{\mathcal{P}}_\Upsilon|=O(h\log n)$.

For each path $Q$ and a parameter $\delta > 0$, let $\mathtt{Portalize}(Q,\delta)$ be a $\delta$-net of $Q$. Vertices in $\mathtt{Portalize}(Q,\delta)$ are called \emph{$\delta$-portals} of $Q$. When $\delta$ is clear from the context, we drop the prefix $\delta$.  For a collection of paths $\mathcal{P}$ we denote $\mathtt{Portalize}(\mathcal{P}, \delta) = \cup_{Q\in \mathcal{P}}  \mathtt{Portalize}(Q,\delta)$.

\paragraph{Algorithm} We now describe a tree decomposition $\mathcal{T}$ for the vertex set $V$. For each bag $B$ in the decomposition we
add edges between all the vertices in the bag (where the weight of an edge $u,v$ is $d_G(u,v)$).
However, the bags containing a vertex $v$ does not necessarily induce a connected subgraph of $\mathcal{T}$. Thus our embedding is one-to-many. Specifically, in the sub-tree decomposition of $\mathcal{T}$ induced by bags containing $v$, each connected component corresponds to a different copy of $v$.
Note that the portal vertices have only a single copy, as all the bags containing such a vertex are connected.
It is thus straightforward that the embedding is dominating. What needs to be proven is that the additive distortion is at most
$\eps$ times the diameter of the input graph  and that the embedding clique preserving.

For a  node $\Upsilon$, consider the graph $\tilde{G}_{\Upsilon}=G[\Upsilon\cup (\cup\tilde{\mathcal{P}}_{\Upsilon})]$ induced by $\Upsilon$ and all the fundamental cycles in all its ancestors (note that the graph $G_{\Upsilon}$ defined in \Cref{sec:oneVortex} is only a subgraph of $\tilde{G}_{\Upsilon}$).  For each node $\Upsilon\in \tau$, we create a bag $B_\Upsilon$ that contains $\mathtt{Portalize}(\tilde{\mathcal{P}}_\Upsilon,  \eps D/2)$. The bags are connected in the same way as the $\tau$ nodes they represent. In addition, each leaf node $\Upsilon_l\in \tau$, $B_{\Upsilon_l}$ also contains all the vertices in $\Upsilon_l$.  For every maximal clique \footnote{A set $Z$ of vertices is a maximal clique if $Z$ is a clique, and there is no clique that strictly contains $Z$. Since $\tilde{G}_{\Upsilon_l}$ is $O(h)$-degenerated, all maximal cliques of $\tilde{G}_{\Upsilon_l}$ can be found in time $O_h(n)$~\cite{ELS10}.} $Z$ in  $\tilde{G}_{\Upsilon_l}$, we create a bag $B_Z$, connected in $\mathcal{T}$ only to  $B_{\Upsilon_l}$, where $B_Z$ contains vertices in $Z\cup\mathtt{Portalize}(\tilde{\mathcal{P}}_{\Upsilon_l},  \eps D/2)$. This concludes the construction.

First observe that the maximal size of a clique in $G$ is $O(h)$. Furthermore, according to the construction of $\tau$, every leaf node $\Upsilon_l$ contains at most $O(h)$ vertices. Finally, for every node $\Upsilon\in\tau$,  $ \tilde{\mathcal{P}}_{\Upsilon}$ contains at most $O(h\log n)$ paths and every path in $\tilde{\mathcal{P}}_{\Upsilon}$ has weight at most $D$. Thus, it holds that $|\mathtt{Portalize}(\tilde{\mathcal{P}}_{\Upsilon},  \eps D/2)|\le O(h\cdot\log n)\cdot O(\frac1\eps)=O(\frac{h\log n}{\eps})$. From the definition we immediately have, 

\begin{observation}\label{obs:tw-leaf} The decomposition $\mathcal{T}$ is a valid tree-decomposition of width  $O(\frac{h\log n}{\eps})$.
\end{observation}

Next we argue that our embedding is clique-preserving.  By induction, in every level of $\tau$, there is a node $\Upsilon$ such that $Z\subseteq\tilde{G}_\Upsilon$. That is, every vertex $v\in Z$ either belongs to $\Upsilon$, or belongs to a fundamental vortex cycle $C_{\Upsilon'}$ of an ancestor $\Upsilon'$ of $\Upsilon$. In particular, there is a leaf node $\Upsilon_l$ such that $Z\subseteq\tilde{G}_{\Upsilon_l}$. Let $Z'$ be some maximal clique in $\tilde{G}_{\Upsilon_l}$ containing $Z$. By the construction, there is a bag $B_{Z'}$ containing all the vertices in $Z'$, and in particular $Z$.

Finally we bound the distortion. Consider a pair of vertices $u,v$. We then show that for every two copies $u',v'$ it holds that $d_H(u',v')\le d_G(u,v)+\eps D$.
Consider first the case where there is a leaf node $\Upsilon_l$ containing both $u,v$. In this case there is a single copy of $u,v$, which belongs to the same bag and therefore it holds that $d_H(u,v)=d_G(u,v)$.
Else, let $P_{u,v}$ be some shortest path between $u$ and $v$ in $G$.
Let $\Upsilon_{u,v}\in\tau$ be the first node such that its fundamental vortex cycle $C_{\Upsilon_{u,v}}$ intersects $P_{u,v}$ at some vertex $z$. Specifically, there is a path $Q_z\in \mathcal{P}(C_{\Upsilon_{u,v}})$ such that $z\in Q_z\cap P_{u,v}$. There is some portal $z'\in \Pt_{Q_z}$ such that $d_G(z,\hat{z})\le\frac\eps2 D$. Further, by the construction of $H$, any bag associated with a descendent of $\Upsilon_{u,v}$ contains $\hat{z}$. In particular, every bag $B$ containing either a copy of $u$ or $v$, also contains $\hat{z}$.  We conclude
\begin{align*}
d_{H}(u',v') & \leq d_{H}(u',\hat{z})+d_{H}(\hat{z},v')=d_{G}(u,\hat{z})+d_{G}(\hat{z},v)\\
& \leq d_{G}(\hat{z},z)+d_{G}(u,z)+d_{G}(z,v)+d_{G}(z,\hat{z})\leq d_{G}(u,v)+\eps D~.
\end{align*}
Note that this argument applies also to two copies $u_1,u_2$ of the same vertex $u$, where $P_{u,u}=\{u\}$.

\subsection{A Cutting Lemma}\label{sec:cut}
Let $H$ be a connected subgraph of an arbitrary graph $G$.
\commentforlater{
  As discussed with Hung this is really a problematic part.
  I am fine with defining this for general graph, but then this
  is used on planar or bounded genus graphs and when used
  it is used in a way that \textbf{must} preserve the genus of the
  surface. Then, it has to be proven that this operation can be
  implemented so as to achieve that.
  If you want to prove this formally you would have to use
  combinatorial embeddings, I think...
}
Let $I(H)$ be the set of all edges incident to vertices in $H$ that do not belong to $E(H)$. Let $LE(H)$ (left edges) and $RE(H)$ (right edges) be a partition of  $I(H)$.   We say a graph obtained from $G$ by \emph{cutting along $H$}, denoted by $G \cut H$, is the graph obtained by: (1) removing all the vertices of $H$ from $G$, (2) making two copies of $H$, say $H^l$ and $H^r$, (3) 
adding an edge between  two copies of the endpoints of  an edge $e$ in $H^l$ ($H^r$) for every edge $e \in LE(H)$ ($e \in RE(H)$), and (4)   adding an edge  $uv^{l}$ ($uv^r$) for each edge $uv \in LE(H)$  ($uv \in RE(H)$) where $v^{l}$ ($v^r$) is the copy of $v$ in $H^l$ ($H^r$). We say that $H$ is \emph{separating} if $G\cut H$ is disconnected, and \emph{non-separating} otherwise.  The following lemma will be useful in for our embedding framework. 

\begin{lemma}\label{lm:cutting}  Let $H$ be a non-separating connected subgraph of $G$.  Then:
	\begin{equation*}
	\dm(G\cut H) \leq 4\dm(G) + 2\dm(H)
	\end{equation*}
\end{lemma}
\begin{proof}
	Let $\Delta = \dm(G)$ and $L = \dm(H)$.  For every vertex $v\in V$ $d_G(v,H)\le\Delta$, which implies $\min(d_{G\cut H}(v,H^l), d_{G\cut H}(v,H^r)) \leq \Delta$ (go along the path realizing the distance $d_G(v,H)$ until we meet a vertex of $H^l\cup H^r$). Since $H$ is non-separating, there is a path $P=v_0,\dots,v_s$ from $v_0\in H^l$ to $v_s\in H^r$. Since  $d_{G\cut H}(v_0,H^l)=0$ and $d_{G\cut H}(v_s,H^r)=0$, there is an index $i$ such that $d_{G\cut H}(v_i,H^l)\le\Delta$ and $d_{G\cut H}(v_{i+1},H^r)\le\Delta$. It follows that $d_{G\cut H}(H^l,H^r)\le 2\Delta+d_G(v_i,v_{i+1})\le3\Delta$. By triangle inequality, for any two vertices $x,y \in H^l \cup H^r $, $d_{G\cut H}(x,y) \leq 2L + 3\Delta$. 
	 
	 Let $u$ and $v$ be two vertices in $G\cut H$ and a  shortest path $P$ between $u$ and $v$ in $G$. If $P \cap H = \emptyset$, then $d_{G\cut H}(u,v) = w(P) \leq \dm(G)$. Otherwise, let $v_i$ and $v_j$ be the first and last vertices belonging to $H_l\cup H_r$ when we follow $P$ from $u$ to $v$ in $G\cut H$. Then $d_{G\cut H}(u,v)\le d_{G\cut H}(u,v_i)+d_{G\cut H}(v_i,v_j)+d_{G\cut H}(v_j,v)\le\Delta+d_{G\cut H}(v_i,v_j)\le4\Delta+2L$.
\end{proof}	 

\subsection{Step (2.1): Planar graphs with more than one vortex, proof of \Cref{lm:em-vortexReduce}}\label{sec:EmbedManyVortices}
We begin by restating the main lemma of the section:
\emVortexReduce*

	The embedding algorithm is presented in \Cref{alg:emb-planar-mult-vortex}. Parameter $s$ represents the step number of the recursion. Initially $s = 0$ and $G_0$ is the input graph with diameter $D$.

\begin{algorithm}[]
	\caption{\texttt{EmbedPlanarMultipleVortices}}	\label{alg:emb-planar-mult-vortex}
	\DontPrintSemicolon
	\SetKwInOut{Input}{input}\SetKwInOut{Output}{output}
	\Input{Graph $G=(V,E,w)=G_\Sigma\cup W_1\cup\dots\cup W_{v(G)}$, parameter $\eps$, and step number $s$}
	\Output{An embedding $f$ to a graph $H$ with tree decomposition $\mathcal{T}$ with  additive distortion $\eps D$}
	\BlankLine
	\If{ $v(G)= 1$}
	{
		$\{f,H,\mathcal{T}\}\leftarrow \mathtt{EmbedPlanarOneVortex}(G,\frac{\eps}{10^{v(G_0)-1}})$\label{line:EmbedVortexBaseCase}\;
		return $\{f,H,\mathcal{T}\}$;
	}
	Let $P$ be the shortest proper vortex path in $G$ between $x_i\in W$ and $y_j\in W'$ and $\mathcal{P}$ be vortex path $X_i\cup Y_j\cup P$\label{line:EmbedVortexPdeff}\;
	$K$ be $G[X_i]\cup G[Y_j]\cup P$\label{line:EmbedVortexKdeff}\;
	$G'\leftarrow G\cut K$\;
	$\{f',H',\mathcal{T}'\} \leftarrow \mathtt{EmbedPlanarMultipleVortices}(G',\eps, s+1)$\label{line:EmbedVortexRecursion}\;
	Add  $\Pt = \mathtt{Portalize}(\mathcal{P},\eps D/2)$ to every bag of $\mathcal{T}'$\label{line:EmbedVortexPortalize} \;
	\For{each maximal clique $Q$ of $G$ such that $Q\cap K \neq\emptyset$\label{line:EmbedVortexFor}}{
		Let $B$ be the bag that contains an image of $Q\setminus K$\label{line:EmbedVortexBagB}\;
		Create a bag $B_Q = (Q\cap K) \cup B$ and make $B_Q$ adjacent to $B$\label{line:EmbedVortexNewBag}\;
	}
	Let $\mathcal{T} \leftarrow \mathcal{T}'$ and $\{f,H,\mathcal{T}\}$  be the resulting embedding;
	\Return $f, H$ and $\mathcal{T}$\;
\end{algorithm}	
	
	Recall that a \emph{proper vortex path} $P$ is a path in $G$ between a vertex $v$ in a vortex $W$ to a vertex $u$ in a vortex $W'$ such all the vertices on $P$ belong to the planar part $G_\Sigma$, and other than the first and last vertices, $P$ do not contain any vortex vertices. 	The path $P$ picked in \cref{line:EmbedVortexPdeff} is the minimal path having its endpoints in two different vortices. By minimality, $P$ is necessarily a proper vortex path. 
	
\begin{figure}[]
		\centering{\includegraphics[width=1.0\textwidth]{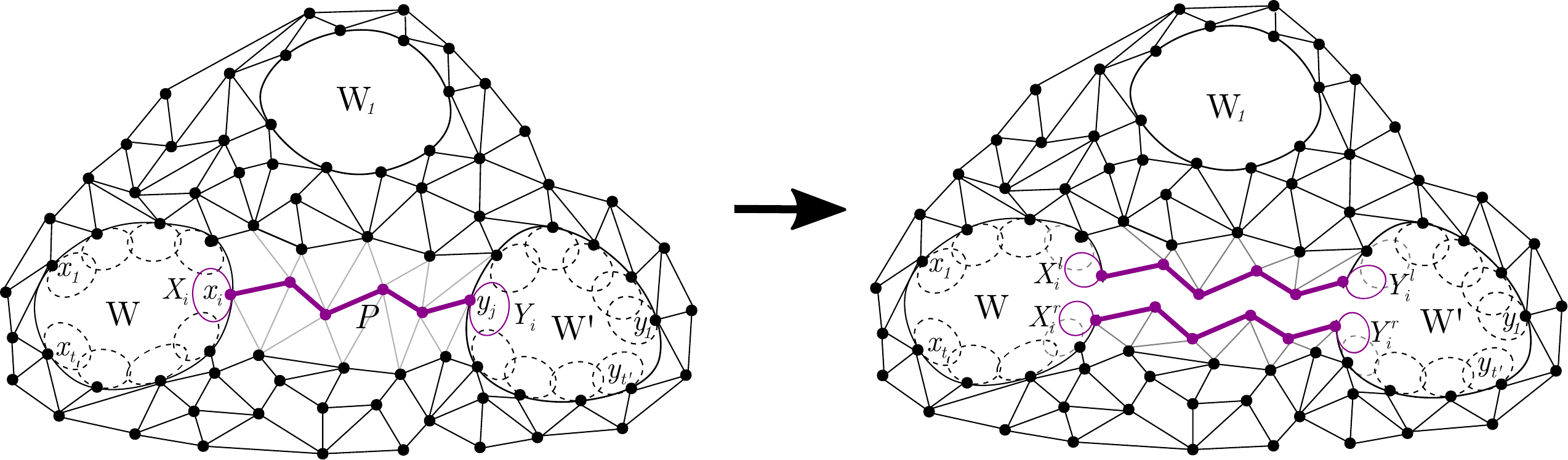}}
		\caption{\label{fig:embeddingMerge}\small \it 
			Cutting the left graph along a vortex path $X_i\cup P\cup Y_j$ (hilighted purple) to obtain the graph on the right figure.
		}
	\end{figure}

	In \cref{line:EmbedVortexKdeff}, we cut $G$ along $K = G[X_i]\cup G[Y_j] \cup P$. In this cutting procedure, we define $LE(K), RE(K)$ as folows. For $P$, we define left edges $LE(P)$  and right edges $RE(P)$ of $P$ w.r.t the drawing of the planar part of $G$ as follows:  $LE(P)$  ($RE(P)$) contains edges drawn  on the  (right) left side of the path as we walk on the path from $x_i$ to $y_j$\footnote{It could be that an edge $ e = (u,v)$ where $u,v \in P$ such that it is drawn on the left side of $v$ and  on the right side of $u$. In this case, we simply assign $e$ to $LE(P)$.}.  For $X_i$, assume that the path decomposition of the vortex containing $X_i$ is $\{X_1,\ldots, X_i, X_{i+1}, \ldots, X_t\}$ where the perimeter vertex $x_{i+1}$ of $X_{i+1}$ is incident to a right edge of $P$; in this case, we define  $RE(X_i)$ (respectively $LE(X_i)$) to be the set of edges incident to vertices of $X_{i}$ in $G[X_{i+1} \cup \ldots \cup X_{t}]$ (respectively $G[X_{1} \cup \ldots \cup X_{i-1}]$).  Similarly, assume that the path decomposition of the vortex containing $Y_j$ is $\{Y_1,\ldots, Y_j, Y_{j+1}, \ldots, Y_{t'}\}$ where the perimeter vertex $y_{j+1}$ of $Y_{j+1}$ is incident to a right edge of $P$; in this case, we define  $RE(Y_j)$ ($LE(Y_j)$) be the set of edges incident to vertices of $Y_{j}$ in $G[Y_{j+1} \cup \ldots \cup Y_{t'}]$ ( $G[Y_{1} \cup \ldots \cup Y_{i-1}]$).  Finally, we define $LE(K) = LE(X_i) \cup LE(P) \cup LE(Y_j)$ and  $RE(K) = RE(X_i) \cup RE(P) \cup RE(Y_j)$.   By cutting $G$ along $H$ we essentially \emph{merge} two vortices of $G$ into a single vortex (see \Cref{fig:embeddingMerge}). 
	
	Let $P^l = \{x_i^l = p^l_0, p^l_1, \ldots, p^l_k = y_j^l\}$ be the left copy of $P$ and $P^r = \{x_i^r = p^r_0, p^r_1, \ldots, p^r_k = y_j^r\}$ be the right copy of $P$. The path decomposition of the new vortex is 
	$$\{X_1,\ldots, X_{i-1}, X^{l}_{i}, p^l_1, \ldots, p^l_{k-1}, Y^l_{j}, Y_{j-1}, \ldots Y_1, Y_{t'}, \ldots, Y_{j+1}, Y^r_{j}, p^r_{k-1}, \ldots, p^r_1, X^r_{i}, X_{i+1}, \ldots, X_t\}$$
	that  has width at most $h$; here $X^l_i, X^r_i$ ($Y^l_j, Y^r_j$) are left and right copies of $X_i$ ($Y_j$), respectively. 
	
	Since the vortex merging step reduces the number of vortices by $1$, $v(G') = v(G) - 1$. Note that cutting $G$ along $K$ does not destroy the connectivity of $G$ as we can walk from the left copy of $x_i$ to a right copy of $x_i$ along the face where $W$ is attached to.   We then recursively apply \Cref{alg:emb-planar-mult-vortex} to $G'$. To account for the damage caused by cutting $G$ along $K$, we add all the portals of each shortest path in $\mathcal{P}$ to every bag of $\mathcal{T}'$. In the for loop at \cref{line:EmbedVortexFor}, we make the embedding clique-preserving. We will show later by induction that $f'(\cdot)$ is clique-preserving and thus, bag $B$ in \cref{line:EmbedVortexBagB} indeed exists. 
	
	The base case is when $v(G)= 1$; in this case, we use the embedding in \Cref{lm:emb-planar-vortex} (\cref{line:EmbedVortexBaseCase}). This completes the embedding procedure.

	\begin{claim}\label{clm:Diam-Gj} For any step $s$, $\dm(G) \leq 10^{s} D$. Furthermore, $s \leq v(G_0)-1$.
	\end{claim}
	\begin{proof}
		The path $P$ in \cref{line:EmbedVortexPdeff} has weight $w(P) \leq \dm(G)$ since it is a shortest path, and since every edge between two vertices in the vortex has length at most $D$, $\dm(H) \leq \dm(G) + 2D\leq 3\dm(G)$. Thus, by \Cref{lm:cutting},  $\dm(G') \leq 4\dm(G) + 2(3\dm(G)) = 10\dm(G)$. Hence, by induction, at step $s$, we have:
		\begin{equation}\label{eq:dm-cutting-vortex}
		\dm(G) \leq 10^{s} \dm(G_0) = 10^{s}D \qedhere
		\end{equation}
		Finally, observe that the recursion has $v(G_0)-1$ steps since after every step, we reduce the number of vortices by $1$.  
	\end{proof}
	
	\begin{proof}[Proof of~\Cref{lm:em-vortexReduce}]
		We first bound the width of the tree decomposition output by the algorithm. Let $\tw(\mathcal{T}')$ be the treewidth of $\mathcal{T}'$ in \cref{line:EmbedVortexRecursion}. By \Cref{clm:Diam-Gj}, the size of the portal set $\Pt$  in \cref{line:EmbedVortexPortalize} is $O(h \frac{10^{s-1}}{\eps} ) = \frac{h 2^{O(v(G_0))}}{\eps}$. Since $G$ is $O(h)$-degenerate,  every clique $Q$ in \cref{line:EmbedVortexFor} has size $O(h)$. Thus, $|B_Q| = \tw(\mathcal{T}') + \frac{h 2^{O(v(G_0))}}{\eps} + O(h)$; this implies
		\begin{equation}
		\tw(\mathcal{T}) = \tw(\mathcal{T}') + \frac{h 2^{O(v(G_0))}}{\eps} + O(h) = \tw(\mathcal{T}') + \frac{h 2^{O(v(G_0))}}{\eps} 
		\end{equation}
		In the base case, the treewidth of the embedding in \cref{line:EmbedVortexBaseCase} is $O(\frac{h2^{O(v(G_0))}\log n }{\eps})$. Thus, the total treewidth after $v(G_0)-1$ steps of recursion is $O(v(G_0))\cdot O(\frac{h2^{O(v(G_0))}\log n }{\eps}) = O(\frac{h2^{O(v(G_0))}\log n }{\eps})$.
		
		We next argue that the embedding is clique-preserving by induction. The base case where $v(G) = 1$,  $f(\cdot)$ is clique-preserving by \Cref{lm:emb-planar-vortex}. By the induction hypothesis, we assume that $f'(\cdot)$ in \cref{line:EmbedVortexRecursion} is clique-preserving. Let $Q$ be a clique of $G$. If $K\cap Q = \emptyset$, then $Q$ is a clique in $G'$ and hence $f(\cdot)$ preserves $Q$ as $f'(\cdot)$ preserves $Q$. Otherwise, since $Q\setminus K$ is a clique in $G'$, there is a bag of $\mathcal{T}'$ containing an image of $Q\setminus K$. Thus there exist a bag $B\in\mathcal{T}'$ containing $Q\setminus K$ (\cref{line:EmbedVortexBagB}). This implies that $\mathcal{T}$ has a bag $B_Q$ containing an image of $Q$ (\cref{line:EmbedVortexNewBag}); We conclude that $f(\cdot)$ is clique-preserving.
		
		It remains to bound the distortion. For the base case, since the diameter is at most $10^{v(G_0)-1} D$, the  distortion is at most $\frac{\eps}{10^{v(G_0)-1}} 10^{v(G_0)-1} D = \eps D$. For the inductive step, suppose that we have additive distortion $\eps D$ for $G'$.
		The set $\Pt$ is contained in every bag of $\mathcal{T}$ (\cref{line:EmbedVortexPortalize}), where each vertex in $v\in Q$ has a portal vertex $t\in\Pt$ at distance at most $\frac\eps2 D$.   Thus, if a shortest path between $u$ and $v$ intersects $K$, then by rerouting the path through the nearest portal in $\Pt$, we obtain a new path with length at most $w(P(u,v)) + 2\eps D/2 = w(P(u,v)) + \eps D$ for every two copies $u',v'$ of $u,v$. Note that this argument holds in particular for all $K$ vertices. Otherwise, using the induction hypothesis the distance between any two copies of $u$ and $v$ is preserved  with additive distortion $\eps D$ (note that no new copies are created in this step).
        \end{proof}

\subsection{Step (2.2): Cutting out genus, proof of \Cref{lm:embed-genus-vortex}}\label{sec:genus-minor}

We begin by restating the main lemma of the section:
\embedGenusVortex*

The main tool we will use in this section is to cut along a vortex cycle to reduce the genus of the surface embedded part of $G$ without disconnecting $GG$.  We assume that every face of $G_{\Sigma}$ is a simple cycle since we can always add more edges to $G_{\Sigma}$ without destroying the shortest path metric. 

\subsubsection{Cutting Along a Vortex Cycle}

Let $F_i$ be the face that $W_i$ is glued to, $1\leq i \leq v(G)$. Let $K_{\Sigma}$ be the graph obtained from $G_{\Sigma}$ by, for any $i \in [1,v(G)]$, adding virtual vertex $f_i$ to each face $F_i$ and connect $f_i$ to every other vertex of $F_i$ by an edge of length $0$. Let $U = \{f_i\}_{i=1}^{v(G)}$ and $B = \cup_{i=1}^{v(G)}F_i$. For simplicity of presentation, we assume that every edge of $G$ has positive length. Pick a vertex $r\in V(G)\setminus (U\cup B)$ and compute a shortest path tree $T$ of $K_{\Sigma}$ from $r$. 

Let $K^*_{\Sigma}$ be the \emph{dual graph} of $K_{\Sigma}$. A \emph{tree-cotree decomposition} w.r.t. $T$ is a partition of $E(K_{\Sigma})\setminus T$ into two sets $(C, X)$ such that $C^*$ is a spanning tree of $K^*_{\Sigma}$; by Euler formula,  $|X| = 2g$ of $\Sigma$ is orientable and $|X| = g$ if  $\Sigma$ is non-orientable. For each $e \in E(K_{\Sigma})\setminus T$, define $C_e$ be the fundamental cycle of $T$ w.r.t $e$. By Lemma 2 of~\cite{Eppstein03}, there is an edge $e \in X$ such that cutting $\Sigma$ along $C_e$    does not disconnect the resulting surface and that $\Sigma\cut C_e$ has smaller Euler genus.  Since  every face $K_{\Sigma}$ is a simple cycle, we have:

\begin{claim} \label{clm:cut-nonseparating}
	There is an edge $e \in X$ such that $K_{\Sigma}\cut C_e$ is connected and that $g(K_{\Sigma}\cut C_e) \leq g(K_{\Sigma})-1 $.
\end{claim}

In cutting $K_{\Sigma}$ along $C_e$, we define "left edges" and "right edges" w.r.t the embedding of $C_e$ on $\Sigma$ (see page 106-107 in the book by Mohar and Thomassen~\cite{MT01} for details on how to define left and right sides of a cycle embedded on a surface-embedded graph.). 

Fix an edge $e$ in Claim~\ref{clm:cut-nonseparating}. (Such an edge can be found in polynomial time by trying all possible edges.) Let $C_e = T[r_0,u] \circ (u,v) \circ T[v,r_0]$ where $(u,v) = e$ and $r_0$ is the lowest common ancestor of $u$ and $v$ in $T$.  We have the following claim whose proof is deferred to \Cref{app:cycle-intersect-vortex-face}.

\begin{restatable}[Single Vortex with Bounded Diameter]{claim}{CycleIntersectVortexFace}
	\label{clm:cycle-intersect-vortex-face}
	For $i \in [1,v(G)]$, $|C_e \cap F_i| \leq 2$ and if $|C_e \cap F_i| = 2$, then ${x_1,f_i,x_2}$ is a subpath of $C_e$ where $C_e\cap F_i = \{x_1,x_2\}$.	
\end{restatable}

\paragraph{Induced vortex cycles}  Fix an orientation of $C_e$ from $r_0$ to $u$, then $v$ and back to $r_0$. Let $F_{i_1}, \ldots, F_{i_k}$ be a sequence of faces such that $C_e \cap F_{i_j} \not= \emptyset$  for all $1\leq j \leq k$and that is ordered by the direction from $r_0$ along $C_e$ back to $r_0$.  Let $\{x_j,y_j\}$ be two vertices of $C_e\cap F_{i_j}$; if $|C_e\cap F_{i_j}| = 1$, then we let $x_{j} = y_{j} = C_e\cap F_{i_j}$.  Cycle $C_e$ \emph{induces} a \emph{vortex cycle} $\mathcal{C}_e = Y_1\cup P_1 \cup \ldots \cup X_k \cup Y_k\cup P_k \cup X_1$ where:
\begin{itemize}
	\item $P_j$ is $C_e[y_j,x_{j+1}]$, $1\leq j \leq k$ with convention $x_{k+1} = x_1$.
	\item $X_j$ is the bag of $W_{i_j}$ attached to $x_j$ and $Y_j$ is the bag of $W_{i_j}$ attached to $y_j$.
\end{itemize}

\paragraph{Cutting along a vortex cycle} First, we perform a cut along each path $P_j$, $1\leq j \leq k$; each vertex of the path will have two copies. The left edges and right edges are defined w.r.t the embedding of $G_{\Sigma}$. Now, for each vortex $W_{i_j} = \{X_1,\ldots, X_t\}\}$ that has (at most) two perimeter vetices $x_j$ and $y_j$, let $X_p$ and $X_q$ be two bags containing $x_j$ and $y_j$ respectively.  Assume that $ p \not = q$ (the case $p = q$ is handled in the same way.) We makes two copies $X^1_p$ and $X^2_p$ of $X_p$ and two copies $X^1_q$ and $X^2_q$ of $X_q$. Let $x^1_j, x^2_j$ be two copies of $x_j$ such that the neighbor of $x^1_j$ in $F_j$ is the perimeter vertex of $X_{p-1}$.  Let $y^1_j, y^2_j$ be two copies of $y_j$ such that the neighbor of $y^2_j$ in $F_j$ is the perimeter vertex of $X_{q+1}$. We make a new path decomposition $W'_{i_j} = \{X_1,\ldots, X^1_p,X^2_p ,\ldots, X^1_q, X^2_q, \ldots, X_t\}$ where:

\begin{itemize}[noitemsep]
	\item $X^i_p$ ($Y^i_q$) is attached to $x_j^i$ ($y_i^i$) for $i = 1,2$. 
	\item Each vertex $x \in X_p$ appears in $\{X_1, \ldots, X_{p-1}\}$ will be replaced by the copy of $x \in X^1_p$. Each vertex $y \in X_q$ appears in $\{X_{q+1}, \ldots, X_{t}\}$ will be replaced by the copy of $y \in X^2_q$. 
	\item For each vertex $x \in X_p $, we replace the occurrence of $x$ in $\{X_p^2,\ldots, X_q^1\}$ by the copy of $x$ in $X_p^2$. For each vertex $y \in X_q\setminus X_p$, we replace the occurrence of $y$ in $\{X_p^2,\ldots, X_q^1\}$ by the copy of $y$ in $X_q^1$.
\end{itemize}

\begin{figure}[]
	\centering{\includegraphics[width=0.8\textwidth]{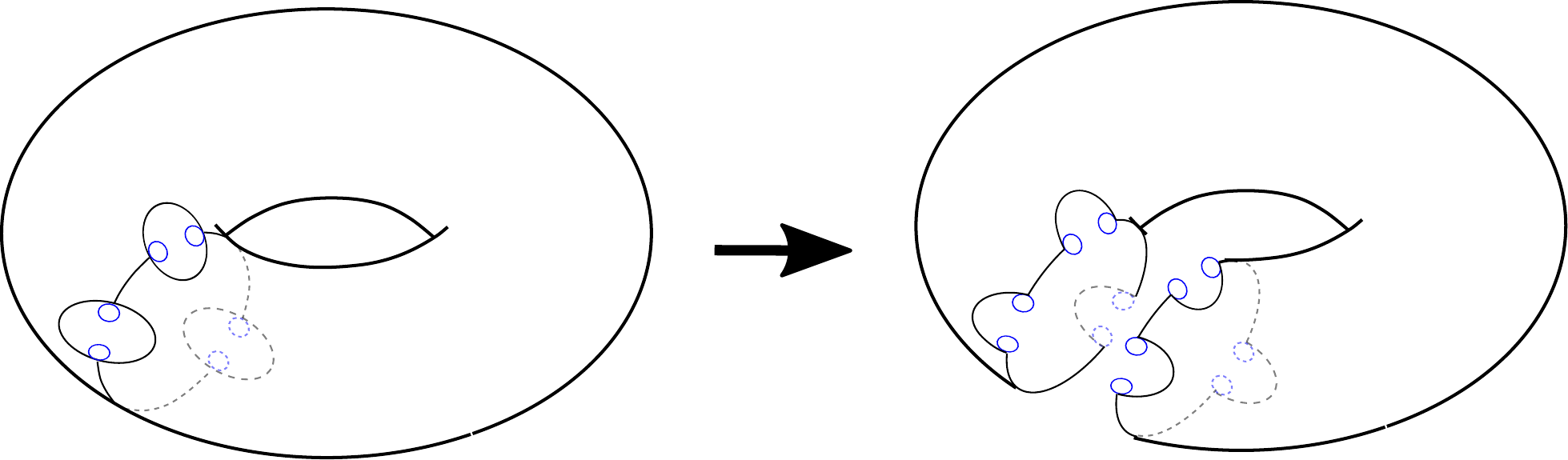}} 
	\caption{\label{fig:cutting-vortex-path}\small \it 
		Cutting along a fundamental vortex cycle to reduce the genus of the graph. The cutting operation induces two new vortices of the resulting graph.
	}
\end{figure}

See \Cref{fig:cutting-vortex-path} for an illustration. We denote the graph afer cutting $G$ along $\mathcal{C}_e$ by $G\cut \mathcal{C}_e$. We show that:

\begin{lemma}\label{lm:cutting-formal}
	There is a vortex cycle $\mathcal{C} = Y_1\cup P_1 \cup X_2 \cup \ldots \cup X_k \cup Y_k\cup P_k \cup X_1$ such that  (a) $G\cut \mathcal{C}$ is connected, (b) $g(G\cut \mathcal{C}) \leq g(G)-1$, (c) $v(G\cut \mathcal{C})  \leq v(G) +1 $, (d) $w(P_i) \leq 3\dm(G)$ for every $1\leq i\leq k$ and (e) $\dm(G\cut \mathcal{C}) \leq 2^{O(v(G))}\dm(G)$.
\end{lemma}
\begin{proof}
	Let $G' = G\cut \mathcal{C}$.	Property (a) and (b) follow directly from \Cref{clm:cut-nonseparating}. Since cutting long a vortex cycle can creat at most independent two copies of the vortex cycle which become two new vortices, $v(G') \leq v(G+1)$.
	
	For property (d), we observe that  $P_i$ is either a shortest path of $G$, or is composed of two shortest path of $G$ share the same endpoint $r_0$ or three shortest paths of $G$ containing edge $(u,v)$. Thus, $w(P_i) \leq 3\dm(G)$.  
	
	To prove (e), it is helpful to think of the process of cutting along $\mathcal{C}_e$ as cutting along each vortex path $\{Y_i \cup P_2\cup X_{i+1}\}$ for $i \in [1,v(G)]$  one by one.  Since $\dm(G[Y_i] \cup P_2 \cup G[X_{i+1}])\leq 5\dm(G)$, by \Cref{lm:cutting}, each cut increases the diameter of the graph by a constant factor (of $14$). Thus, cutting along at most $v(G)$ vortex paths increases the diameter of $G$ by a factor of $2^{O(v(G))}$.
\end{proof}

\subsubsection{The proof}

The embedding algorithm is presented in \Cref{alg:embed-genus-vortex}. Parameter $s$ represents the depth of the recursion. Initially $s = 0$ and $G_0$ is the input graph with diameter $D$. 	In \cref{line:EmbedGenusCut}, we cut $G$ along $\mathcal{C}$; the goal is to reduce the genus of the surface embedded part of $G$ using \Cref{lm:cutting-formal}. In \cref{line:portalizeVortexCycle} we \emph{portalize} the vortex cycle $\mathcal{C}$; we regard $\mathcal{C}$ is a collection of paths where each vertex in a vortex bag is a singleton path. In the for loop at \cref{line:EmbedGenusFor} we make the embedding clique-preserving.

\begin{algorithm}[t]
	\caption{\texttt{GenusReduction}}	\label{alg:embed-genus-vortex}
	\DontPrintSemicolon
	\SetKwInOut{Input}{input}\SetKwInOut{Output}{output}
	\Input{Graph $G=(V,E,w)$ with genus $g(G)$ and $v(G)$ vortices, parameter $\epsilon < 1$, and step number $s$}
	\Output{An embedding $f$ to a graph $H$ with tree decomposition $\mathcal{T}$ with additive distortion $\epsilon D$}
	\BlankLine
	\If{ $g(G)= 0$}
	{
		$\{f,H,\mathcal{T}\}\leftarrow  \mathtt{EmbedPlanarMultipleVortices}(G,\frac{\epsilon}{2^{c (g(G_0)v(G_0))}})$ for some big constant $c$\label{line:EmbedGenusBaseCase}\;
		return $\{f,H,\mathcal{T}\}$;
	}
	Let $\mathcal{C}$ be the vortex cycle guaranteed by \Cref{lm:cutting-formal} w.r.t. $G$\;
	$\Pt\leftarrow \mathtt{Portalize}(\mathcal{C},\epsilon D/2)$\label{line:portalizeVortexCycle}\;
	$G'\la G\cut \mathcal{C}$\label{line:EmbedGenusCut}\;
	$\{f',H',\mathcal{T}'\}\leftarrow  \mathtt{GenusReduction}(G',\epsilon, s+1)$\label{line:genusReduction}\;
	Add $\Pt$ to every bag of $\mathcal{T}'$\label{line:EmbedGenusPortalize}\;
	\For{each maximal clique $Q$ of $G$ such that $Q\cap \mathcal{C}  \neq\emptyset$\label{line:EmbedGenusFor}}{
		Let $B$ be the bag that contains an image of $Q\setminus \mathcal{C}$ \label{line:bagClique}\;
		Create a bag $B_Q = \left(Q\cap \mathcal{C}\right) \cup B$ and make $B_Q$ adjacent to $B$\;
	}
	Let $\mathcal{T} \leftarrow \mathcal{T}'$ and $\{f,H,\mathcal{T}\}$  be the resulting embedding\;
	\Return $f, H$ and $\mathcal{T}$\;
\end{algorithm}

We first bound the treewidth of the embedding. Let $s_{m}$ be the maximum of recursion depth of the algorithm.  We have $s_{m} \leq g(G_0)$. By~\Cref{lm:cutting-formal}, $ v(G') \leq v(G)+1$and hence for every step $s$, 
\begin{equation}\label{eq:diam-G-everystep}
\dm(G) \leq  2^{O(\sum_{i=1}^{s} v(G_0) + i)} \dm(G_0) = 2^{O(v(G_0)g(G_0))}D
\end{equation}
Thus, we have:
\begin{equation*}
\begin{split}
|\Pt| &= |\mathtt{Portalize}(\mathcal{C},\epsilon D/2)| \\
& = 2 \left(O(v(G)h) + \frac{v(G) 2^{O(v(G_0)g(G_0))}}{\epsilon}\right) = \frac{h 2^{O(v(G_0) g(G_0))}}{\epsilon}\\
\end{split}
\end{equation*}
Since $G$ is $O(h + g(G))$-degenerate,  every clique $Q$ in \cref{line:EmbedGenusFor} has size $O(v(G) + g(G))$. Thus, $|B_Q| = \tw(\mathcal{T}') +  \frac{h 2^{O(v(G_0) g(G_0))}}{\epsilon} + O(h + g(G))$; this implies
\begin{equation}\label{eq:tw-recursion}
\tw(\mathcal{T}) = \tw(\mathcal{T}') + \frac{h 2^{O(v(G_0) g(G_0))}}{\epsilon}
\end{equation}

For the base case when $s = s_{\max}$,  $v(G) \leq v(G_0) + g(G_0)$ and hence according to \Cref{lm:em-vortexReduce} the treewidth of $\mathcal{T}$ in \cref{line:EmbedGenusBaseCase} is $O(\frac{h2^{O(v(G_0) g(G_0))}\log n }{\epsilon})$.  Thus, the total treewidth after $g(G_0)$ steps of recursion by equation (\ref{eq:tw-recursion}) is $O(g(G_0) \frac{h2^{O(v(G_0) g(G_0))}\log n }{\epsilon}) = O(\frac{h2^{O(v(G_0))}\log n }{\epsilon})$.

To show that the embedding is clique-preserving, we use iduction.  The base case where $g(G) = 0$,  $f(\cdot)$ is clique-preserving by \Cref{lm:em-vortexReduce}. By the induction hypothesis, we assume that $f'(\cdot)$ in \cref{line:genusReduction}  is clique-preserving. Let $Q$ be a clique of $G$. If $\mathcal{C}\cap Q = \emptyset$, then $Q$ is a clique in $G'$ and hence $f(\cdot)$ preserves $Q$ as $f'(\cdot)$ preserves $Q$. Otherwise, since $Q\setminus \mathcal{C}$ is a clique in $G'$, there is a bag of $\mathcal{T}'$ containing an image of $Q\setminus \mathcal{C}$; bag $B$ in \cref{line:bagClique} exists.  This implies that $\mathcal{T}$ has a bag $B_Q$ containing an image of $Q$, and thus $f(\cdot)$ is clique-preserving.

We now bound the distortion. In the base case, the diameter of $G$ is at most $ 2^{O(v(G_0)g(G_0))}D$ by equation (\ref{eq:diam-G-everystep}). Thus, the  distortion is at most $\frac{\epsilon}{2^{c(v(G_0)g(G_0))}} 2^{O(v(G_0)g(G_0))}D = \epsilon D$ when $c$ is sufficiently big; recall that in \cref{line:EmbedGenusBaseCase} we apply \Cref{lm:em-vortexReduce} with parameter $\frac{\epsilon}{2^{c(v(G_0)g(G_0))}}$.
For the inductive case, we observe that $\Pt$ is contained in every bag of $\mathcal{T}$ since in \cref{line:EmbedGenusPortalize}, we add $\Pt$ to every bag of $\mathcal{T}'$. Thus, if a shortest path between $u$ and $v$ intersects $\mathcal{C}$, then by rerouting the path through the nearest portal in $\Pt$, we obtain a new path with length at most $w(P(u,v)) + 2\epsilon D/2 = w(P(u,v)) + \epsilon D$. Otherwise,  by induction $d_G(u,v) \leq d_{H'}(u,v) + \eps D =  d_{H}(u,v) + \eps D$.

\subsection{Step (2.3): Removing apices, proof of \Cref{lm:embed-nearly-embeddable}}\label{sec:EmbedApices}
Recall that a nearly $h$-embeddable graph $G$ has an apex set $A$ of size at most $h$ such that $G\setminus A = G_{\Sigma}\cup \{W_1 \ldots\cup W_h\}$.  In this section, we will devise a \emph{stochastic} embedding of $G$ into a small treewidth graph with \emph{expected} additive distortion $\eps D$.

\embedNearlyEmbeddable*

The main tool we use in the proof of \Cref{lm:embed-nearly-embeddable} is \emph{padded decompositions}.
Given a partition $\mathcal{Q}$ of $V(G)$, $\mathcal{Q}(v)$ denotes the cluster containing $v$. 
A partition $\mathcal{Q}$ of $V(G)$ is $\Delta$-bounded if for every cluster $Q\in\mathcal{Q}$, $\dm(C) \Delta$. Similarly, a distribution $\mathcal{D}$ is  $\Delta$ bounded if every partition $\mathcal{Q}\in\supp(\mathcal{D})$ is $\Delta$-bounded.
We denote by $B_G(v,r)=\{u\in V(G)\mid d_G(u,v)\le r\}$ the ball of radius $r$ around $v$.
\begin{theorem}[Theorem 15~\cite{Fil19padded}]\label{thm:padded-minor} Consider a $K_r$-minor-free graph $G$, and parameter $\Delta$. There is an $O(r\Delta)$-bounded distribution over partitions $\mathcal{D}$, such that for every $v\in V(G)$ and $\gamma\in(0,1)$, 
\begin{equation}\label{eq:padded-prob}
\Pr_{\mathcal{Q}\sim\mathcal{D}}[B_G(v,\gamma\Delta)\subseteq \mathcal{Q}(v)]\ge e^{-\gamma}
\end{equation}	
\end{theorem}

Let $G^{-} = G\setminus A$.  Note that the diameter of $G^{-}$ may be unboundedly greater than $D$.
Note that nearly $h$-embeddable graphs exclude $K_{r(h)}$ as a minor where $r(h)$ is some constant depending on $h$ only (indeed, $r(h) = O(h)$). 
Thus we can apply \Cref{thm:padded-minor} with parameter $\Delta=\frac {8D}{\eps}$ to obtain an $O(h\cdot\frac D\eps)$-bounded partition $\mathcal{Q}$ of $G^{-}$.
For each cluster $C\in\mathcal{Q}$, let $N_G(C)$ be the set of vertices with a neighbor in $C$. Let $G^{-}_C \leftarrow G[C \cup N_G(C)]$ be the graph induced by $C \cup N_G(C)$. We apply  \Cref{lm:embed-genus-vortex} to $G^{-}_C$ with accuracy parameter $\Omega(\frac{\eps^2}{h})$. Thus we obtain a one-to-many, clique preserving embedding $f_C$ into $H_C$ which has tree decomposition $\mathcal{T}_C$ of width $O_h(\frac{\log n}{\eps^2})$, and additive distortion $\Omega(\frac{\eps^2}{h})\cdot O(h\cdot\frac D\eps)\le \frac\eps2 D$.
Next we combine all the graphs $\cup_{C\in\mathcal{Q}}H_C$ into a single graph $H$. This is done by adding the set of apices $A$ with edges towards all the other vertices (where the weight of the edge $\{u,v\}$ is $d_G(u,v)$). 
Note that if a vertex $v$ belongs to both $G^{-}_{C}$ and $G^{-}_{C'}$, its copies in $H_{C}$ and $H_{C'}$ will not be merged.
The tree decomposition $\mathcal{T}$ of $H$ can be constructed by adding $A$ to all the bags, and connecting the tree decompositions of all the clusters $\cup_{C\in\mathcal{Q}}\mathcal{T}_C$ arbitrarily.
Finally we define the embedding $f$. For apex vertices it is straightforward, as there is a single copy. For a vertex $v\in V(G)\setminus A$, $f(v)=\cup_{C\mbox{ s.t. }v\in C}f_C(v)$.
See \Cref{alg:emb-nearly-embeddable} for illustration.

\begin{algorithm}[t]
	\caption{\texttt{EmbedNearlyEmbeddableGraphs}}	\label{alg:emb-nearly-embeddable}
	\DontPrintSemicolon
	\SetKwInOut{Input}{input}\SetKwInOut{Output}{output}
	\Input{A nearly $h$-embeddable graph $G=(V,E,w)$}
	\Output{A stochastic embedding $f$ to a graph $H$ with tree decomposition $\mathcal{T}$ of width $O_h(\frac{\log n}{\eps^2})$ and expected additive distortion $\eps D$}
	\BlankLine
	Let $  G^- \leftarrow G\setminus A$ and $\Delta \leftarrow \frac{8D}{\eps}$\;	
	Sample a partition $\mathcal{Q}$ from the distribution promised by  \Cref{thm:padded-minor} with parameter $\Delta$\;
	\ForEach{cluster $C \in \mathcal{Q}$}
	{$G^{-}_C \leftarrow G[C \cup N_G(C)]$ \hspace{87pt} {\color{Darkgreen}\small /** subgraph induced by $C$ and its neighbors in $G$}\label{line:extendNeighbors}\;
		$\{f_C,H_C,\mathcal{T}_C\}\leftarrow \mathtt{GenusReduction}(G^{-}_{S},\Omega(\eps^2))$ \hspace{87pt} {\color{Darkgreen}\small /** \Cref{alg:embed-genus-vortex}}\;
		Add $A$ to all bags of $\mathcal{T}_C$\;
	}
	$\mathcal{T}$ obtained by connecting the tree decompositions $\mathcal{T}_C$ of all the clusters in $\mathcal{Q}$ arbitrarily\;	
	$H\leftarrow$ the graph induced by $\mathcal{T}$\;
	Set $f(v) \leftarrow \cup_{C\in\mathcal{Q}}\mathcal{T}_C$ for  each $v\in V$\;
	\Return  $f, H$ and $\mathcal{T}$\;
\end{algorithm}

	It is straightforward that $\mathcal{T}$ has treewidth $O_h(\frac{\log n}{\eps^2})+|A|=O_h(\frac{\log n}{\eps^2})$. Next, we argue that $f$ preserves cliques. Consider a clique $Q$. If $Q\subseteq A$ then every bag contains $Q$ and we are done. Otherwise, let $C$ be some cluster that contains a vertex $v\in Q\setminus A$. By definition of $N_G(C)$, it holds that $Q\setminus A\subseteq C\cup N_G(C)$. Thus there is a bag in $\mathcal{T}_C$ containing (copies of) all the vertices in $Q\setminus A$. In particular, by construction, there is a bag in $\mathcal{T}$ containing (copies of) all the vertices in $Q$.
	
	It remains to bound the expected distortion of $f$.
	First note that as all the vertices are connected to all the vertices in $A$ by edges of weight at most $D$, $H$ has diameter at most $2D$. 
	Let $u,v$ be two  vertices of $G$ (it could be that $u=v$). Let $P(u,v)$ be (some) shortest path in $G$ between $u$ and $v$.
	We consider two cases:
	
	\textbf{Case 1.}  $P(u,v)\cap A \neq \emptyset$: Let $t\in P(u,v)\cap A$. Then for every $u'\in f(u),v'\in f(v)$, it holds that $d_{H}(u',v')\le d_{H}(u',t)+d_{H}(t,v')=d_{G}(u,t)+d_{G}(t,v)=d_{G}(u,v)$.
	
	\textbf{Case 2.}  $P(u,v)\cap A = \emptyset$: In this case, $P(u,v)\subseteq G^{-}$ and hence $d_{G^{-}}(u,v)=d_{G}(u,v) \leq D$. 
	By equation (\ref{eq:padded-prob}),
	$$\Pr_{\mathcal{Q}\sim\mathcal{D}}[B_{G^{-}}(v,2D)\subseteq \mathcal{Q}(v)]=\Pr_{\mathcal{Q}\sim\mathcal{D}}[B_{G^{-}}(v,\frac\eps4\Delta)\subseteq \mathcal{Q}(v)]\ge e^{-\frac\eps4}\ge1-\frac\eps4~.$$
	Denote the event that $B_{G^{-}}(v,2D)\subseteq \mathcal{Q}(v)$ by $\Phi$.
	If $\Phi$ occurred, then both $u,v$, all their neighbors, and all the vertices in $P(u,v)$ belong to $\mathcal{Q}(v)$. It follows that $u$ and $v$ do not belong to any graph $G^{-}_C$ for $C\neq \mathcal{Q}(v)$, and further that $d_{G^{-}_{\mathcal{Q}(v)}}(u,v)=d_G(u,v)$. By the guarantee of  \Cref{lm:embed-genus-vortex}, it holds that
	$$\max_{u'\in f(u),v'\in f(v)}d_H(u',v')\le \max_{u'\in f_{\mathcal{Q}(v)}(u),v'\in f_{\mathcal{Q}(v)}(v)}d_{H_{\mathcal{Q}(v)}}(u',v')\le d_{G^-_{\mathcal{Q}(v)}}(u,v)+\frac\eps2 D= d_G(u,v)+\frac\eps2 D~.$$
	From the other hand, if $\Phi$ did not occurred, the maximal distance between every two copies of $u$ and $v$ is at most $2D$ (the diameter of $H$). We conclude:
	\begin{align*}
	& \mathbb{E}[\max_{u'\in f(u),v'\in f(v)}d_{H}(u',v')]\\
	& \qquad=\Pr\left[\Phi\right]\cdot\mathbb{E}[\max_{u'\in f(u),v'\in f(v)}d_{H}(u',v')\mid\Phi]+\Pr\left[\bar{\Phi}\right]\cdot\mathbb{E}[\max_{u'\in f(u),v'\in f(v)}d_{H}(u',v')\mid\bar{\Phi}]\\
	& \qquad\le1\cdot\left(d_{G}(u,v)+\frac{\eps}{2}D\right)+\Pr\left[\bar{\Phi}\right]\cdot2D\\
	& \qquad\le d_{G}(u,v)+\frac{\eps}{2}D+\frac{\eps}{4}\cdot2D=d_{G}(u,v)+\eps D
	\end{align*}

\subsection{Step (2.4): General minor-free graphs, proof of \Cref{thm:embedding-minor}}\label{sec:EmbedGeneralMinor}

In this section, we prove the following lemma that together with \Cref{lm:embed-nearly-embeddable} imply \Cref{thm:embedding-minor}. 
Let $h(r)$ be a function of $r$, such that every $K_r$-minor-free graph can be decomposed into a clique-decomposition of nearly $h(r)$-embeddable graphs (\Cref{thm:RS}). 

\begin{lemma}\label{lm:emb-clique-sum} If nearly $h$-embeddable graphs  have a stochastic one-to-many embedding $f$ into  treewidth-$\tw(h,\eps)$ graphs such that:
	\begin{enumerate} [noitemsep]
		\item[(1)] $f$ is clique-preserving
		\item[(2)] the expected additive distortion of $f$ is at most $\eps D$.
	\end{enumerate}
	then there is a stochastic embedding of $K_r$-minor-free graphs into graphs of treewidth at most $\tw(h(r),\eps) + h(r)\cdot\log n$ with expected additive distortion $\eps D$.\\
	Here $\tw(h,\eps)$ is some function depending only on $h$ and $\eps$.
\end{lemma}

\begin{proof}[Proof of \Cref{lm:emb-clique-sum}]	
	Consider a $K_r$-minor-free graph $G$, and let $\mathbb{T}$ be its clique-sum decomposition. That is
	$G = \cup_{(G_i,G_j) \in E(\mathbb{T})}G_i \oplus_h G_j$
	where each $G_i$ is a nearly $h(r)$-embeddable graph.  
	We call the clique involved in the clique-sum of $G_i$ and $G_j$ the \emph{joint set} of the two graphs.
	The embedding of $G$ is defined recursively. Specifically, we will prove by induction that $G$ can be stochastically embedded into $\left(\tw(h(r),\eps) + h(r)\cdot \log |\mathbb{T}|\right)$-treewidth graphs with expected additive distortion $\eps D$.
	
	Note that $\mathbb{T}$ is a tree. Let $\tilde{G}_i\in\mathbb{T}$ be the \emph{central piece} of $\mathbb{T}$ chosen using the following lemma.
	\begin{lemma}[\cite{Jordan69}]\label{lm:tree-sep} Given a tree $T$ of $n$ vertices, there is a vertex $v$ such that every connected component of $T\setminus \{v\}$  has at most $\frac{n}{2}$ vertices. 
	\end{lemma}
	Let $G_1,\dots,G_p$ be the neighbors of $\tilde{G}$ in $\mathbb{T}$. Note that $\mathbb{T}\setminus \tilde{G}$ contains $p$ connected components $\mathbb{T}_1,\dots,\mathbb{T}_p$, where $G_i\in \mathbb{T}_i$, and  $\mathbb{T}_i$ contains at most $|\mathbb{T}|/2$ pieces. 
	We will abuse notation and refer to $\mathbb{T}_i$ also as the graph graph induced by the vertices in all the pieces in $\mathbb{T}_i$. Note that $\mathbb{T}_i$ is $K_r$-minor-free. Further, for every $u,v\in  \mathbb{T}_i$ (or $u,v\in\tilde{G}$) it holds that $d_{\mathbb{T}_i}(u,v)=d_{G}(u,v)$ 
	($d_{\tilde{G}}(u,v)=d_{G}(u,v)$)\footnote{To see this consider a shortest path $P(u,v):u=z_0,\dots,z_q=v$ from $u$ to $v$ in $G$. If $P(u,v)\subseteq \mathbb{T}_i$ then we are done. Else, let $a$ ($b$) be the minimal (maximal) index s.t. $z_a\notin \mathbb{T}_i$ ($z_b\notin \mathbb{T}_i$). Necessarily $z_{a-1},z_{b+1}\in C_i\subseteq\mathbb{T}_i$ where $C_i$ is the joint set between $G_i$ and $\tilde{G}$. Moreover, they are neighbors. Thus $z_0,\dots,z_{a-1},z_{b+1},\dots,z_q$ is a path from $u$ to $v$ in $\mathbb{T}_i$ of length $d_G(u,v)$. Similarly (with some additional steps) one can prove $d_{\tilde{G}}(u,v)=d_{G}(u,v)$ for $u,v\in\tilde{G}$.}.	
	In particular, the diameter of the graphs $\mathbb{T}_1,...,\mathbb{T}_p,\tilde{G}$ is bounded by $D$. 
	
	First we use the assumption on $\tilde{G}$, and sample a clique preserving, one-to-many embedding $\tilde{f}$ of $\tilde{G}$ into graph $\tilde{H}$ with tree decomposition $\tilde{\mathcal{T}}$ of width at most $\tw(h,\eps)$, and expected additive distortion $\eps D$.
	Next, for every $i$, we use the inductive hypothesis on $\mathbb{T}_i$, and sample a clique preserving, one-to-many embedding $f_i$ of $G_i$ into graph $H_i$ with tree decomposition $\mathcal{T}_i$ of width at most $\tw(h,\eps)+h(r)\cdot\log|\mathbb{T}_i|$, and expected additive distortion $\eps D$.

	Next, we create a single one-to-many embedding $f$ of $G$ into a graph $H$ with tree decomposition $\mathcal{T}$.
	We combine the $p+1$ different embeddings as follows. Initially, we just take a disjoint union of all the graphs $\tilde{H},H_1,\dots, H_p$, keeping all copies of the different vertices separately. Next, we will identify some copies, and add some edges.
	For each $i$, let $C_i$ be the joint set of $\tilde{G}$ and $G_i$, i.e., the clique used for their clique sum. As both $\tilde{f}$ and $f_i$ are clique-preserving, there are bags $\tilde{B}_{C_i}\in\tilde{\mathcal{T}}$ and $B_{C_i}\in\mathcal{T}_i$ containing copies of $C_i$. Identify the copies of $C_i$ in $\tilde{B}_{C_i}$ and $B_{C_i}$. Denote this copy by $\bar{C}_i$.
	Add an edge in $H$ between every vertex $v'\in \bar{C}_i$ to every other vertex $u'\in H_i$. Here if $v'$ ($u'$) is a copy of $v\in V(G)$ ($u\in V(G)$) the weight of the edge $\{u',v'\}$ will be $d_G(u,v)$.
	This finished the construction of $H$. The embedding $f$ is defined naturally. For $v\in \mathbb{T}_i\setminus\tilde{G}$ let $f(v)=f_i(v)$. For $v\in\tilde{G}$ let $\mathcal{I}_v=\{i\mid v\in\mathbb{T}_i\}$ be the indices of the joint sets $v$ belongs to (it might be an empty set). Then $f(v)=\tilde{f}(v)\cup\bigcup_{i\in\mathcal{I}_v}f_i(v)$ (note that we identified the copies $\bar{C}_i$ for each $i$ previously). 
	Finally, the tree decomposition $\mathcal{T}$ of $H$ is constructed by first taking $\tilde{\mathcal{T}}$, and for every $i$, adding $\mathcal{T}_i$ to $\tilde{\mathcal{T}}$ via an edge between the bags $\tilde{B}_{C_i}$ and $B_{C_i}$. Further, the vertices $\bar{C}_i$ will be added to all the bags in $\mathcal{T}_i$. 
	
	It is straightforward to verify that $\mathcal{T}$  is a legal tree decomposition for $H$. The width of every bag in the central part $\tilde{\mathcal{T}}$ is at most $\tw(h,\eps)$. While for every bag $B$ from $\mathcal{T}_i$, using the induction hypothesis its width is bounded by 
	\begin{align*}
	\tw(h,\eps)+h(r)\cdot\log|\mathbb{T}_{i}|+|\bar{C}_i| & \le\tw(h,\eps)+h(r)\cdot\left(\log|\mathbb{T}|-1\right)+h(r)\\
	& =\tw(h,\eps)+h(r)\cdot\log|\mathbb{T}|~.
	\end{align*}
	
	It is straightforward that the one-to-many embedding $f$ is dominating (as $\tilde{f},f_1,\dots,f_p$ were dominating, and the newly add edges dominate the original distances). It is left to prove that $f$ has expected additive distortion $\eps D$.
	Consider a pair of vertices $u,v\in G$. We proceed by case analysis:
	\begin{itemize}
		\item If there is an index $i$ such that $u,v\in\mathbb{T}_i\setminus\tilde{G}$. 
		Then by the induction hypothesis, 
		\[
		\mathbb{E}[\max_{u'\in f(u),v'\in f(v)}d_{H}(u',v')]\le\mathbb{E}[\max_{u'\in f_{i}(u),v'\in f_{i}(v)}d_{H_{i}}(u',v')]\le d_{\mathbb{T}_{i}}(u,v)+\eps D=d_{G}(u,v)+\eps D
		\]
		\item If both $u,v$ belong to $\tilde{G}$.
		Consider some general vertex $z\in\tilde{G}$. If $z\in\mathbb{T}_i$, then there is some copy $\bar{z}_i\in\bar{C}_i$ such that we added an edge of weight $0$ from  $\bar{z}_i$ to any copy in $f_i(z)$.
		In other words, for every copy $z'\in f(z)$ there is a copy $\bar{z}\in\tilde{f}(z)$ at distance $0$.

		Considering $u,v\in\tilde{G}$, by the assumption on $\tilde{f}$ it holds that, 
		\[
		\mathbb{E}[\max_{u'\in f(u),v'\in f(v)}d_{H}(u',v')]\le\mathbb{E}[\max_{\bar{u}'\in\tilde{f}(u),\bar{v}'\in\tilde{f}(v)}d_{\tilde{H}}(\bar{u}',\bar{v}')]\le d_{\tilde{G}}(u,v)+\eps D=d_{G}(u,v)+\eps D~.
		\]
		\item If there is an index $i$ such that $u\in\mathbb{T}_{i}\setminus\tilde{G}$ and $v\in \tilde{G}$ (the case $v\in\mathbb{T}_{i}\setminus\tilde{G}$ and $u\in \tilde{G}$ is symmetric). Then necessarily, there is a vertex $z\in C_i$ laying on a shortest path from $v$ to $u$ in $G$.
		By construction, we added an edge between the copy $\bar{z}\in\bar{C}_i$ to any copy $u'\in f(u)$. It follows that 
		\begin{align*}
		\mathbb{E}[\max_{u'\in f(u),v'\in f(v)}d_{H}(u',v')] & \le\mathbb{E}[\max_{u'\in f(u),v'\in f(v)}d_{H}(u',\bar{z})+d_{G}(\bar{z},v)]\\
		& \le d_{G}(u,z)+\mathbb{E}[\max_{z'\in f(z),u'\in f(u)}d_{H}(z',v')]\\
		& \le d_{G}(u,z)+d_{G}(z,v)+\eps D=d_{G}(u,v)+\eps D~,
		\end{align*}
		where the last inequality follows by the previous case.

		\item If there are indices $i_u\ne i_v$ such that $u\in\mathbb{T}_{i_u}\setminus\tilde{G}$ and $v\in\mathbb{T}_{i_v}\setminus\tilde{G}$. Then necessarily, there are vertices $z_u\in C_{i_u}$ and $z_v\in C_{i_v}$ laying on a shortest path from $v$ to $u$ in $G$.		
		By construction, we added an edge between $\bar{z}_u\in \bar{C}_{i_u}$ ($\bar{z}_v\in \bar{C}_{i_v}$) to every copy in $f(u)$ ($f(v)$). It follows that
		\begin{align*}
		\mathbb{E}[\max_{u'\in f(u),v'\in f(v)}d_{H}(u',v')] & \le\mathbb{E}[\max_{u'\in f(u),v'\in f(v)}d_{H}(u',\bar{z}_{u})+d_{H}(\bar{z}_{u},\bar{z}_{v})+d_{H}(\bar{z}_{v},v')]\\
		& \le d_{G}(u,z_{u})+\mathbb{E}[\max_{z_{u}'\in f(z_{u}),z_{v}'\in f(z_{u})}d_{H}(z_{u}',z_{v}')]+d_{G}(z_{v},v)\\
		& \le d_{G}(u,z_{u})+d_{G}(z_{u},z_{v})+\eps D+d_{G}(z_{v},v)=d_{G}(u,u)+\eps D~,
		\end{align*}
		where the second inequality followed by the second case.
	\end{itemize}	
\end{proof}

\paragraph{Remark} The stochastic embedding $f$ we constructed is one-to-many, but we can keep it one-to-one by simply retaining (any) one vertex in the image of $u$.

\subsection{Corollaries} \label{sec:cor}

In the case where there are no vortices, we can use our constructions, combined with the $\poly(\frac1\eps)$-treewidth embedding of \cite{FKS19}
to generalize their bound to graphs of bounded genus, and to apex graphs (for the latter the embedding is stochastic).

 \begin{theorem}[Theorem 1.3~\cite{FKS19}]\label{thm:embeddinng-planar}
 	Given a  planar graph $G$ of diameter $D$ and a parameter $\eps < 1$, there exists a deterministic embedding $f$ from $G$ to a graph $H$ of treewidth $O(\poly(\frac{1}{\epsilon}))$ such that for every $x, y \in G$:
 	\begin{equation}
 	d_G(x,y) \leq d_H(x,y) \leq d_G(x,y)  +  \eps D
 	\end{equation}
 \end{theorem}

By using the same cutting approach (using \Cref{lm:cutting-formal})) in \Cref{alg:embed-genus-vortex}, we can reduce the problem to the case where the graph $G$ in \cref{line:EmbedGenusBaseCase} \Cref{alg:embed-genus-vortex} is planar. Thus, we can invoke the embedding algorithm in \Cref{thm:embeddinng-planar} to embed $G$ with additive distortion $\frac{\eps D}{2^{cg(G)}}$ and treewidth $O_{g}(\poly(\frac1\eps))$.

\begin{corollary}\label{cor:embedding-genus}
	Given a graph $G$  with diameter $D$ embedded on a surface with genus $g$ and a parameter $\eps < 1$,
	there exists an embedding $f$ from $G$ to a graph $H$ of treewidth at most $O_{g}(\poly(\frac{1}{\epsilon}))$ such that for every $x, y \in G$:
	\begin{equation}
	d_G(x,y) \leq d_H(f(x),f(y))  \leq d_G(x,y)  + \eps D
	\end{equation}
\end{corollary}

For a bounded genus graphs with a constant number of apices, we can design a  stochastic embedding with constant treewidth using padded decomposition~\cite{Fil19padded} as in \Cref{alg:emb-nearly-embeddable}. In this case, since we do not need to preserve triangles and vortices, we do not need to extend $G^-_C$ to include neighbors $N_G(C)$ as in \cref{line:extendNeighbors} of \Cref{alg:emb-nearly-embeddable}.

\begin{corollary}\label{cor:embedding-apex}
	Let $G$ be a graph of diameter $D$ that has a set of vertices $A$ called \emph{apices} such that $G\setminus A$ can be (cellularly) embedded on a surface of genus $g$. Given any parameter $\eps < 1$, there exists a stochastic embedding of $G$ into treewidth $O_g(\poly(\frac1\eps) )+ |A|$ graphs with expected additive distortion $\eps D$.
\end{corollary}

\section{Algorithmic Applications}\label{apendix:AlgApplications}
\newcommand{\ins}{\text{in}}
\newcommand{\cost}{\text{cost}}
\newcommand{\out}{\text{out}}
\newcommand{\Z}{\mathbb{Z}}
\newcommand{\set}[1]{\{#1\}}
\newcommand{\distance}{\text{distance}}

\subsection{EPTAS for subset TSP}\label{appendix:subsetTSP}

 Given a subset spanner with lightness $\Psi(\epsilon)$, we can design an efficient PTAS for Subset TSP in time $2^{O(\Psi(\epsilon)/\epsilon)}n^{O(1)}$ using the contraction decomposition framework by Demaine \etal \cite{DHK11} (originally introduced by Klein \cite{Klein05} for the planar case). The framework has four steps:
 
 \begin{itemize}
 	\item \textbf{Step 1} Find a subset spanner $H$ with lightness $\Psi(\epsilon)$.
 	\item  \textbf{Step 2} Partition the edge sets of $H$ into $s = \Theta_h(\Psi(\epsilon)/\eps)$ sets $E_1,E_2,\ldots E_s$ such that for any set $E_i$, the graph $H_i = \nicefrac{H}{E_i}$ obtained by contracting all edges in $E_i$ has treewidth at most $O(s)$.
 	\item \textbf{Step 3} For each $i$, solve the Subset TSP problem in treewidth-$O(s)$ graphs in time $2^{O(s)}n^{O(1)}$ (see Appendix D in the full version of \cite{Le20}).
 	\item \textbf{Step 4} Lift the solution found in Step 3 for each $i \in [1,s]$ to a solution $G$ and retun the minimum solution over all $i$ (see Section 3 in \cite{Le20} for details of the lifting procedure). 
 \end{itemize}
 
The overall running time is hence $2^{O(s)}n^{O(1)} = 2^{O(\Psi(\epsilon)/\epsilon)}n^{O(1)}$. This in combination with \Cref{thm:tsp-spanner} implies \Cref{thm:tsp-eptas}.

\subsection{Approximation schemes for vehicle routing}\label{appendix:vhr}

We consider a uniform-capacity vehicle-routing problem.
An instance consists of:
\begin{itemize}
\item a graph $G=(V,E)$ and cost function $\cost: E \longrightarrow \R_+$,
\item a \emph{capacity} $Q \in \Z_+$,
\item a function $d:\ V \longrightarrow \Z_+$, called the
  \emph{delivery requirement function},
 and
\item a vertex $r$, called the \emph{depot}.
\end{itemize}
To describe a solution, we introduce some terminology that will also
be helpful later, in describing the algorithm.

For each pair $u,v$ of vertices of $G$, fix a min-cost $u$-to-$v$ path.
We define a \emph{route} to be a sequence that alternates between
nonnegative integers and vertices of $G$.  The \emph{start} of the route is the
first vertex, and the \emph{end} of the route is the last.  For a
route $R$ consisting of vertices $v_0, \ldots, v_k$, and an edge
$e$, the multiplicity $m(e, R)$ of  $e$ in $R$ is
$|\set{i\ :\ e \text{ is in the min-cost $v_{i-1}$-to-$v_i$
    path}}|$ The \emph{cost} of $R$ is
$\sum_{e\in E} m(e,R) \cost(e)$, or equivalently
$\sum_{i=1}^k \distance_G(v_{i-1}, v_i)$.  The integers are used to
indicate deliveries: the integer appearing immediately after a vertex
$v$ indicates the number of deliveries made at that vertex.  An
integer before the first vertex of a route is naturally interpreted as
representing deliveries that a vehicle makes \emph{before} following
that route but, for technical reasons to be apparent later, this is
not enforced.  We refer to the integer in a route before the first
vertex as the route's \emph{pre-delivery number}.  We refer to the sum
of the other integers as the route's \emph{internal delivery number}.
We refer to the sum of pre-delivery and internal delivery numbers as
the route's \emph{total delivery number}.  A route is \emph{feasible}
if the total delivery number is at most $Q$, and if either the
internal delivery number is positive or the start or end is the depot.

We say a feasible route is a \emph{tour} if it starts and ends at the
depot.
A \emph{solution} is a multiset
of tours, each starting and ending at the depot, such that
for each vertex $v$ the total number of deliveries made at $v$ is $d(v)$.
The objective is to find a solution of minimum total cost.

We prove the following theorem in
Section~\ref{subsec:fptas-vehiclerouting}.
\begin{theorem}
  \label{thm:vehiclerouting}
  Let $\eps>0$. There exists an algorithm that, for any instance of
  the vehicle routing problem $(G,\cost(\cdot),Q,d(\cdot),r)$ outputs a
  $(1+O(\eps))$-approximate solution in time
  $(Q\eps^{-1}\log n)^{O(wQ/\eps)} n^3$ where $n=|V(G)|$ and $w$ is
  the treewidth of $G$.
\end{theorem}

Note that, for bounded width $w$ and bounded capacity $Q$, this is an
efficient PTAS.\footnote{Note that in Theorem~\ref{thm:vehiclerouting}
  exponential dependencies on $w$ and $Q$ are needed to get a
  polynomial time approximation scheme since the problem with
  arbitrary capacity $Q$ is APX-hard on trees~\cite{Becker18} and the problem with
  bounded capacity is APX-hard graphs on graphs of arbitrary
  treewidth.} 
 Previously no efficient PTAS was known
for bounded width and bounded capacity.

From there, an immediate application of the framework of Becker,
Klein, Schild~\cite{BKS19} yields a serie of corollaries.
The framework of Becker, Klein, Schild implies that for any
graph $G$ in some graph class $\mathcal{G}$,
if there exists a polynomial time algorithm that computes
a stochastic embedding 
into a dominating graph $H$ with treewidth $w$ such that:
\begin{equation*}
  E_{f\sim\mathcal{D}}[d_H(f(u),f(v))] \leq d_G(u,v) +
  \epsilon (d_G(s,u) +  d_G(s,v)),
\end{equation*} 
then there exists an approximation scheme for vehicle routing
with bounded capacity for $\mathcal{G}$ with running time
$n^{O(1)} + T(w, \epsilon, n)$, where $T(w, \epsilon, n)$ is the
best running time for an approximation scheme for $n$-vertex graphs of
treewidth at most $w$.

As a consequence, we obtain the following corollary:
\begin{corollary} There is an efficient PTAS for bounded-capacity
  vehicle routing in planar metrics.
\end{corollary}

Moreover, again as a consequence of the results of Becker, Klein, and
Saulpic~\cite{BKS18}, we have:
\begin{corollary} There is an efficient PTAS for bounded-capacity
  vehicle routing in metrics of bounded highway dimension.
\end{corollary}

As corollaries of Lemma~\ref{lm:root-embedding} and
Lemma~\ref{cor:root-embedding-gunus} below, we obtain the two following
results.

\begin{corollary} There is a QPTAS for bounded-capacity
e  vehicle routing in minor-free metrics. 
\end{corollary}

\begin{corollary} There is an efficient PTAS for bounded-capacity
  vehicle routing in bounded genus metrics.
\end{corollary}

\subsubsection{Reduction to the bounded treewidth case}
We first prove the following lemma. 
\begin{lemma}\label{lm:root-embedding}Given an $n$-vertex edge-weighted $K_r$-minor-free graph $G$ and a distinguished vertex $s$, in polynomial time, one can embed $G$ stochastically into a dominating graph $H$ with  treewidth $O_r(2^{\poly(\frac{1}{\epsilon})}\log n)$ such that:
	\begin{equation}
	E_{f\sim\mathcal{D}}[d_H(f(u),f(v))] \leq d_G(u,v) + \epsilon (d_G(s,u) +  d_G(s,v))~.
	\end{equation} 
\end{lemma}
\begin{proof}
	The proof closedly follows the proof of Theorem 2 in Becker et al.~\cite{BKS19}. Assume that the minimum edge weight is $1$.  Choose a random $x\in [0,1]$. We partition $V(G)$ into bands $\mathcal{B} = \{B_0,B_1,\ldots B_{m}\}$ such that:
	\begin{itemize}[noitemsep]
		\item $B_0 = \{u | d_G(s,u) \leq  \left(\frac{1}{\epsilon}\right)^{\frac{x}{\epsilon}}\}$.
		\item $B_i = \{ u |   \left(\frac{1}{\epsilon}\right)^{\frac{i-x}{\epsilon}} \leq d_G(s,u) \leq  \left(\frac{1}{\epsilon}\right)^{\frac{i + x}{\epsilon}}\}$ for $1\leq i\leq m$.
		\item  $\cup_{i=0}^m B_i = V(G)$.
	\end{itemize}

Let $\mathcal{B}(u)$ be the band containing vertex $u$. The key property of this random partition is that (Lemma 2~\cite{BKS19}):
\begin{equation}\label{eq:diffBandProb}
\Pr[\mathcal{B}(u) \not= \mathcal{B}(v)] \leq \epsilon \qquad \mbox{for $u,v$ s.t }\epsilon d_G(s,v)  \leq d_G(s,u) \leq d_G(s,v) 
\end{equation}					
	
Let $G_i$ be the graph that contains all pairwise shortest paths between vertices in $\{r\} \cup B_i$. Let  $L_0 = 1$ and $L_i = \left(\frac{1}{\epsilon}\right)^{\frac{i-x}{\epsilon}}$ for $i \in [1,m]$ and $U_i = \left(\frac{1}{\epsilon}\right)^{\frac{i+x}{\epsilon}} $ for $i \in [0,m]$. Let $\delta_i = U_i/L_i$. Observe that for every $i \in [0,m]$

\begin{equation}
\delta_i  \leq  \left(\frac{1}{\epsilon}\right)^{\frac{2x}{\epsilon}} \leq  \left(\frac{1}{\epsilon}\right)^{\frac{2}{\epsilon}}
\end{equation}

and that each graph $G_i$ has $\dm(G_i) \leq 2U_i$. Since $G_i$ is a $K_r$-minor-free graph, we apply \Cref{thm:embedding-minor} to stochastically embed $G_i$ into $H_i$ with additive distortion $\frac{\epsilon}{\delta_i}U_i$ and of treewidth:

\begin{equation}\label{eq:tw-Hi}
\tw(H_i) = O_r(\frac{2\delta_i\log n}{\epsilon}) = O_r(2^{\poly(\frac{1}{\epsilon})} \log n )
\end{equation}

We can construct the graph $H$ and the corresponding embedding by adding an edge of weigth $d_G(s,v)$ from $s$ to every vertex $v$ of $\cup_{i=1}^m H_i$ that has (one) preimage in $G$. Clearly $H$ has treewidth $\max_{i=1}^m \tw(H_i) +1 = O_r(2^{\poly(\frac{1}{\epsilon})} \log n )$.

It remains to bound the distortion betweeu $u,v \in V(G)$. We assume w.l.o.g that $d_G(s,u) \leq d_G(s,v)$. 
\begin{itemize}[noitemsep]
	\item \textbf{Case 1} $ d_G(s,u) < \eps d_G(s,v)$, then 
		\begin{equation*}
		\begin{split}
				\mathbb{E}[d_H(u,v)] &\leq 		\mathbb{E}[d_H(s,u) + 	d_H(s,v)] = 	d_G(s,u) + 	d_G(s,v) \\
			&\leq 	d_G(s,u) + 	d_G(s,u)  + d_G(u,v) \leq d_G(u,v) + 2\epsilon d_G(s,v) 
		\end{split}
		\end{equation*}
	\item \textbf{Case 2} $ \epsilon d_G(s,v) \leq d_G(s,v)$. Let $\Phi$ be the event that $u$ and $v$ are in the same band $B_i$ for some $i \in [0,m]$. By \Cref{eq:diffBandProb}, $\Pr[\bar{\Phi}] \leq \epsilon$.	We have:
	\begin{equation*}
	 	\mathbb{E}[d_{H}(u,v)|\Phi] = 	\mathbb{E}[d_{H_i}(u,v)] \leq d_{G_i}(u,v) + \frac{\epsilon}{\delta_i}U_i = d_{G}(u,v) + \epsilon L_i \leq d_{G}(u,v) + \epsilon d_{G}(s,u)
	\end{equation*} 
  and:
  \begin{equation*}
  \mathbb{E}[d_{H}(u,v)|\bar{\Phi}] \leq  	\mathbb{E}[d_{H}(s,u) + d_{H}(s,v)] = d_{G}(s,u) + d_{G}(s,v)
  \end{equation*} 
  That implies:
  \begin{equation*}
  \begin{split}
 \mathbb{E}[d_{H}(u,v) &= \Pr[\Phi] \mathbb{E}[d_{H}(u,v)|\Phi] +  \Pr[\bar{\Phi}]\mathbb{E}[d_{H}(u,v)|\bar{\Phi}] \\
 &\leq \mathbb{E}[d_{H}(u,v)|\Phi] + \epsilon \mathbb{E}[d_{H}(u,v)|\bar{\Phi}] \\
 & \leq d_{G}(u,v) + \epsilon d_{G}(s,u) + \epsilon (d_{G}(s,u) + d_{G}(s,v))\\
 &\leq d_{G}(u,v) + 2\epsilon (d_{G}(s,u) + d_{G}(s,v))
\end{split} 
 \end{equation*}  
\end{itemize}
In both cases, $ \mathbb{E}[d_{H}(u,v)  \leq q d_{G}(u,v) + 2\epsilon (d_{G}(s,u) + d_{G}(s,v))$. By scaling $\epsilon \leftarrow \epsilon/2$ we obtain the desired distortion with the same treewidth bound.
\end{proof}

By applying the same argument to bounded genus graphs and using \Cref{thm:embedding-genus} to embed $G_i$, we obtain the following corollary:

\begin{corollary}\label{cor:root-embedding-gunus} Given an $n$-vertex edge-weighted graph $G$ of genus $g$ and a distinguished vertex $s$, in polynomial time, one can embed $G$ stochastically into a dominating graph $H$ with  treewidth $O_g(2^{\poly(\frac{1}{\epsilon})})$ such that:
	\begin{equation}
	E_{f\sim\mathcal{D}}[d_H(f(u),f(v))] \leq d_G(u,v) + \epsilon (d_G(s,u) +  d_G(s,v))~.
	\end{equation} 
\end{corollary}

\begin{proof}[Proof of \Cref{thm:vhr-qptas}] 
	To obtain an approximation scheme for vehicle routing problem with
	bounded capacity, namely where $Q$ is considered a fixed constant,
	using~\Cref{lm:root-embedding}, 	the idea is to embed an instance of the vehicle routing problem to
	a graph $H$ with parameter $\hat{\epsilon} = \epsilon/Q$  and hence
	treewidth $\tw = O_r(\log n \cdot 2^{\poly(Q/\epsilon)})$ by
	\Cref{lm:root-embedding}.
	We then solve the instance of the vehicle routing problem on $H$ and
	"lift" the solution to the solution of the original graph: for each tour $T_H$ in $H$ starting from $\Phi(s)$ covering an order sequence of points $f(v_1) ,f(v_2), \ldots, f(v_q)$ where $v_i$ is a vertex of $G$, $i \in [1,q]$, and $q\leq Q$,  we convert it into a tour $T_H$ starting from $s$ covering points $v_1,\ldots, v_q$ in this order. The expected total cost of the lifted solution is  at most $(1+\epsilon)$ times the optimal cost by Lemma 5 in~\cite{BKS19}. The running time, by \Cref{thm:vehiclerouting-tw-dp}, is:
	\begin{equation}
	 \left( Q\eps^{-1} \log n\right)^{O_r(\log n) 2^{\poly(\frac{Q}{\epsilon})}Q/\epsilon}n^{O(1)} = n^{O_{\eps,r,q}(\log \log n)}.
	\end{equation}
\end{proof}

\begin{proof}[Proof of \Cref{thm:vhr-eptas}] The algorithm for genus-$g$ graphs is exactly the same. However, the treewidth of $H$ in this case is $O_g(2^{\poly(\frac{Q}{\epsilon})})$. Thus, the running time, by \Cref{thm:vehiclerouting-tw-dp}, is:
	\begin{equation}
	\left( Q\eps^{-1} \log n\right)^{O_g(2^{\poly(\frac{Q}{\epsilon})}Q/\epsilon)}n^{O(1)} = 2^{O_g(\poly(\frac{Q}{\epsilon}))} n^{O(1)}
	\end{equation} 

\noindent In the above equation, we use the following inquality:

\begin{equation*}
     (\log n)^d  n^{c} \leq n^{c+1} 2^{d^2} \qquad \mbox{when $n$ is sufficiently big}
\end{equation*}

\noindent which can be proved as follows. If $d \geq \log \log n$, then:

\begin{equation*}
d\log \log n + c\log n \leq d^2 + c\log n \leq (c+1)\log n + d^2
\end{equation*}
Otherwise, 
\begin{equation*}
d\log \log n + c\log n \leq (\log \log n)^2+ c\log n \leq (c+1)\log n  \leq (c+1)\log n + d^2
\end{equation*}
when $n$ is sufficiently big. 
\end{proof}

\subsubsection{An FPT-approximation scheme for vehicle routing with bounded capacity in bounded treewidth graphs}
\label{subsec:fptas-vehiclerouting}

\paragraph{Preliminaries}

For simplicity, we work with branch decompositions instead of tree
decompositions.  A branch decomposition of a graph $G$ is a maximal
collection $\mathcal C = \set{C_1, C_2, \ldots, C_k}$  of nonempty
mutually noncrossing sets of edges.  Equivalently, there is a binary
tree each node of which is a set $C$ of edges, where each leaf
is a singleton set and each nonleaf $C$ has two children, $C'$ and $C''$,
such that $C'\cap C''=\emptyset$ and $C=C'\cup C''$.  We refer to
the sets $C_i$ forming the branch decomposition as
\emph{branch clusters}\footnote{This is not (yet) standard
  terminology.} or simply \emph{clusters}.
Given a cluster $C$, we denote by $\partial_G(C)$ the set of vertices
$v$ such that $v$ is incident to some edge in $C$ and some edge not in
$C$.  We refer to the vertices of $\partial_G(C)$ as \emph{portals} of $C$.\footnote{This is not (yet) standard terminology.}
We refer to other vertices incident to edges of $C$ as \emph{internal} vertices.
The \emph{width} of $\set{C_1, C_2, \ldots, C_k}$ is
$\max_i |\partial_G(C_i)|$.  The branchwidth of a graph is the minimum
width of a branch decomposition.
It is known that a graph of branchwidth
$w$ has treewidth $O(w)$ and vice versa.

Now we outline concepts underlying the algorithm.
The input consists of the graph $G$, cost
function $\cost(\cdot)$, capacity $Q$, delivery requirement function $d(\cdot)$,
and depot $r$.  We assume that $G$ comes equipped with a branch
decomposition.\footnote{An optimal branch decomposition can be found
  in time exponential in the width.}  We assume moreover that the
depth of the branch decomposition tree is at most $c\log n$ where
$c$ is a constant and $n=|V(G)|$.\footnote{One can transform a branch decomposition to one of depth $O(\log n)$
while at most doubling the width.  See~\cite{EKM12}.}
We assume that the depot is \emph{not} a portal of any
cluster.\footnote{This can be achieved by introducing a zero-cost edge
  $uv$ where $v$ is the original depot and $u$ is a new vertex and is
  the new depot.}
Specify for each vertex a leaf cluster that contains an edge incident
to that vertex.  The vertex is said to be \emph{inside} this cluster
and inside every ancestor of this cluster in the branch
decomposition.

We say a route $R$ is \emph{good} with respect to a
cluster $C$ if every vertex of $R$ is incident to $C$ or is the depot,
and if the start and end of $R$ are portals of $C$ or the depot, and
if every vertex $v$ that appears in $R$ followed immediately by a
positive integer is a vertex that is inside $C$.

We now define \emph{concatenation} of routes, a binary operation denoted by $\circ$.
If
$$R_1=j\ u\ s_1\ v\ k$$
 (where $j$ and $k$ are integers, $u$ and $v$ are vertices, and $s_1$ is a sequence)
is a route
with start $u$ and end $v$ and pre-delivery number $j$ and total delivery number $q$, and
$$R_2 = q\ v\ \ell\ s_2$$
(where $q$ and $\ell$ are integers and $v$ is a vertex and $s_2$ is a sequence)
is a route with start $v$ and pre-delivery number $q$ then
$$R_1 \circ R_2 = j \ u\ s_1\ v\ (k+\ell)\ s_2$$
is a route with pre-delivery number $j$.

A partial solution for $C$ is supposed to capture how a solution might
intersect $C$.  A \emph{partial solution} for $C$ is
a multiset $\mathcal R$ of feasible routes, each good with
respect to $C$, such that, for each vertex $v$ inside $C$, the number
of deliveries to $v$ is $d(v)$.

Because we seek an \emph{efficient} dynamic program, we want
to consider only partial solutions that are constrained in a way that 
allows for more efficient computation.
We will prove that there is a near-optimal solution that satisfies
these constraints.

Let $\mathcal R$ be a partial solution for $C$.  For a portal $p$ of
$C$ and integer $q\in \set{0,1,\ldots, Q}$, we define the
\emph{outflow of $\mathcal R$ at $p$ with delivery $q$} to be the
number of routes $R$ in $\mathcal R$ that end at $p$ and that have
total delivery number $q$.  Similarly, the \emph{inflow of
  $\mathcal R$ at $p$ with delivery $q$} is the number of routes
in $\mathcal R$ that start at $p$ and that have pre-delivery number $q$.

A \emph{configuration} $\kappa$ for $C$ specifies, for each portal $p$ of $C$
and each $q\in \set{0,1,\ldots, Q}$, an integer $f^{\ins}_{p,q}(\kappa)$ and
an integer $f^{\out}_{p,q}(\kappa)$.   A partial solution for $C$ \emph{induces}
a configuration $\kappa$ for $C$ if, for each portal $p$ and
$q\in\set{0,1,\ldots, Q}$, $f^{\ins}_{p,q}(\kappa)$ is the inflow at $p$
with delivery $q$ and $f^{\out}_{p,q}(\kappa)$ is the outflow.
The configuration is \emph{constrained} if each integer is the ceiling of a power of
$1+\frac{\epsilon}{cQ\log n}$.

\paragraph{Algorithm}

The algorithm first computes all-pairs shortest paths for the input
graph $G$ with respect to the given cost function.  The algorithm then
executes a dynamic program that processes the clusters in some
leaf-to-root order (e.g. in nondecreasing order of the size of cluster).
For each cluster $C$, it computes a table $T_C[\cdot]$ that maps each
configuration $\kappa$ in a subset of the constrained configurations
for $C$ to a partial solution for $C$ that induces $\kappa$.
Once the algorithm has constructed the table $T_{\hat C}$ for the root
cluster $\hat C$, it outputs the sole entry in that table.  There is
only one configuration $\hat \kappa$ for $\hat C$ because $\hat C$ has no portals.
Furthermore, because $T_{\hat C}[\hat \kappa]$ is good with respect to
$\hat C$, which has no portals, it follows that every route in
$T_{\hat C}[\hat \kappa]$ is a tour and that these tours jointly
handle the delivery requirements.

Now we describe how the algorithm populates the table $T_C[\cdot]$ for $C$.
For each constrained configuration $\kappa$ of $C$, the algorithm
proceeds as follows.  If $C$ is a leaf cluster, it suffices to
consider a constant number of partial solutions that are good with
respect to $C$ and that induce $\kappa$.  The algorithm enumerates
these, find the partial solution $\mathcal R$ of minimum cost, and
assigns it to $T_C[\kappa]$.

Suppose $C$ has children $C_1$ and $C_2$.  In this case, the algorithm
iterates over 
pairs $(\kappa_1,\kappa_2)$ of configurations for which
$T_{C_1}[\kappa_1]$ and $T_{C_2}[\kappa_2]$ are defined.  For each pair, the
algorithm constructs an instance of a min-cost flow problem with no capacities.
The algorithm finds the minimum solution and
uses it (as described below) to construct a partial solution $\mathcal
R$ for $C$.  Finally, the algorithm assigns to $T_C[\kappa]$ the
cheapest partial solution for $C$ that it found.

Now we describe the instance of the transportation problem.  For
convenience, we let $C_0$ denote $C$ and we let $\kappa_0$ denote $\kappa$.
The flow network consists of a subnetwork for  each $q\in
\set{0,1,\ldots, Q}$, as follows:
\begin{itemize}
\item For  each $p\in \partial(C_1)\cup
\partial(C_2)$, for, the network has a node $v_{p,q}$.
For $i=0,1,2$, define $g(i,p,q)$ to be $f^\ins_{p,q}(\kappa_i) - f^\out_{p,q}(\kappa_i)$
if $p \in \partial(C_i)$ and zero if $p\not \in
\partial(C_i)$.  The \emph{supply} of $v_{p,q}$ is $g(0,p,q)+g(1,p,q)+g(2,p,q)$.

\item The network also has a node $s_q$ and a node $t_q$, both representing
the depot.  The supply of $s_q$ is required only to be nonnegative,
and the supply of $t_q$ is required only to be nonpositive.

\item For each pair $p,\bar p \in \partial(C_1) \cup \partial(C_2)$,
  there is an arc $v_{p,q} \longrightarrow v_{\bar p,q}$ whose cost is
  the $p$-to-$\bar p$ distance.
\item For each $p\in \partial(C_1) \cup \partial(C_2)$, there is an
  arc $s_q \longrightarrow v_{p,q}$ and an arc
  $v_{p,q}\longrightarrow t_q$, each of whose cost is the $p$-to-depot
  distance.
\end{itemize}
A multiset $\mathcal S$ of arcs is a feasible solution if the
following holds for each node $v$:
\begin{itemize}
  \item If the supply of $v$ is positive then the number of arcs in $\mathcal
S$ with tail $v$ is equal to the supply.
\item If the supply of $v$ is negative (in which case we treat it as a
  demand) then the number of arcs in $\mathcal S$ with head $v$ is
  equal to the negative of the supply.
\end{itemize}
The 
the algorithm finds the minimum-cost feasible solution $\mathcal S$,
and calls {\sc CombineSolutions}$_C(C, \mathcal S, T_{C_1}(\kappa_1),
T_{C_2}[\kappa_2])$ to combine recursively computed solutions to
form a partial solution for $C$.
      
\begin{tabbing}
{\sc CombineSolutions}$_C(\mathcal S, \mathcal R_1, \mathcal R_2)$: \\ 
\quad \= initialize $\mathcal R$ to consist of $\mathcal R_{C_1}\cup \mathcal R_{C_2}$\\
 \> for each copy in $\mathcal S$ of an arc $s_q \longrightarrow
v_{p,q}$, add to $\mathcal R$ a route $q\ r\ 0\ p\ 0$ \\
\> for each copy in $\mathcal S$ of an arc $v_{p,q} \longrightarrow
t_q$, add to $\mathcal R$ a route $q\ p\ 0\ r\ 0$\\
\> while possible\\
\> \quad \= find two routes $R_1, R_2\in \mathcal R$ such that $R_1
\circ R_2$ is defined,\\
\>  \> and replace them in $\mathcal R$ with $R_1 \circ R_2$.
\end{tabbing}

\begin{lemma} The cost of the solution obtained by {\sc
    CombineSolutions}$_C(\mathcal S, \mathcal R_1, \mathcal R_2)$ is
  the sum of costs of $\mathcal S, \mathcal R_1, \mathcal R_2$.
\end{lemma}

Because the solution to the min-cost flow problem achieves the given
supplies, and because the partial solutions $T_{C_1}(\kappa_1)$
and $T_{C_2}(\kappa_2)$ induce the configurations $\kappa_1$ and
$\kappa_2$ respectively, one can prove the following:

\begin{lemma}\label{lm:dp-feasible}
  The multiset found by {\sc CombineSolutions} is (1) a partial
  solution (2) that is good with respect to $C$ and (3) that induces
  $\kappa$.
\end{lemma}

\paragraph{Overview of analysis}

Now we show that the
algorithm returns a solution of cost at most $1+\epsilon$ times
optimal. First we outline the analysis.  Say a route is
\emph{degenerate} if it has the form $q\ r\ 0\ v\ 0$ or
$q\ v\ 0\ r\ 0$.  Let $\mathcal T^*$ be the optimal solution.  We
derive, for each cluster $C$, a partial solution
$\tilde {\mathcal R}_C$ that induces a constrained configuration
$\kappa_C$, and a set $\mathcal X_C\subset \tilde {\mathcal R}_C$ of
degenerate routes with the following property:

\textbf{Property 1:} The cost of
$\tilde {\mathcal R}_C$ equals $\sum_{e \in C} m(e, \mathcal R^*)
\cost(e)$ plus the cost of
\begin{equation} \label{eq:excess}
\bigcup \set{\mathcal X_{\bar C}\ :\ \bar C \text{ a descendant of
  }C}
\end{equation}
\medskip

In particular, for $C = \hat C$, the root cluster, the total cost of 
$\tilde {\mathcal R}_{\hat C}$ exceeds that of $\mathcal R^*$ by the
total cost of (\ref{eq:excess}).
  We will show that the latter is
at most $c'\epsilon$ times the cost of $\mathcal R^*$ where $c'$ is a
constant.

For each nonleaf cluster $C$, the partial solution $\tilde {\mathcal R}_C$ is
the output of {\sc CombineSolutions}$_C(\mathcal S, \tilde {\mathcal R}_{C_1},
\tilde {\mathcal R}_{C_2})$ where $C_1$ and $C_2$ are the children of
$C$ and $\mathcal S$ is a feasible solution to the min-cost flow instance.

Therefore, it follows by induction that, in the dynamic program, for
each cluster $C$ the cost of $T_C[\kappa_C]$ is at most the cost of
$\tilde {\mathcal R}_C$.  In particular, for $C=\hat
C$, the root cluster, we will show that the cost of (\ref{eq:excess})
is at most $c'\epsilon$ times the cost of $\mathcal R^*$ where $c'$ is a
constant, so the cost of $T_{\hat C}[\kappa_{\hat C}]$ is at most
$1+c'\epsilon$ times optimal.

To bound the cost of (\ref{eq:excess}) for $C=\hat C$, we will show
how to construct a
rooted forest $\mathcal F$.  Each leaf is a triple $(R, C, v)$ where
$R$ is a route in $\mathcal R^*$, $C$ is a cluster, and $v$ is a
vertex inside $C$ at which $R$ does a delivery.  Because each such
route $R$ delivers to at most $Q$ vertices $v$, each vertex $v$ is
inside at most $c\log n$ clusters $C$, each tour $R\in \mathcal R^*$
appears in a leaf at most $Q c \log n$ times.  Thus, if we define the
cost of $(T,C,v)$ to be the cost of $T$,
$$(cQ\log n)\ \cost(\mathcal T^*) \geq \text{total cost of the leaves}$$

Each nonleaf of $\mathcal F$ is an element of (\ref{eq:excess}), and
has at least $\epsilon^{-1}cQ\log n$ children, each of which has
cost at least that of its parent.  Hence
$$\text{total cost of the leaves} \geq \left(\epsilon^{-1}cQ\log n -
  1\right) \cdot \text{total cost of nonleaf tours}$$
which implies that the sum of costs of (\ref{eq:excess}) is at most
$\epsilon\left(1-\frac{\epsilon}{cQ \log n}\right)$ times the optimal value.

\paragraph{Construction}

For each leaf cluster $C=\set{e}$, one can derive a partial solution
${\mathcal R}_C$ good for $C$ such that
$m(e, {\mathcal R}_C)$ is the number of occurrences of $e$ in tours in $\mathcal R^*$
where a delivery is made to an endpoint of $e$ inside $C$. However, the induced configuration is not necessarily constrained.  We
obtain a solution $\tilde {\mathcal R}_C$ from ${\mathcal R}_C$ by
adding a set $\mathcal X_C$ of degenerate paths:
\begin{mdframed}
\begin{tabbing}
initialize $\mathcal X_C$ to empty\\
for each vertex $p\in e$, for each $q=0,1,\ldots, Q$,\\
\quad \= let $f^{\out}_{p,q}$ be the outflow of ${\mathcal R}_C$ at $p$ with delivery $q$\\[1ex]
\> let $\Delta^\out_{p,q}$ be the smallest nonnegative integer  such that $f^{\out}_{p,q}+\Delta^\out_{p,q}$ is the ceiling \\\hspace{1cm}of a power of
$1+\frac{\epsilon}{cQ\log |V|}$\\[1ex]
\> add $\Delta^\out_{p,q}$ copies of the route $q\ r\ 0\ p$ to $\mathcal X_C$\\[1ex]
\> let $f^{\ins}_{p,q}$ be the inflow of ${\mathcal R}_C$ at $p$ with delivery $q$\\[1ex]
\> let $\Delta^\ins_{p,q}$ be the smallest nonnegative integer  such that $f^{\ins}_{p,q}+\Delta^\ins_{p,q}$ is the ceiling  \\\hspace{1cm}of a power of
$1+\frac{\epsilon}{cQ\log |V|}$\\[1ex]
\> add the route $q\ p\ 0\ \ r$ to $\mathcal X_C$
\end{tabbing}
\end{mdframed}
The construction ensures that $\tilde {\mathcal R}_C$ induces a
constrained configuration $\kappa_C$.  Furthermore, 
for each $p\in e$, for each
$q=0,1,\ldots, Q$, $\Delta^\out_{p,q} \leq \frac{\epsilon}{cQ\log n} f^{\out}_{p,q}$
so each of the corresponding $\Delta^\out_{p,q}$ degenerate routes can be assigned 
$\epsilon^{-1} cQ\log n$ routes that pass through $p$ and do
deliveries to one of the vertices in $C$.  Thus the lowest levels of
the forest $\mathcal F$ are defined.

The construction for a nonleaf cluster $C$ is similar.
There is a natural partial solution for $C$ that is derived from
$\tilde {\mathcal R}_{C_1}$ and $\tilde {\mathcal R}_{C_1}$ and
$\mathcal R^*$ but the configuration induced is not necessarily
constrained, so degenerate routes are added to bump the integers in
the configuration up to ceilings of powers of
$1+\frac{\epsilon}{cQ\log |V|}$.  Each degenerate route
starting/ending at portal $p$ can be assigned $\epsilon^{-1} cQ\log n$
routes in $\tilde {\mathcal R}_{C_1}$ and $\tilde {\mathcal R}_{C_2}$
that pass through $p$; these are the children of the degenerate route
in $\mathcal F$.

\section{Lower bound for deterministic embedding into bounded treewidth graphs}\label{sec:emblowerbound}

Given an unweighted graph $H=(V,E)$, denote by $H_k$ the $k$-subdivision of $H$ (the graph where each edge is replaced by a $k$-path).
Our proof is based on the following lemma by Carrol and Goel \cite{CG04}:
\begin{lemma}[\cite{CG04} Lemma 1]\label{lem:CG04}
	Let $G$ be a (possibly weighted) graph that excludes $H$ as a minor. Every dominating embedding from $H_k$ to $G$ has multiplicative distortion at least $\frac{k-3}{6}$.
\end{lemma}

We now prove~\Cref{thm:LB}; we start by restating the theorem.

\EmLowerBound*

\begin{proof}
	Set $k=90$. For every $n\in \N$, let $H^n$ be the unweighted graph consisting of an $n\times n$ grid, with an additional vertex $\psi$ who is a common neighbor of all other vertices. 
	Define $\mathcal{H}=\{H^n_k\mid n\in \N\}$. 
	Note that $H^n_k$ has $|H^n_k|=\Theta(n^2\cdot k)=\Theta(n^2)$ vertices.
	In addition, all the graphs in $\mathcal{H}$ are $K_6$ free.
	Furthermore, for each $n$, $H^n$ has diameter $2$, and thus $H^n_k$ has diameter $2k+2\left\lfloor \frac{k}{2}\right\rfloor \le3k$.
	
	Consider some $H^n_k\in\mathcal{H}$, as $H^n_k$ includes the $n\times n$ as a minor, it has treewidth at least $n$ (see e.g. \cite{RS86}). Using \Cref{lem:CG04}, every embedding $f$ of $H^n_k$ into a graph $G$ with treewidth at most $o(\sqrt{|V(H_k^n)|})<n-1$ will have multiplicative distortion at least $\frac{k-3}{6}$. In particular, there is an edge $(u,v)$ in $H^n_k$ with such a multiplicative distortion (as otherwise using the triangle inequality all pairs of vertices will have distortion smaller than $\frac{k-3}{6}$). We conclude:
	\[
	d_{G}(f(u),f(v))\ge\frac{k-3}{6}=d_{H_{k}^{n}}(u,v)+\frac{k-9}{6}\ge d_{H_{k}^{n}}(u,v)+\frac{k-9}{18k}\cdot D=d_{H_{k}^{n}}(u,v)+\frac{1}{20}\cdot D~.
	\]
\end{proof}

\section{Conclusion}

We have proved two structural results for minor-free graphs: (a) a subset spanner with constant lightness exists and (b) there is a stochastic embedding of diameter $D$ minor-free graphs into treewidth $O(\frac{\log n}{\epsilon^2})$ graphs  and additive distortion $\epsilon D$. 
The results are obtained from a new multi-step framework for designing algorithms in minor-free graphs, which we believe is of independent interest. There are two major algorithmic applications of our structural results: an EPTAS for TSP and the first QPTAS for the vehicle routing with bounded capacity problem, both in minor-free graphs. 

We also provide an efficient FPT approximation scheme for the vehicle routing with bounded capacity problem in bounded treewidth graphs. As corollaries, we obtain EPTASes for the same problem in planar graphs, bounded genus graphs and graphs with bounded highway dimension. Major open problems from our work are:

\begin{enumerate}
	\item  Can a minor-free graph of diameter $D$ be stochastically embedded into a graph with treewidth $c(\epsilon)$  and distortion $\epsilon D$ in polynomial time, where $c(\epsilon)$ only depends on $\epsilon$? If the answer to this question is positive, one can immediately get a PTAS for the vehicle routing problem with bounded capaciy in minor-free graphs.
	\item Can one design a PTAS or QPTAS for Steiner tree, Steiner forest, surviviable network design problems in minor-free graphs?
\end{enumerate}

\paragraph{Acknowledgement} Part of this work was initiated at the workshop \emph{The Traveling Salesman Problem: Algorithms $\&$ Optimization} at Banff International Research Station; Hung Le and Vincent Cohen-Addad thank the organizers of the workshop and Banff for their hospitality.
Arnold Filtser is supported by the Simons Foundation.
Philip Klein is supported by NSF Grant CCF-1841954. Hung Le is supported by an NSERC grant, a
 PIMS postdoctoral fellowship, and a start-up grant from Umass Amherst.

	\bibliographystyle{alphaurlinit}
	\bibliography{spanner}
	
	\pagebreak
	\appendix
\section{Additional Notation}\label{appendix:additionalNotation}
	
	\begin{figure}[]
		\centering{\includegraphics[width=0.8\textwidth]{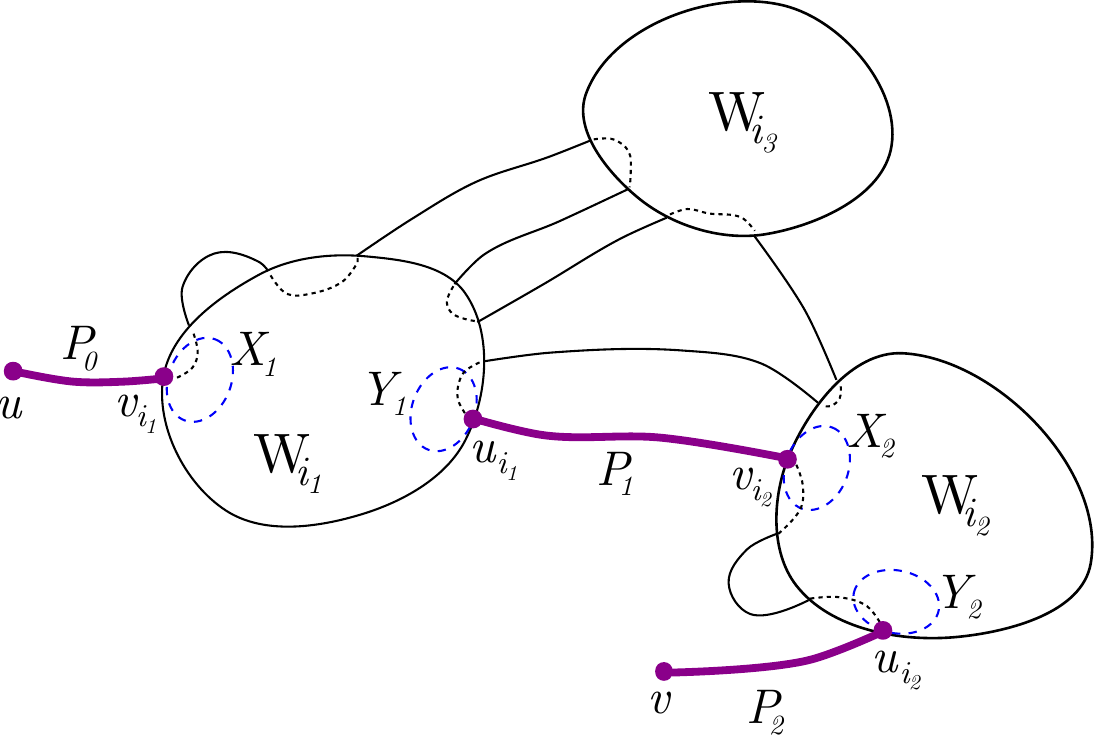}} 
		\caption{\label{fig:vortex-path-complex}\small \it 
			An example of a vortex path (purpule-higlighted)  $\mathcal{V}[u,v]=P_0\cup X_1\cup Y_1\cup P_1 \cup X_2 \cup Y_2\cup P_{2}$ induced by a path $P[u,v]$ between $u$ and $v$. It could be that a vortex $W_{i_3}$ contains a vertex of $P[u,v]$ but is disjoint from the vortex path  $\mathcal{V}[u,v]$
		}
	\end{figure}	
	
	\paragraph{Tree decomposition}  	\emph{A tree decomposition} of $G(V,E)$, denoted by $\mathcal{T}$, satisfying the following conditions: 
	\begin{enumerate} [noitemsep,nolistsep]
		
		\item Each node $i \in V(\mathcal{T})$ corresponds to a subset of vertices $X_i$ of $V$  (called bags), such that $\cup_{i \in V(\mathcal{T})}X_i = V$.
		\item For each edge $uv \in E$, there is a bag $X_i$ containing both $u,v$.
		\item For a vertex $v \in V$, all the bags containing $v$ make up a subtree of $\mathcal{T}$.  
	\end{enumerate}

	The \emph{width} of a tree decomposition $\mathcal{T}$ is $\max_{ i \in V(\mathcal{T})}|X_i| -1$ and the treewidth of $G$, denoted by $\tw$, is the minimum width among all possible tree decompositions of $G$.  A \emph{path decomposition} of a graph $G(V,E)$ is a tree decomposition where the underlying tree is a path. The pathwidth of $G$, denoted by $\pw$, defined accordingly.
	
	\begin{multicols}{2}
		\footnotesize 
		\subsection{Key notation for \Cref{sec:oneVortex} }\label{appendix:key}
		\vspace{-3pt}
		\begin{description}
			\item[$G=(V,E,w)$] : planar graph with a single vortex.
			\item[$K$] : terminal set of size $k$.
			\item[$D=O_h(L)$] : the diameter of $G$.
			\item[$G_\Sigma$] : the embedded part.
			\item[$W$] : vortex.
			\item[$\{X_1,\dots,X_t\}$] : path decomposition of $W$ of width $h$.
			\item[$\{x_1,\dots,x_t\}$] : perimeter vertices.
			\item[{$\mathcal{V}\left[u,v\right]=P_0\cup X\cup Y\cup P_1$}] : vortex path.
			\item[{$\bar{\mathcal{V}}[u,v]$}] : projection of a vortex path.
			\item[$\tilde{x}$] : auxiliary perimeter vertex with bag $\tilde{X}=\{\tilde{x}\}$, $\tilde{x}$ is a neighbor of all the other vertices in $W$.
			\item[$T_{\Sigma}$] : shortest path tree of $G_\Sigma$ rooted at $\{x_1,\dots,x_t\}$.
			\item[$T_{\tilde{x}}=T_\Sigma\cup\{(\tilde{x},v)\mid v\in W\setminus\{\tilde{x}\}\}$] : spanning tree of $G$.
			\item[{$C=\mathcal{V}_1[r,u]\cup \mathcal{V}_2[r,v]$}] : fundamental vortex cycle.
			\item[$\mathcal{P}(C)$] : set paths constituting $C$ ($|\mathcal{P}(C)|\le2(h+1)+1$).
			\item[$\bar{C}$] : closed curve induced by $C$.
			\item[$\mathcal{I},\mathcal{E}$] : interior and exterior of $C$.
			\item[$\tau$] : hierarchical partition tree of $V$.
			\item[$\Upsilon$] : subset of $V$, and node of $\tau$.
			\item[$G_\Upsilon$] : graph associated with $\Upsilon$.
			\item[$W_\Upsilon$] : the vortex of $G_\Upsilon$.
			\item[$T_\Upsilon=T_{\tilde{x}}\cap G_\Upsilon$] : spanning tree of $G_\Upsilon$ rooted in $\tilde{x}$.
			\item[$C_\Upsilon$] : a fundamental vortex cycle of $\Upsilon$ w.r.t. $T_\Upsilon$.
			\item[$\bar{C}_\Upsilon$] : closed curve induced by $C_\Upsilon$.
			\item[$\Upsilon^{\mathcal{E}}$, $\Upsilon^{\mathcal{I}}$] : interior and exterior of 	$C_\Upsilon$ Also the children of $\Upsilon$ in $\tau$.
			\item[{$\mathcal{P}(C)$}] : set of paths constituting fund. vor. cycle $C$.
			\item[{$\mathcal{C}_\Upsilon$}] : the set of all the fundamental vortex cycles removed from the ancestors of $\Upsilon$ in $\tau$.
			\item[$\bar{\mathcal{C}}_\Upsilon$] : the set of paths constituting $\mathcal{C}_\Upsilon$.
			\item[$\mathcal{P}_\Upsilon\subseteq \mathcal{C}_\Upsilon$] : subset of shortest paths that is added to $G_\Upsilon$.
			\item[$v_Q$] : representative vertex of a path $Q\in \mathcal{P}_\Upsilon$.
			\item[$\omega$] : weight function over the vertices.
			\item[$K_\Upsilon=\Upsilon\cap K$] : the set of terminals in $\Upsilon$.
		\end{description}	
	\end{multicols}

\section{Missing Proofs}

\subsection{Proof of \Cref{clm:path-intersect-vortex-face}}\label{app:cycle-intersect-vortex-face}

We begin by restating the claim

\CycleIntersectVortexFace*	
	
\begin{proof}
First, we observe that 
	\begin{equation}\label{dist:boundary}
	d(x,y) = 0 \qquad \forall x,y \in F_i, 1\leq i \leq v(G)
	\end{equation}
	since edges $(f_i,x),(f_i,y)$ have weight $0$. 	
	
	\begin{observation}\label{clm:path-intersect-vortex-face}
		For any $u \in V(K_{\Sigma})$ and any $i \in [1,v(G)]$, $|T[r,u] \cap F_i| \leq 2$ and if $|T[r,u] \cap F_i| = 2$, then ${x_1,f_i,x_2}$ is a subpath of $T[r,u]$ where $T[r,u]\cap F_i = \{x_1,x_2\}$.
	\end{observation}
	\begin{proof}
		Suppose that $|T[r,u] \cap F_i|  \geq 3$, then there must be two vertices $x_1,x_2 \in F_i$ such that $T[x_1,x_2] \subseteq T[r,u]$ does not go through $f_i$. Since $T[x_1,x_2]$ is a shortest path of positive length, this contradicts \Cref{dist:boundary}. This argument also implies that if $T[r,u]\cap F_i = \{x_1,x_2\}$, then $\{x_1,f_i,x_2\}$ must be a subpath of $T[r,u]$.
	\end{proof}

		If $r_0 \not= f_i$, then by Claim~\ref{clm:path-intersect-vortex-face}, there is only one path among $T[r_0,u], T[r_0,v]$ that can contain a vertex of $F_i$; otherwise, both paths share the same vertex $f_i$ which contradicts that they are vertex disjoint. If $r_0 = f_i$, then $|T[r_0,u]\cap F_i| = |T[r_0,v]\cap F_i| = 1$. Thus, the claim holds. \qed 
\end{proof}

\end{document}